\documentclass[%
reprint,
superscriptaddress,
amsmath,amssymb,
aps,
prd,
noeprint,
floatfix,
longbibliography
]{revtex4-2}
\usepackage{color}
\usepackage[normalem]{ulem}
\usepackage[utf8]{inputenc}
\usepackage{graphicx}% Include figure files
\usepackage{siunitx}
\DeclareSIUnit\ppm{ppm}
\usepackage{dcolumn}% Align table columns on decimal point
\usepackage{bm}% bold math
\usepackage{physics}
\usepackage{stmaryrd}% intervals
\usepackage{upgreek}
\usepackage{pifont}
\usepackage{hyperref}% add hypertext capabilities
\usepackage[super]{nth}
\usepackage[capitalise]{cleveref}
\usepackage[caption=false]{subfig}
\usepackage{booktabs}
\graphicspath{{./figs/}}

\newcommand{\intInterval}[2]{\llbracket #1 ; #2 \rrbracket}

\newcommand{\tritium}{\mathrm{T}}

\newcommand{\bohr}{{a_0}}
\newcommand{\phaseSpace}[2]{{#1} \sqrt{{#1}^2 - #2^2}}
\DeclareMathOperator{\betaRate}{R}
\DeclareMathOperator{\heavyside}{\Theta}
\newcommand{\neutrinoEnergy}{\varepsilon_f}
\newcommand{\transitionMatrix}{T}

\newcommand{\matrixElementGen}[1]{\abs{#1}^2}
\newcommand{\matrixElement}[1]{\matrixElementGen{\transitionMatrix_f^\mathrm{#1}}}
\newcommand{\reducedEndpoint}{{E_0}}
\newcommand{\cabibbo}{{\Theta_\mathrm{C}}}
\newcommand{\couplingGen}[2]{{#1}_\mathrm{#2}}

\newcommand{\Coupling}[1]{\couplingGen{G}{#1}}
\newcommand{\couplingSqGen}[2]{\couplingGen{#1}{#2}^2}

\newcommand{\CouplingSq}[1]{\couplingSqGen{G}{#1}}
\newcommand{\TT}{{\tritium_2}}
\newcommand{\HT}{{\mathrm{HT}}}
\newcommand{\DT}{{\mathrm{DT}}}
\newcommand{\HeT}{${}^3\mathrm{HeT}^+$}
\newcommand{\HeH}{${}^3\mathrm{HeH}^+$}
\newcommand{\HeD}{${}^3\mathrm{HeD}^+$}
\newcommand{\He}{${}^3\mathrm{He}^+$}

\newcommand{\ezero}{\mbox{$E_0$}}
\newcommand{\mtwonue}{\mbox{$m^2_\nu$}}
\newcommand{\mnue}{\mbox{$m_\nu $}}
\newcommand{\mnuesq}{\mbox{$m_\nu^2 $}}

\newcommand{\mnui}{\mbox{$m_{i}$}}

\newcommand{\me}{\mbox{$m_\mathrm{e}$}}
\newcommand{\bspec}{$\upbeta$ spectrum}
\newcommand{\belec}{$\upbeta$ electron}
\newcommand{\bdecay}{$\upbeta$ decay}
\newcommand{\bspech}{$\upbeta$-spectrum}
\newcommand{\belech}{$\upbeta$-electron}
\newcommand{\bdecayh}{$\upbeta$-decay}%\newcommand{\krm}{$^\mathrm{83m}$Kr}
\newcommand{\krm}{\textsuperscript{83m}Kr}

\newcommand{\rbg}{$R_\mathrm{bg}$}

\newcommand{\ssyst}{$\sigma_{\mathrm{syst}}$}
\newcommand{\sstat}{$\sigma_{\mathrm{stat}}$}

\newcommand{\rbeta}{$R_{\upbeta}$(E)}
\newcommand{\params}{\boldsymbol{\theta}}
\newcommand{\paramsbest}{\boldsymbol{\hat{\theta}}}
\newcommand{\paramsdata}{\boldsymbol{\hat{\theta}_{\text{data}}}}

\newcommand{\paramsmcstat}{\boldsymbol{\hat{\theta}_{\text{sample}}^{\text{stat}}}}
\newcommand{\paramsmcsyst}{\boldsymbol{\hat{\theta}_{\text{sample}}^{\text{syst}}}}
\newcommand{\paramsmctot}{\boldsymbol{\hat{\theta}_{\text{sample}}^{\text{tot}}}}
\newcommand{\nuisance}{\boldsymbol{\eta}}
\newcommand{\nuisancemc}{\boldsymbol{\eta}_{\text{sample}}}
\newcommand{\likelihood}{\mathcal{L}}
\newcommand{\asi}[4]{#1_{-#2}^{+#3}\,\si{#4}}

\makeatletter
\newcommand*{\transpose}{%
  {\mathpalette\@transpose{}}%
}
\newcommand*{\@transpose}[2]{%
  \raisebox{\depth}{$\m@th#1\intercal$}%
}
\makeatother

\begin{document}
%\linenumbers\relax % Commence numbering lines

%    \preprint{APS/123-QED}

    \title{Analysis methods for the first KATRIN neutrino-mass measurement}

    \newcommand{\berlin}{Institut f\"{u}r Physik, Humboldt-Universit\"{a}t zu Berlin, Newtonstr. 15, 12489 Berlin, Germany}
\newcommand{\bonn}{Helmholtz-Institut f\"{u}r Strahlen- und Kernphysik, Rheinische Friedrich-Wilhelms-Universit\"{a}t Bonn, Nussallee 14-16, 53115 Bonn, Germany}
\newcommand{\cmu}{Department of Physics, Carnegie Mellon University, Pittsburgh, PA 15213, USA}
\newcommand{\cwru}{Department of Physics, Case Western Reserve University, Cleveland, OH 44106, USA}
\newcommand{\etp}{Institute of Experimental Particle Physics~(ETP), Karlsruhe Institute of Technology~(KIT), Wolfgang-Gaede-Str. 1, 76131 Karlsruhe, Germany}
\newcommand{\fulda}{University of Applied Sciences~(HFD)~Fulda, Leipziger Str.~123, 36037 Fulda, Germany}
%
%%% BEGIN: KIT institutes
%
%\newcommand{\iap}{Institute for Nuclear Physics~(iap), Karlsruhe Institute of Technology~(KIT), Hermann-von-Helmholtz-Platz 1, 76344 Eggenstein-Leopoldshafen, Germany}
\newcommand{\iap}{Institute for Astroparticle Physics~(IAP), Karlsruhe Institute of Technology~(KIT), Hermann-von-Helmholtz-Platz 1, 76344 Eggenstein-Leopoldshafen, Germany}
\newcommand{\ipe}{Institute for Data Processing and Electronics~(IPE), Karlsruhe Institute of Technology~(KIT), Hermann-von-Helmholtz-Platz 1, 76344 Eggenstein-Leopoldshafen, Germany}
\newcommand{\itep}{Institute for Technical Physics~(ITEP), Karlsruhe Institute of Technology~(KIT), Hermann-von-Helmholtz-Platz 1, 76344 Eggenstein-Leopoldshafen, Germany}
\newcommand{\tlk}{Tritium Laboratory Karlsruhe~(TLK), Karlsruhe Institute of Technology~(KIT), Hermann-von-Helmholtz-Platz 1, 76344 Eggenstein-Leopoldshafen, Germany}
\newcommand{\ppq}{Project, Process, and Quality Management~(PPQ), Karlsruhe Institute of Technology~(KIT), Hermann-von-Helmholtz-Platz 1, 76344 Eggenstein-Leopoldshafen, Germany    }
%
%%% END: KIT Institutes
%
\newcommand{\inr}{Institute for Nuclear Research of Russian Academy of Sciences, 60th October Anniversary Prospect 7a, 117312 Moscow, Russia}
\newcommand{\lbnl}{Institute for Nuclear and Particle Astrophysics and Nuclear Science Division, Lawrence Berkeley National Laboratory, Berkeley, CA 94720, USA}
\newcommand{\madrid}{Departamento de Qu\'{i}mica F\'{i}sica Aplicada, Universidad Autonoma de Madrid, Campus de Cantoblanco, 28049 Madrid, Spain}
\newcommand{\mainz}{Institut f\"{u}r Physik, Johannes-Gutenberg-Universit\"{a}t Mainz, 55099 Mainz, Germany}
\newcommand{\mpp}{Max-Planck-Institut f\"{u}r Physik, F\"{o}hringer Ring 6, 80805 M\"{u}nchen, Germany}
\newcommand{\massit}{Laboratory for Nuclear Science, Massachusetts Institute of Technology, 77 Massachusetts Ave, Cambridge, MA 02139, USA}
\newcommand{\mpik}{Max-Planck-Institut f\"{u}r Kernphysik, Saupfercheckweg 1, 69117 Heidelberg, Germany}
\newcommand{\muenster}{Institut f\"{u}r Kernphysik, Westf\"alische Wilhelms-Universit\"{a}t M\"{u}nster, Wilhelm-Klemm-Str. 9, 48149 M\"{u}nster, Germany}
\newcommand{\npi}{Nuclear Physics Institute of the CAS, v. v. i., CZ-250 68 \v{R}e\v{z}, Czech Republic}
\newcommand{\unc}{Department of Physics and Astronomy, University of North Carolina, Chapel Hill, NC 27599, USA}
\newcommand{\washington}{Center for Experimental Nuclear Physics and Astrophysics, and Dept.~of Physics, University of Washington, Seattle, WA 98195, USA}
\newcommand{\wuppertal}{Department of Physics, Faculty of Mathematics and Natural Sciences, University of Wuppertal, Gau{\ss}str. 20, 42119 Wuppertal, Germany}
\newcommand{\saclay}{IRFU (DPhP \& APC), CEA, Universit\'{e} Paris-Saclay, 91191 Gif-sur-Yvette, France }
\newcommand{\tum}{Technische Universit\"{a}t M\"{u}nchen, James-Franck-Str. 1, 85748 Garching, Germany}
\newcommand{\tunl}{Triangle Universities Nuclear Laboratory, Durham, NC 27708, USA}
%
%%% BEGIN: other institutions
%
\newcommand{\ornl}{Also affiliated with Oak Ridge National Laboratory, Oak Ridge, TN 37831, USA}
%
%\newcommand{\swansea}{Department of Physics, Swansea University, Singleton Park, Swansea SA2 8PP, United Kingdom}
%\newcommand{\ucsb}{Department of Physics, University of California at Santa Barbara, Santa Barbara, CA 93106, USA}
%
%%% END: other institutions
%

\affiliation{\tlk}
\affiliation{\tum}
\affiliation{\saclay}
\affiliation{\ipe}
\affiliation{\etp}
\affiliation{\iap}
\affiliation{\inr}
\affiliation{\muenster}
\affiliation{\mpik}
\affiliation{\mpp}
\affiliation{\unc}
\affiliation{\tunl}
\affiliation{\madrid}
\affiliation{\wuppertal}
\affiliation{\washington}
\affiliation{\npi}
\affiliation{\itep}
\affiliation{\massit}
\affiliation{\cmu}
\affiliation{\lbnl}
\affiliation{\fulda}
\affiliation{\berlin}
\affiliation{\ppq}

%\affiliation{\bonn}
%\affiliation{\cwru}
%\affiliation{\mainz}
%\affiliation{\ppq}
%\affiliation{\mainz}

\author{M.~Aker}\affiliation{\tlk}
\author{K.~Altenm\"{u}ller}\affiliation{\tum}\affiliation{\saclay}
\author{A.~Beglarian}\affiliation{\ipe}
\author{J.~Behrens}\affiliation{\etp}\affiliation{\iap}
\author{A.~Berlev}\affiliation{\inr}
\author{U.~Besserer}\affiliation{\tlk}
\author{B.~Bieringer}\affiliation{\muenster}
\author{K.~Blaum}\affiliation{\mpik}
\author{F.~Block}\affiliation{\etp}
\author{B.~Bornschein}\affiliation{\tlk}
% ...

\collaboration{KATRIN Collaboration}\noaffiliation

    \date{\today}

    \begin{abstract} 
        We report on the data set, data handling, and detailed analysis techniques of the first neutrino-mass measurement by the Karlsruhe Tritium Neutrino (KATRIN) experiment, which probes the absolute neutrino-mass scale via the \bdecayh{} kinematics of molecular tritium. The source is highly pure, cryogenic T$_2$ gas. The \belec{}s are guided along magnetic field lines toward a high-resolution, integrating spectrometer for energy analysis. A silicon detector counts \belec{}s above the energy threshold of the spectrometer, so that a scan of the thresholds produces a precise measurement of the high-energy spectral tail. After detailed theoretical studies, simulations, and commissioning measurements, extending from the molecular final-state distribution to inelastic scattering in the source to subtleties of the electromagnetic fields, our independent, blind analyses allow us to set an upper limit of \SI{1.1}{\electronvolt} on the neutrino-mass scale at a 90\% confidence level. This first result, based on a few weeks of running at a reduced source intensity and dominated by statistical uncertainty, improves on prior limits by nearly a factor of two. This result establishes an analysis framework for future KATRIN measurements, and provides important input to both particle theory and cosmology.
    \end{abstract}

%    \keywords{Suggested keywords}
    \maketitle

    \tableofcontents

    \section{Introduction} 
    \label{sec:introduction}

    The absolute mass scale of the neutrino remains a key open question in contemporary physics, with far-reaching implications from cosmology to elementary particle physics. Despite numerous efforts along three complementary lines of approach (observational cosmology, the search for neutrinoless double-\bdecay{}, and direct searches using the kinematics of weak-interaction processes such as single \bdecay{} or electron capture), only upper bounds on the neutrino mass have been found so far (see, \textit{e.g.},~\cite{Lattanzi:2017ubx,Dolinski:2019nrj,Drexlin2013} for reviews on these subjects). Meanwhile, neutrino flavor-oscillation experiments (\textit{e.g.},~\cite{Fukuda:1998mi,Ahmad:2002jz}) have firmly established the existence of non-zero neutrino masses. 
    
    With the advent of precision cosmology, corresponding bounds on neutrino masses have been dramatically improved, and now form the tightest constraints available. Yet, cosmological bounds on $\sum\mnui$ (the sum of the distinct neutrino-mass eigenvalues $m_i$) are derived using the paradigm of the cosmological standard model ($\Lambda$CDM), and the values obtained vary with the selection of data sets included in the analysis. The Planck collaboration has inferred robust bounds from cosmic-microwave-background power spectra alone: $\sum \mnui <$ \SI{0.26}{\electronvolt} (\SI{95}{\percent} confidence level, CL), which can be further improved to $\sum \mnui <$ \SI{0.12}{\electronvolt} (\SI{95}{\percent} CL) by including lensing and baryon-acoustic-oscillation data~\cite{Aghanim:2018eyx}. Meanwhile, laboratory searches for neutrinoless double-\bdecay{} are sensitive to the neutrino-mass scale, under the assumption that neutrinos are Majorana particles that make the dominant contribution to the decay mechanism. Here, the observable is the coherent sum of weighted neutrino mass values $\langle m_{\beta\beta} \rangle = |\sum U_{ei}^2 m_i|$, where $U_{\mathrm{e}i}$ denotes the electron-flavor element coupled to the $i^{\mathrm{th}}$ neutrino-mass state in the neutrino mixing matrix. Presently, the most sensitive limits on $\langle m_{\beta\beta} \rangle$ are set by searches in $^{76}$Ge (GERDA, 0.07 - 0.16 eV)~\cite{GERDA:2020} and in $^{136}$Xe (KamLAND-Zen, 0.06 - 0.17 eV)~ \cite{KamLANDZen:2016}. The ranges of these 90\% confidence limits arise from uncertainties in nuclear-matrix elements.
        
    Direct laboratory-based measurements are an indispensable model-independent probe of the neutrino-mass scale, resting solely on the determination of kinematic parameters. Two weak processes particularly suitable for this quest are the electron capture of $^{163}$Ho~\cite{Gastaldo:2017edk,Giachero:2016xnn} and the \bdecay{} of tritium:
    \begin{equation}
        \mathrm{T} \rightarrow {}^3\mathrm{He}^+ + \upbeta^- + \bar{\nu}_e .
    \label{eq:beta_decay}
    \end{equation}

    The kinematics of these decays provide access to the effective neutrino-mass square value, an incoherent sum over the weighted squares of the mass values $m_i$ ($i = 1, 2, 3$):
        \begin{equation}
        \mnuesq = \sum_i{|U_{\mathrm{e}i}}|^2 m_i^2.
    \end{equation}

    Historically, the Mainz and Troitsk experiments used tritium to set the previous most stringent direct upper limit at $\mnue <$ \SI{2}{\electronvolt} (\SI{95}{\percent} CL)~\cite{Kraus2005,Aseev2011} with a high-accuracy shape measurement of the \bdecayh{} spectrum in the vicinity of its kinematic endpoint ($E_0 =$ \SI{18.57}{\kilo\electronvolt} for molecular tritium, T\textsubscript{2}). Meanwhile, the mass splittings measured in oscillation experiments impose a lower limit on this observable.
Depending on the ordering of the pattern of neutrino-mass eigenstates $\nu_i$, this floor is either approximately \SI{8}{\milli\electronvolt} (normal ordering) or \SI{50}{\milli\electronvolt} (inverted ordering) -- see, \textit{e.g.}, Ref.~\cite{Esteban:2018azc}.

    The Karlsruhe Tritium Neutrino (KATRIN) experiment~\cite{Osipowicz:2001sq,KDR2004} is further improving this approach to target a neutrino-mass sensitivity of \SI{0.2}{\electronvolt} (\SI{90}{\percent} CL) after five years of measurement time; note the change to 90\% confidence level. This goal requires an improvement of about two orders of magnitude in the $\mnuesq{}$ observable.
    To accomplish this challenging measurement, KATRIN relies on the proven technology of the MAC-E filter (Magnetic Adiabatic Collimation with an Electrostatic filter, developed for neutrino-mass measurements by the Mainz and Troitsk groups~\cite{Lobashev:1985mu, Picard1992}) and a large \bdecayh{} luminosity provided by a gaseous molecular tritium source (following pioneering work at the Los Alamos experiment~\cite{Robertson:1991vn}). After commissioning and characterizing the complex \SI{70}{\metre}-long electron beamline, initially with monoenergetic calibration sources~\cite{Arenz:2018kma} and subsequently with first-tritium \belec{}s~\cite{Aker:2019qfn},  the KATRIN collaboration has recently reported an improved upper limit on the neutrino mass of $\mnue <$ \SI{1.1}{\electronvolt} (\SI{90}{\percent} CL) based on an initial four-week science run~\cite{Aker:2019uuj}. This result yields an improvement of about a factor of two with respect to the best previous direct bound.

    In this work, we present a detailed account of the data set acquired, data-handling and analysis techniques applied, and statistical inference methods employed to derive this result. In the following we will use the term ``KATRIN Neutrino Mass run 1'' (KNM1) to label the inaugural four-week science campaign that marks the first operation of KATRIN at high tritium purity, at about a quarter of the nominal tritium source strength. During KNM1, an integrated \bspec{} was acquired over a ``full'' energy interval stretching from about \SI{90}{eV} below to about \SI{50}{eV} above the endpoint \ezero{}. The actual neutrino-mass analysis was performed in a narrower interval, [\ezero --\SI{37}{eV}, \ezero +\SI{49}{eV}], in which the measurement is statistics-dominated. Within this \SI{86}{\electronvolt} analysis interval, the data set comprises a total ensemble of \SI{2.03e6}{events} after data-quality selection cuts. The ensemble was collected over a measurement time of \SI{521.7}{\hour} and is composed of \num{1.48e6} \bdecay{} electrons below $E_0$ and \SI{0.55e6}{events} in a flat background over the entire analysis interval.

    We begin this paper with an overview of the experimental setup (Sec.~\ref{sec:setup}) and the configuration in which the KATRIN beamline was operated, including data handling and measurement strategy (Sec.~\ref{sec:knm1_campaign}). (For reference, Table~\ref{tab:glossary} lists  abbreviations frequently used in this paper.) Two key ingredients of the analysis, the \bspech{} model and the instrument response function, are presented in Secs.~\ref{sec:spectrum_modelling} and~\ref{sec:responsefunctionmodel}. Section~\ref{sec:background} summarizes relevant sources of background and their characteristics.  General underlying principles of the analysis, in which data from the individual detector pixels and \bspech{} scans are combined into a single spectrum for fitting, are given in Sec.~\ref{sec:general_analysis}.

        \begin{table}[tb]
        \centering
        \caption{Acronyms and abbreviations used in this work.}
        \label{tab:glossary}
        \begin{tabular}{ll}
            \toprule
            		ADC & Analog-to-Digital Converter \\
			CL & Confidence Level \\
			cps & counts per second \\
			DAQ & Data-Acquisition System \\
			d.o.f. & degree(s) of freedom \\
			e-gun & electron gun \\
			FSD & molecular Final-State Distribution \\
			FPD & Focal-Plane Detector \\
			HV & High Voltage \\
			KNM1 & KATRIN Neutrino Mass run 1 \\
			LARA & LAser RAman spectroscopy system \\
			$\Lambda$CDM & $\Lambda$ Cold-Dark Matter model   \\
			&	\phantom{ss} (cosmological standard model) \\
			MAC-E filter & Magnetic Adiabatic Collimation with  \\
			&	\phantom{ss}Electrostatic filter \\
			MC & Monte Carlo \\
			ppm & part per million \\
			p-value & Probability of achieving a result as extreme \\
			&	\phantom{ss} as the one found, through statistical \\
			&	\phantom{ss} fluctuation \\
			Q-value & Kinetic energy released in tritium \bdecay{}  \\
			&	\phantom{ss}(for zero neutrino mass) \\
			ROI & Region of Interest \\
			TOF & Time of Flight \\
			WGTS & Windowless Gaseous Tritium Source \\
            \bottomrule
        \end{tabular}
    \end{table}

    Section~\ref{sec:systematic_uncertainties} presents a detailed assay of individual systematic uncertainties.  Section~\ref{sec:spectralfit} documents the strategy employed for blind analysis, describes two complementary methods 
    employed to propagate the systematic uncertainties into the neutrino-mass fit, and shows the resulting spectral fit and uncertainty breakdown. Section~\ref{sec:upperlimit} details the construction of the confidence belt and the derivation of the neutrino-mass upper limit via the Feldman-Cousins~\cite{Feldman:1997qc} and the Lokhov-Tkachov~\cite{Lokhov:2015zna} approaches. Our Lokhov-Tkachov result of $\mnue <$ \SI{1.1}{\electronvolt} (\SI{90}{\percent} CL), presented in Ref.~\cite{Aker:2019uuj}, was obtained using Frequentist methods. In this work we also present a derivation of the upper limit based on Bayesian methods, yielding a limit of $\mnue<$ \SI{0.9}{\electronvolt} (90\% C.I.) (Sec.~\ref{sec:bayesian_analysis}). This method uses a different approach to deal with the unphysical region of negative neutrino-mass squared. 

    In Sec.~\ref{sec:endpoint_measurement}, as a consistency check of KATRIN's absolute energy scale, we show that the effective endpoint value $E_0$ obtained from the fit to the \bspec{} agrees with independent measurements of the Q-value through the $^3$He-T mass difference.

    We conclude by summarizing our findings (Sec.~\ref{sec:discussion}) and discussing them in the wider context of contemporary neutrino-mass probes (Sec.~\ref{sec:conclusion}).

    \section{KATRIN experimental setup}
    \label{sec:setup}

    \begin{figure*}[t!]
        \centering
        \includegraphics[width=0.8\textwidth]{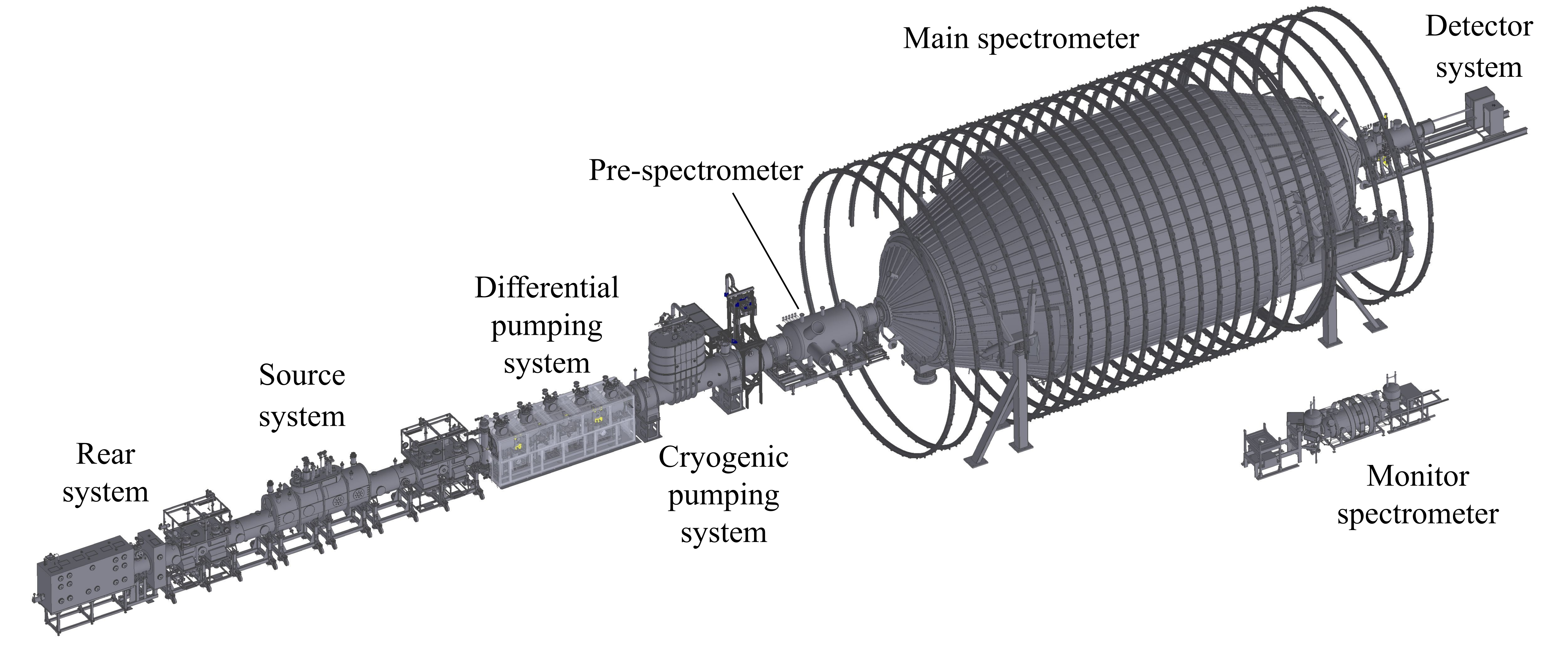}
        \caption{Overview of the \SI{70}{\metre} KATRIN beamline. Moving downstream, from left to right, the major components are: the rear system, the source system, the differential pumping system, the cryogenic pumping system, the pre-spectrometer, the main spectrometer, and the detector system. The monitor spectrometer monitors the retarding potential of the main spectrometer.}
        \label{fig:beamline}
    \end{figure*}

    Figure~\ref{fig:beamline} gives an overview of the KATRIN apparatus.  Briefly, in order to ensure sufficient statistics, a bright tritium source produces some \num{2.45e10} decays each second in the KNM1 configuration. In order to perform a fine-grained energy analysis near the tritium endpoint, the energies of the resulting \belec{}s are analyzed by a pair of MAC-E-filter spectrometers~\cite{Lobashev:1985mu, Picard1992}. These basic functions require the support of extensive systems for handling the tritium gas, maintaining vacuum conditions, ensuring adiabatic electron transport, mitigating or eliminating backgrounds, detecting \belec{}s, and calibrating and monitoring the apparatus as a whole. The resulting \SI{70}{\metre} beamline is described in detail in Ref.~\cite{KATRIN:hw2020}; here, we offer a brief summary. 
    
    T$_2$ gas from a temperature- and pressure-controlled buffer vessel at \SI{313}{\kelvin} is cooled to \SI{30}{\kelvin} and continuously injected via a capillary into the center of the source system. The resulting Windowless Gaseous Tritium Source (WGTS) freely streams to both ends of the system, where it is continuously pumped away with turbomolecular pumps. This results in a stable pressure distribution inside the source beam tube~\cite{Kuckert:2018kao}. Once the T$_2$ gas is pumped away, it flows over a PdAg membrane filter that is permeable only to hydrogen isotopes. A constant fraction of the circulating gas is also removed at this stage for later purification, and is replaced with highly pure T$_2$ directly after the filter. The purified gas is fed back to the temperature- and pressure-controlled buffer vessel, forming a closed loop. The loop system is integrated with the infrastructure of the Tritium Laboratory Karlsruhe, which provides tritium purification of exhaust gas, tritium storage, and fresh tritium supply for KATRIN~\cite{Kazachenko:FST2008,Priester:2015bfa,Priester:FST2020}.

    Within the \SI{10}{\metre}-long, \SI{90}{\milli\metre}-diameter source beam tube~\cite{Grohmann:2013ifa}, tritium decays produce \belec{}s that are guided along magnetic field lines~\cite{Arenz:2018jpa} through the rest of the experimental beamline. At the upstream end, the WGTS terminates in a gold-plated rear wall, which can be held at a fixed potential and/or illuminated with ultraviolet light to liberate photo\-electrons. At the downstream end, the windowless nature of the source is essential to avoid catastrophic energy loss, but necessitates other means for the confinement of tritium. The \belec{}s are first guided around magnetic chicanes through two pumping stages, namely a differential pumping system and a cryogenic pumping system, which collectively reduce the partial pressure of tritium by more than \num{14}~orders of magnitude~\cite{Friedel2019}. Specially designed electrodes within the differential stage~\cite{KleinPhD2018} prevent the transmission of tritium ions.

	\belec{}s must then pass through a pair of MAC-E-filter spectrometers, operated in tandem. Each MAC-E filter is characterized by strong magnetic fields at the entrance and exit, with a region of weak magnetic field in the center. Since the magnetic moment is conserved in the adiabatic transport of the electrons through the beamline, the electron momenta rotate to become approximately parallel to the magnetic field lines, producing a broad, roughly collimated beam. A longitudinal retarding potential therefore analyzes the total kinetic energy of the electrons at the central ``analyzing plane,'' at which the magnetic field is the weakest. Electrons below the resulting energy threshold are reflected upstream, toward the source; electrons above the energy threshold are transmitted downstream, toward the spectrometer exit. The transmission function of the spectrometers was extensively calibrated prior to the measurement (Sec.~\ref{sec:responsefunctionmodel}).

    The first MAC-E filter in the tandem pair, the pre-spectrometer~\cite{Prall2012}, has a fixed energy threshold at \SI{10}{\kilo\electronvolt} and removes the bulk of the low-energy electrons. Immediately downstream, the main spectrometer is the high-resolution, adjustable-threshold filter that analyzes the integral \bspec{}. Each data-taking ``scan'' (Sec.~\ref{sec:MTD}) consists of a sequence of main-spectrometer retarding-potential settings, with a new threshold of integration at each setting. The electropolished interior stainless-steel surface of the main spectrometer is lined with two layers of inner, wire electrodes, providing fine shaping of the electric fields and, when operated at a negative potential offset from the main-spectrometer vessel, electrostatic rejection of low-energy secondary electrons from the main-spectrometer surface~\cite{Valerius:PPNP2010}. The vessel potential is supplied by a commercial system, with additional regulation and post-regulation designed and built by the collaboration to suppress \SI{50}{\hertz} mains noise and other sources of interference~\cite{Katrin:HV}. Air-cooled magnetic coils, mounted on a framework surrounding the main spectrometer, compensate for the Earth's magnetic field, fringe fields of the solenoids, and residual magnetization~\cite{Glueck2013}. The ultra-high vacuum in the spectrometer is maintained by non-evaporable getter strips and turbomolecular pumps~\cite{Arenz:2016mrh}. Liquid-nitrogen-cooled copper baffles are positioned across the pump ports to suppress background electrons due to radon decay in the main volume~\cite{Drexlin2017, Goerhardt:2018wky}. To mitigate backgrounds from the Penning trap between the two MAC-E filters, a conductive electron catcher is inserted into the inter-spectrometer region at each change in the set voltage of the main spectrometer~\cite{KATRIN-Penning:2020}. This device removes trapped electrons that would produce secondary ions and electrons.

    Electrons that pass through the main spectrometer undergo additional acceleration via the post-acceleration electrode, improving rejection of non-spectrometer backgrounds. When they reach the detector system, they are counted in the focal-plane detector (FPD)~\cite{Amsbaugh:2014uca}, a monolithic silicon \emph{p-i-n} diode segmented into 148 equal-area pixels. The FPD and its readout electronics are elevated to the post-acceleration potential, and preamplified signals are transmitted to the data-acquisition (DAQ) system via optical fiber. Each FPD pulse is digitized in a 12-bit analog-to-digital converter (ADC), and its amplitude and timing are reconstructed on-line by the sequential application of two trapezoidal filters~\cite{jor94, Amsbaugh:2014uca}. These values are then recorded using the Object-oriented Real-time Control and Acquisition (ORCA) framework~\cite{how04}, which can also communicate directly with the main-spectrometer high-voltage system using a web-based database tool~\cite{Chilingaryan10}. Pulse amplitudes are translated into energies in near-time processing (Sec.~\ref{sec:data_pipeline}), based on the results of regular calibration runs with an $^{241}$Am photon source.

    Multiple calibration and monitoring systems provide essential information during both neutrino-mass scans and dedicated runs~\cite{Babutzka:2012xd}. In the tritium loops feeding the source, a laser-Raman spectroscopy system (LARA)~\cite{Sturm2010paper, Schloesser2013a, Zeller:sensors2020} monitors the relative concentrations of hydrogen isotopologs, particularly T$_2$, DT, and HT, within the source gas. In the rear system upstream of the source, an electron gun (e-gun), following the design of a similar e-gun used for testing the main spectrometer~\cite{Behrens:2017cmd}, serves as an angle- and energy-selective calibration source. This e-gun delivers electrons through an aperture in the rear wall at the upstream end of the source. Observed in the FPD, these electrons test the response function of the experiment as a whole. Two radioactive, in-vacuum calibration sources are also available: gaseous $^{83m}$Kr that can be circulated within the source when its temperature is elevated to about \SI{100}{\kelvin}~\cite{Sentkerestiova-GKrS:2018}, and a condensed $^{83m}$Kr source that can be inserted into the cryogenic pumping system~\cite{PhDBauer2014}.

    Upstream of the rear wall, the $\upbeta$-induced x-ray spectroscopy system continuously monitors the source activity: silicon drift detectors view x-rays produced by \belec{}s scattering in the rear wall~\cite{Roellig2015}. Further downstream, within the cryogenic pumping system, a forward beam monitor provides complementary activity monitoring ~\cite{Ellinger2017}. This monitor includes two silicon \emph{p-i-n} diodes for electron rate and spectrum measurements, a Hall sensor, and a temperature gauge. A vacuum manipulator allows these sensors to be positioned radially within the beam; normally, the forward beam monitor is positioned at the outer edge of the \belec{} flux.

    The main-spectrometer retarding potential, which defines the energy analysis, is continuously monitored both by a voltage divider with demonstrated part-per-million (ppm) precision~\cite{Thummler:2009rz, Bauer2013a, Arenz:2018ymp, Rest:2019} and by the refurbished MAC-E filter from the historical Mainz experiment~\cite{Kraus2005}. Now relocated to KATRIN, this monitor spectrometer references the main-spectrometer retarding potential to an atomic standard via synchronous scans of a $^{83m}$Kr conversion line~\cite{Erhard2014}.

    Prior to the KNM1 neutrino-mass run, the full KATRIN beamline was commissioned with photo\-electrons, ions, and $^{83m}$Kr conversion electrons in 2016--2017~\cite{Arenz:2018kma}, and with small amounts of tritium in D$_2$ carrier gas in 2018~\cite{Aker:2019qfn}. Subsequently, in another campaign with D$_2$, the electron gun was commissioned and gas properties of the source were investigated~\cite{Schloesser:FST2020}.    KNM1 marked the first time that the inner surfaces of the injection capillary and source system were exposed to large amounts of tritium. Radiochemical reactions between T$_2$ and these metal surfaces produced both CO and tritiated methane, which condensed on the cold metal surface of the capillary and partially obstructed tritium flow over time. To improve stability during this burn-in period, KATRIN operated at a reduced column density of $\rho d_\mathrm{exp} =$ \SI{1.11e17}{molecules\per\centi\metre\squared}.

    \section{The KNM1 measurement campaign}
    \label{sec:knm1_campaign}
    In this section we describe the operating conditions of the KATRIN experiment during its first high-purity tritium campaign (KNM1), which took place from \nth{10} April to \nth{13} May 2019. In particular, we characterize the system performance in terms of the source-gas isotopic purity (Sec.~\ref{sub:source_composition}) and column density (Sec.~\ref{ch:columnDensityDetermination}), as well as the reproducibility, homogeneity, and stability of the electron starting potential in the source (Sec.~\ref{sec:starting_potential}) and the retarding potential in the analyzing plane (Sec.~\ref{sec:ap-potentials}). We also discuss the detection of \belec{}s and the definition of a region of interest (Sec.~\ref{sec:electron_counting}) as well as the processing and analysis pipeline for the data  (Sec.~\ref{sec:data_pipeline}).

    The requirements for system stability arise from the method adopted to measure the tritium \bspec{} by repeatedly scanning the retarding potential in alternating up and down sweeps (Sec.~\ref{sec:MTD}), and from the fact that KNM1 data from all pixels and all scans are combined into a single spectrum for fitting. In the final analysis, then, experimental parameters are essentially averaged over both space (across the detector) and time (across like scan steps throughout the KNM1 data-taking period). Later on, Sec.~\ref{sec:general_analysis} explores the justification for this analysis method in the statistics-dominated KNM1 data set.

    For the KNM1 campaign, the sequence of scan steps, each consisting of a retarding-potential set point distributed in the interval $\lbrack \ezero - $\SI{91}{\electronvolt}, $\ezero + $\SI{49}{\electronvolt}$\rbrack$, resulted in a typical scan duration of \SI{2.5}{\hour}. Therein, each scan step corresponds to a measurement time varying from \SI{17}{\second} for high-rate points deeper in the spectrum to \SI{576}{\second} near the endpoint region, as will be shown later in Fig.~\ref{fig:MTD}.

    \subsection{Tritium source parameters} 
    \label{sub:source_composition}

    The average source activity during KNM1 neutrino-mass data-taking was about \SI{2.45e10}{\becquerel}, maintained by a column density of \SI{1.11e17}{molecules\per\centi\metre\squared}. This was achieved by a cumulative tritium throughput of \SI{4.9}{\gram\per\day}.

    The gas injected into the source consists mainly of molecular T$_2$. Due to initial impurities and exchange reactions with the stainless-steel piping and vessel, the other hydrogen isotopologs (H$_2$, HD, HT, D$_2$, and DT) are also present in minor fractions.
    A PdAg membrane (permeator) in the tritium loop~\cite{Bornschein:fst2005} continuously filters the circulated tritium gas to prevent the recirculation of built-up impurities.
    The relative fractions $c_x$ of the six hydrogen isotopologs are continuously monitored by LARA, downstream of the permeator. The relative molecular isotopolog fractions $c_x$ and the atomic tritium purity $\varepsilon_{\mathrm{T}}$ are defined as:
    \begin{align}
        c_x &= \frac{N_x}{\sum_i^6 N_i}\ \mathrm{, and}\\
        \varepsilon_{\mathrm{T}} &= \frac{N_{\mathrm{T_2}} +\frac{1}{2} (N_{\mathrm{HT}}+N_{\mathrm{DT}})}{\sum^6_i N_i}\
        \label{eq:tritium_purity}
    \end{align}
\noindent where $N_x$ is the number of molecules of isotopolog $x$ in the source, and the sums are over all six isotopologs. The tritium purity is monitored with better than~\num{e-3} statistical precision~\cite{Zeller:sensors2020}.

\begin{figure}[t!]
        \centering
        \includegraphics[width=0.49\textwidth]{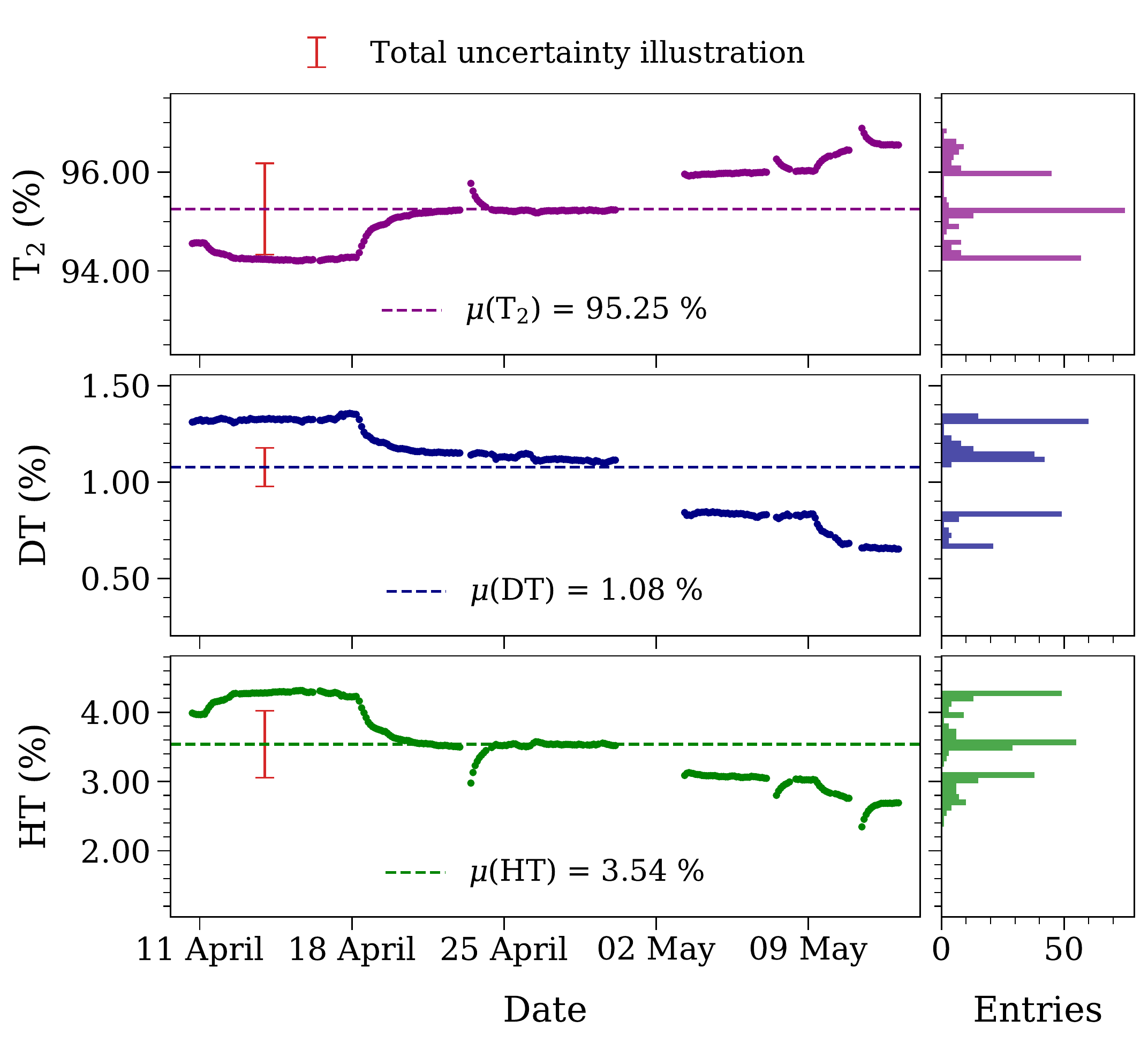}
        \caption{Evolution of the relative fractions of the three tritiated isotopologs injected into the source during KNM1. The dotted lines show the mean values over KNM1, with the red error bars illustrating the total uncertainties (statistical and systematic). The steps and kinks in the trends indicate times at which a new tritium gas batch was fed into the circulation loop. As the tritium is re-processed in several steps at the Tritium Laboratory Karlsruhe~\cite{Welte:FST2017}, its purity varies slightly between batches. }
        \label{fig:data_lara}
    \end{figure}

		The time evolution of the relative fractions of the three tritiated isotopologs injected into the source during KNM1 is shown in Fig.~\ref{fig:data_lara}.  On average, the concentrations of the tritiated species throughout the campaign were $c_{\mathrm{T_2}} = $ \num{0.953}, $c_{\mathrm{HT}} = $ \num{0.035}, and $c_{\mathrm{DT}} = $ \num{0.011}; these values are used in the final neutrino-mass analysis. The resulting tritium purity is $\varepsilon_{\mathrm{T}}= $ \num{0.9758(13)}~\cite{Zeller:sensors2020}. The prominence of HT as a secondary species is due to exchange reactions with H atoms that are naturally present in stainless-steel piping~\cite{Schloesser:FST2020}, and the residual presence of DT is due to the isotope-separation process used to purify the tritium~\cite{Dorr:2005}.
    The inactive species (H$_2$, HD, and D$_2$) are only present in trace amounts, as they are strongly suppressed by shifts of the chemical equilibrium in the presence of high-surplus T$_2$.

    \subsection{Column density} 
    \label{ch:columnDensityDetermination}
    The column density $\rho d$ determines the number of tritium atoms $N_T$ in the source
    \begin{equation}
    N_T = 2 \varepsilon_{\mathrm{T}} \cdot \rho d \cdot A,
    \end{equation}
    where $A$ is the cross-sectional area of the WGTS, and the factor of \num{2} is necessary because $\rho d$ is defined in terms of the number of T$_2$ molecules. The column density further defines the $s$-fold scattering probabilities $P_s$ of electrons, traveling parallel to magnetic field lines through the entire tritium source, with the gas molecules:
     \begin{eqnarray}
     	P_s = \frac{\left(\rho d \sigma \right)^s}{s\mathrm{!}}\mathrm{e}^{-\rho d \sigma}.
	\label{eq:scatterProb}
     \end{eqnarray}
    The product $\rho d\sigma$, where $\sigma$ is the cross section for inelastic scattering of electrons from molecular tritium (Sec.~\ref{ch:InelScattCross}), gives the expected number of scatterings. It must be known with high accuracy for the analysis~\cite{Kuckert:2018kao}.

    The precise absolute value of $\rho d \sigma$ is obtained from measurements with the narrow-angle, quasi-mono\-energetic e-gun located in the rear system. This e-gun produces a high-intensity beam of electrons via the photo\-electric effect according to the principle described in Ref.~\cite{Behrens:2017cmd}. On their path towards the detector, the electrons traverse the source, where they can undergo inelastic scattering and in the process lose energy. Only those electrons with sufficient remaining energy to surpass the spectrometer potential are counted in the detector. By measuring the electron rate at different retarding potentials and fitting a model response function (Sec.~\ref{sec:responsefunctionmodel}) to these data, we may make a precise determination of $\rho d\sigma$.

    E-gun electrons differ from $\upbeta$-electrons with regard to their starting positions and their energy and angular distributions. For this reason a modified response function, including a precise description of the e-gun beam characteristics, is used in the column-density determination. The e-gun electron rate is measured at retarding potentials where the impact of the column density is the strongest. The mean energy of the e-gun electrons is set to \SI{18.78}{\kilo\electronvolt}, allowing a clean separation from \belec{}s that could bias the column-density determination.  During KNM1, $\rho d\sigma$ was determined with the e-gun on a weekly basis, achieving relative uncertainties of less than \SI{0.9}{\percent}. 

  	As described in Sec.~\ref{sec:setup}, the first exposure of the inner loop to T$_2$ resulted in the production of gas species which condensed on the surface of the injection capillary. This obstruction caused the tritium injection flow and column density to drift over time at constant tritium injection pressure. By lowering the column density to be a factor of approximately \num{5} smaller than the nominal column density $\rho d_{\mathrm{nom}}$, and by increasing the tritium injection pressure several times during KNM1, these drifts were kept lower than \SI{3}{\percent}. To ensure precise monitoring of the column density during the whole measurement period, the e-gun measurements were combined with continuous $\rho d$ fluctuation data from a mass-flow meter with \SI{200}{sccm} full-scale range~\cite{FlowMeter}, applied to the tritium injection flow. The reproducibility of the flow meter during KNM1 is conservatively estimated to be \SI{1.5}{\upmu bar\cdot l/s}. Based on simulations that show a linear relation between $\rho d\sigma$ and the tritium injection flow for a narrow throughput range~\cite{Heizmann2018}, a linear calibration function is suitable to relate the measured throughput to $\rho d\sigma$.

\begin{figure}[t!]
        \centering
        \includegraphics[width=0.47\textwidth]{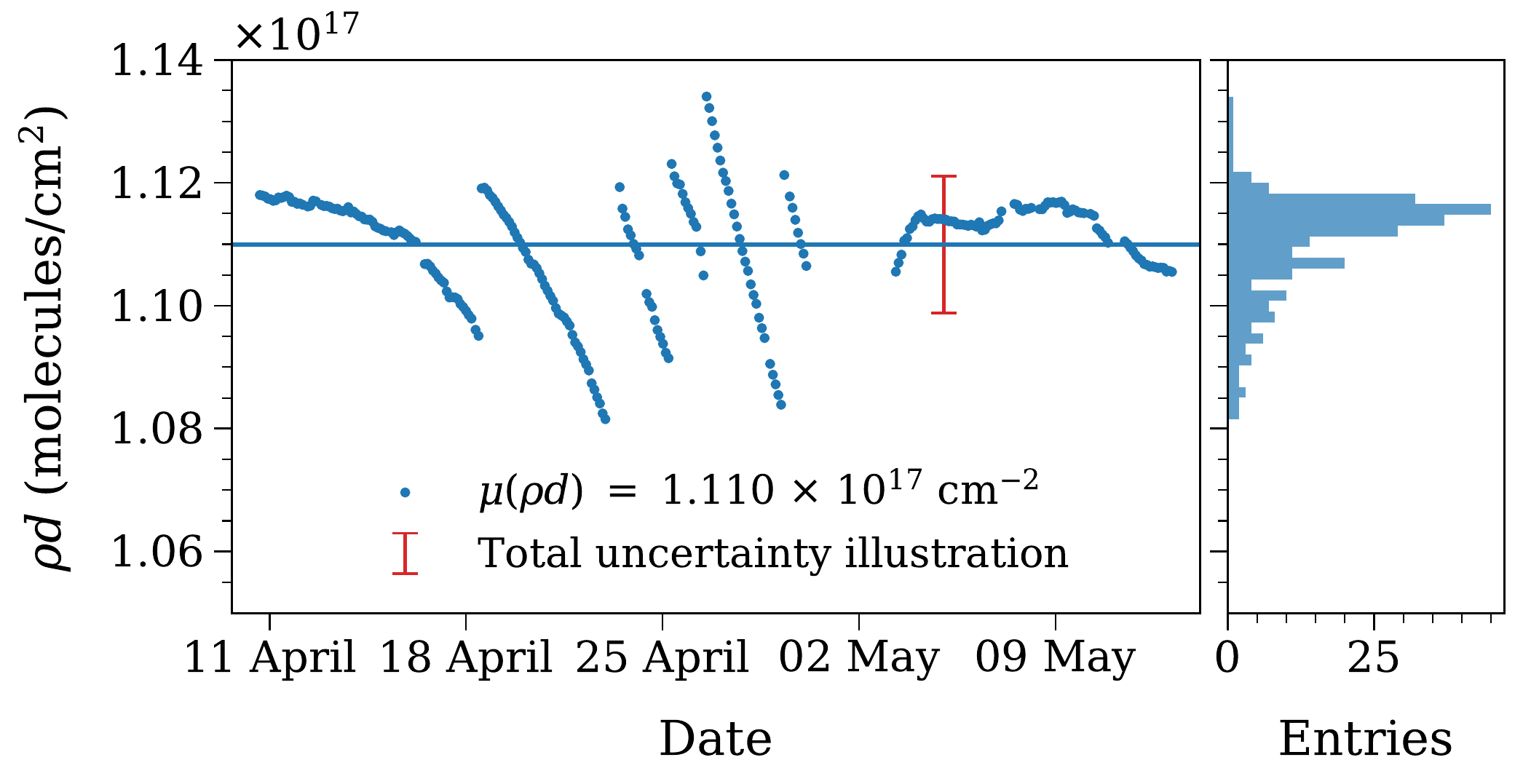}
        \caption{Evolution of the column density during KNM1; the uncertainty is dominated by systematics arising from the relationship of e-gun data to the measured throughput, and from fluctuations in the latter quantity. The visible decrease of the column density over time is caused by conductance changes of the tritium injection capillary. By increasing the tritium injection pressure several times, the column density was stabilized.}
        \label{fig:data_rhod}
    \end{figure}

    With this strategy, we determine the column density with high precision for all tritium data-taking. The time evolution and distribution of the column-density values are shown in Fig.~\ref{fig:data_rhod}; the average value of $\rho d\sigma$ is \num{0.404} at the molecular tritium endpoint. Using the cross-section value from Eq.~\ref{eq:FerencCrossSection} further below, this value translates to an average column density of $\rho d =$ \SI{1.11e17}{molecules\per\centi\metre\squared}.

    \subsection{Electron starting potential} 
    \label{sec:starting_potential}

    The starting potential of the \belec{}s is provided by a cold and strongly magnetized plasma in the WGTS. The magnitude of the potential depends on the boundary conditions at the rear wall and the grounded beam tube. By optimization of the rear-wall set voltage, a homogeneous, stable plasma potential can be created.
    This is important because both spatial inhomogeneities and temporal fluctuations of the plasma potential distort our spectrum in a manner analogous to the neutrino mass. Indeed, the shift in neutrino-mass squared due to an error $\Delta \sigma^2$ in the Gaussian variance of a  continuous variable (such as the starting potential of the \belec{}s) is  given at leading order by~\cite{Robertson:ARNP1988}:
%    For a Gaussian variance $\sigma^2$ of a continuous variable like the starting potential of the \belec{}s, the resulting shift is  given at leading order by~\cite{Robertson:ARNP1988}: %%FLAGFLAG
    \begin{align}
	    \Delta m_\nu^2 = -2\Delta \sigma^2 \; . \label{eq:sigma}
    \end{align}
\noindent Since we combine all pixels and all scans for our KNM1 fits (Sec.~\ref{sec:general_analysis}), our analysis does not account for inhomogeneities or temporal fluctuations, and the full variance of the electron starting potential therefore contributes via Eq.~\ref{eq:sigma}.

    The source plasma is generated by the weakly self-ionizing  tritium gas.
    According to simulation, each \belec\ creates on average \num{36}~secondary electrons, and thus 36 positive ions, through scattering interactions. Throughout the central part of the WGTS, the ions have a mean free path of less than \SI{0.5}{\meter} for momentum transfer with the neutral gas. Consequently, the flow of neutral tritium gas drives the ions toward both ends of the source. The low-energy, secondary electrons follow the ion motion in order to maintain quasi-neutrality, facilitated by their much higher mobility along the magnetic field lines. While the ions quickly become fully thermalized to the \si{\milli\electronvolt} scale, the energy spectrum of secondary and \belec s ranges from \si{\milli\electronvolt} to \si{\kilo\electronvolt}.

     The electric potential inside the plasma depends on the surface potentials at its boundaries. These are determined in turn by their intrinsic work functions $\phi$, which are expected to differ by several \SI{100}{\milli\volt}~\cite{Babutzka2014}, and by the applied bias voltages. As the beam tube is grounded ($U_{\mathrm{bt}} =$ \SI{0}{\volt}), only the rear-wall bias voltage $U_{\mathrm{RW}}$ remains to compensate the work-function differences. At an optimal  $U_{\mathrm{RW}}$, the radial and longitudinal inhomogeneities of the plasma potential both vanish, as expected from simulations~ with the assumption of negligible work-function inhomogeneities \cite{Kuckert2016}.
    \par

    The optimal rear-wall bias voltage was determined by measuring the $\upbeta$-rate at various $U_{\mathrm{RW}}$ settings. Comparing these rates to reference spectra, we extracted the dependence of the spectral endpoint \ezero{} on the FPD ring number --  which correlates to radius in the source.
    
		For $U_{\mathrm{RW}}=$ \SI{-150}{\milli\volt}, a flat radial \ezero{} distribution was found. Also, the measurement of the plasma-induced current on the rear wall showed no drifts and less noise than at other bias voltages. $U_{\mathrm{RW}}$ was therefore set to \SI{-150}{\milli\volt} for the measurement campaign.

    The systematic effect of remaining spatial inhomogeneities and fluctuations of the plasma potential can be constrained by studying the line widths and positions of quasi-monoenergetic conversion electrons from gaseous \krm\ co-circulating in the T$_2$ gas~\cite{Belesev:2008zz}. The L$_3$-$32$ line at \SI{30472.2(5)}{\electronvolt} is particularly interesting for this study. First, it is located above \ezero{}. Second, the \SI{37.8(5)}{\percent} branching ratio into this final state leads to a high signal-to-noise ratio~\cite{Venos:2018tnw}. Third, it possesses a small intrinsic line width of $\Gamma\approx \SI{1}{\electronvolt}$. In a previous campaign using gaseous \krm{} in the absence of tritium gas~\cite{Arenz:2018kma}, the KATRIN experiment measured an L$_3$-$32$ line position of $E_{\mathrm{L_3\text{-}32}}=30472.604\pm0.003_{\mathrm{stat}}\pm0.025_{\mathrm{sys}}\,\mathrm{eV}$ and a Lorentzian line-width of $\Gamma_{\mathrm{L_3\text{-}32}}=1.152\pm0.007_{\mathrm{stat}}\pm 0.013_{\mathrm{sys}}\,\mathrm{eV}$~\cite{Altenmueller:JPG2020}. This effective line position includes a shift arising from the absolute work-function difference between the source and the main spectrometer.

    After the KNM1 neutrino-mass campaign ended, plasma studies were performed for two days with co-circulating \krm\ and T$_2$. It should be noted that the column density during neutrino-mass measurements was only \SI{22}{\percent} of the nominal value of \SI{5.0e17}{molecules\per\centi\metre\squared}, while during the plasma study it was about \SI{30}{\percent} of the nominal value.

		The krypton admixture did not affect general plasma properties, such as charged-particle density or electric potentials, because the partial pressure and activity ($\approx \SI{3}{\mega\becquerel}$) of krypton were several orders of magnitude below those of tritium ($\approx \SI{33}{\giga\becquerel}$).
    However, the plasma was affected by the beam-tube temperature of \SI{100}{\kelvin}, elevated from the nominal \SI{30}{\kelvin}. This higher temperature was necessary to prevent the krypton from freezing, but also increased the temperature of the dominant low-energy part of the electron energy distribution~\cite{Nastoyashchii:FST2005}. 
 The plasma temperature is known to strongly influence the rate of electron-ion recombination at the meV scale. As the recombination rate is much stronger at \SI{30}{\kelvin}, we expect plasma effects at elevated source temperature to be more prominent. We thus use results obtained during the krypton measurement at \SI{100}{\kelvin} to set an upper limit of the scale of possible plasma effects.
		The $\upbeta$-decay electrons and non-thermalized electrons make only minor contributions to the number density, but their dominant role in the energy density of charged particles requires a detailed investigation.

                The intrinsic Lorentzian line width was measured with gaseous \krm{} in the absence of tritium, with the experimental conditions as similar as possible to the L$_3$-$32$ measurements with co-circulating $\TT$/\krm{} (described above). By comparing these two measurements and assuming an energy-independent  background, we find that the presence of $\TT$ results in a Gaussian line broadening of $<$ \SI{80}{\milli\volt} for rear-wall settings in the range \SI{-350}{\milli\volt}$< U_{\mathrm{RW}} <$ \SI{+350}{\milli\volt}. The collaboration is currently investigating the impact of a possible radial-dependent background, which could arise due to detector effects.

	The impact of these findings on the neutrino-mass measurement is discussed in Sec.~\ref{sec:syst_startingpot}.

    \subsection{Analyzing-plane potentials} 
    \label{sec:ap-potentials}

    The threshold energy for electrons to pass through the MAC-E filter is determined by the value of the retarding potential $U$ at the analyzing plane. Any unknown instabilities in the retarding potential directly affect the energy scale of the tritium spectrum and can introduce systematic effects on \mtwonue{}. To first order, significant, unaccounted-for continuous inhomogeneities and fluctuations effectively broaden the spectrum as seen in Eq.~\ref{eq:sigma}. Our KNM1 analysis does not account for inhomogeneities or fluctuations in $U$, so that the full variance is seen in the broadening. 
        For the target sensitivity of KATRIN, the energy scale must be stable to within \SI{60}{\milli\electronvolt} or \SI{3}{\ppm} on a baseline retarding potential of \SI{-18.6}{\kilo\volt}.

    To achieve this, we have constructed a dedicated measurement chain, including precision high-voltage dividers with proven long-term stability on the ppm level over one year~\cite{Arenz:2018ymp}. A custom-built post-regulation system~\cite{Katrin:HV} ensures stability at higher frequencies, up to \SI{1}{\mega\hertz}.

    In order to stack multiple scans for the KNM1 analysis (Sec.~\ref{sec:stacking}), it is not only necessary to have \SI{3}{\ppm} monitoring, but also to achieve comparable precision in both the stability at each scan step and the reproducibility of the retarding potential from scan to scan.

\begin{figure}[t!]
        \centering
        \includegraphics[width=0.45\textwidth]{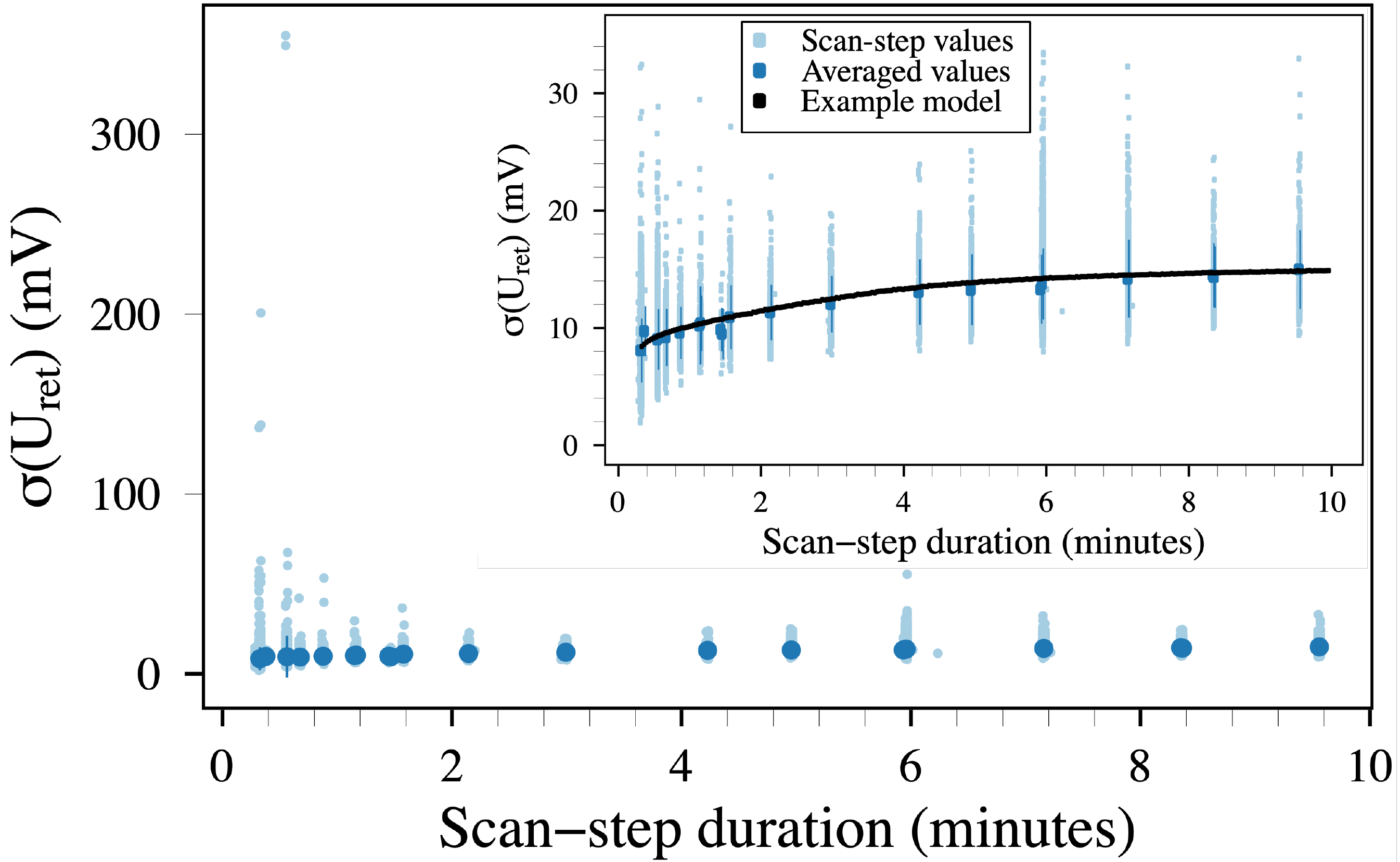}
        \caption{Achieved stability of the retarding potential as a function of the scan-step duration. Light blue points show the standard deviation of the measured retarding potential for each scan step during KNM1. Dark blue points show the mean of this standard deviation for scan steps with the same length; their error bars show the standard deviation of this value. Outliers arise from a brief period in which the change in HV setpoint was incorrectly synchronized with the DAQ, a problem affecting 0.3\% of scan steps. In the inset figure, scans with synchronization errors have been removed to show the performance of the HV subsystem. The black line gives the prediction of the statistical model described in the text.}
        \label{fig:HV_Measurement}
    \end{figure}

	Figure~\ref{fig:HV_Measurement} shows the achieved high-voltage stability while acquiring data at individual scan steps over the full measurement interval. This stability is on average below \SI{15}{\milli\volt}, significantly exceeding requirements. The observed increase in standard deviation as a function of scan-step duration is described well by a simple statistical model that combines a random-walk diffusion process with a feedback loop.
     The reproducibility of retarding potentials from scan to scan follows a Gaussian distribution with a width of $\sigma=$\SI{34\pm1}{\milli\volt}. This limitation of the reproducibility is directly related to the digital-to-analog converter inside the post-regulation setup; for measurement phases after KNM1, finer-grained regulation is in place.

    The retarding potential is continuously monitored during the measurements. Therefore, at each scan step, the time evolution of the retarding potential is known with ppm precision. Neglecting this in the analysis introduces an additional broadening of the energy scale, leading to a neutrino-mass shift of $\Delta m_\nu^2 =$ \SI{-3e-3}{\electronvolt\squared}. This shift is less than half the allotment for the high-voltage-related systematic uncertainty in the KATRIN uncertainty budget for full five-year statistics~\cite{KDR2004}, and can be neglected in the KNM1 analysis.

    \subsection{Electron counting and region of interest}
    \label{sec:electron_counting}

	The FPD records a low-resolution, differential spectrum of electrons that have passed the high-resolution energy threshold set by the main spectrometer. Measuring the integrated tritium \bspec{} for KNM1, and thereby extracting the neutrino mass, requires an accurate count of electrons that arrive at the FPD within an energy region of interest (ROI) during each scan step. The ROI cut allows rejection of backgrounds and noise events generated near or in the FPD.

	When electrons strike the FPD, its pixels are triggered individually, with thresholds set just above the noise floor at around \SI{5}{\kilo\electronvolt}. As described in detail in earlier work~\cite{Amsbaugh:2014uca}, the energy and timing for each pulse are reconstructed online using a double trapezoidal filter and then recorded; FPD waveforms are not saved during normal operations. The shaping length of the trapezoidal-filter pair is set to \num{1.6} $\mu$s, optimizing the energy resolution at around \SI{1.8}{\kilo\electronvolt} (full width at half maximum, FWHM). During $\upbeta$ scans, rates are too low for significant pileup, but severe pileup during high-rate e-gun measurements can result in deadtime when multiple coincident events drive the baseline out of the ADC dynamic range. This effect is mitigated by individually adjusting the gain of each channel to approximately \num{5} ADC counts per \si{\kilo\electronvolt}, preserving good energy resolution while defining a dynamic range (up to \SI{400}{\kilo\electronvolt}) sufficient to accommodate pileup. These settings were implemented in the DAQ firmware prior to the KNM1 measurement. Simulations of the readout chain show that the fraction of time during which the baseline is shifted out of the ADC input range is less than \SI{0.05}{\percent} for \SI{50}{kcps} of \SI{28.6}{\kilo\electronvolt} electrons, a 100-fold improvement compared to previous settings. 

    \begin{figure}[t!]
        \centering
        \includegraphics[width=0.35\textwidth]{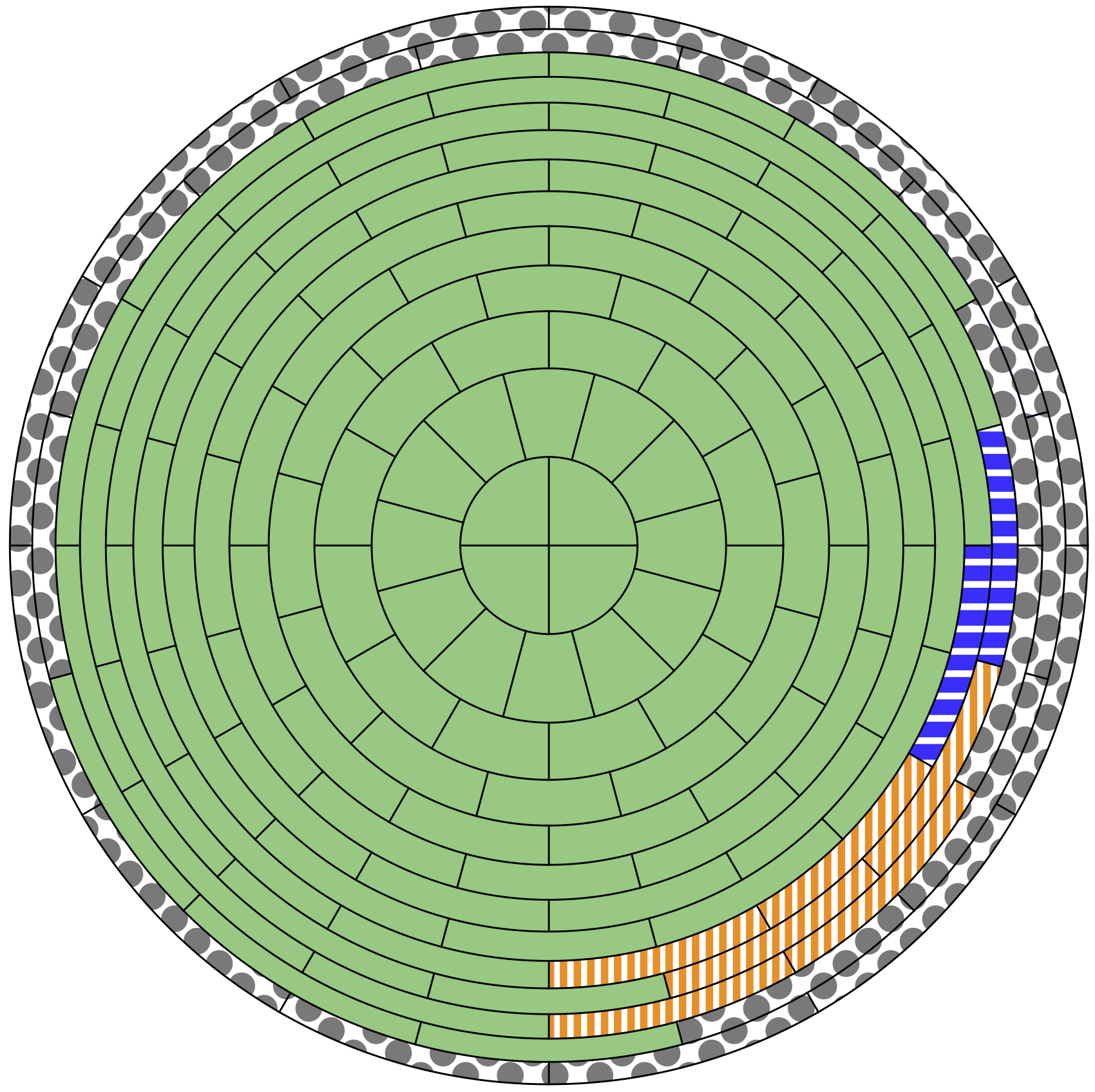}
        \caption{FPD pixel selection for KNM1. All \num{117} selected pixels, colored in solid green, are used for the analysis. The two pixels filled with horizontal blue lines are excluded due to shadowing by the forward beam monitor. The six pixels filled with vertical orange lines are excluded due to intrinsic noisy behaviour. All pixels filled with gray circles are excluded since they are partially shadowed by  beamline components.}
        \label{fig:KNM1PixelSelection}
    \end{figure}

    Out of the \num{148} pixels, we define a list of \num{117} selected detector pixels, distributed as shown in Fig.~\ref{fig:KNM1PixelSelection}. The excluded pixels are either noisy, or shadowed by beamline instrumentation in the \belech{} path along the magnetic flux tube.

    \begin{figure}[t!]
        \centering
        \includegraphics[width=0.48\textwidth]{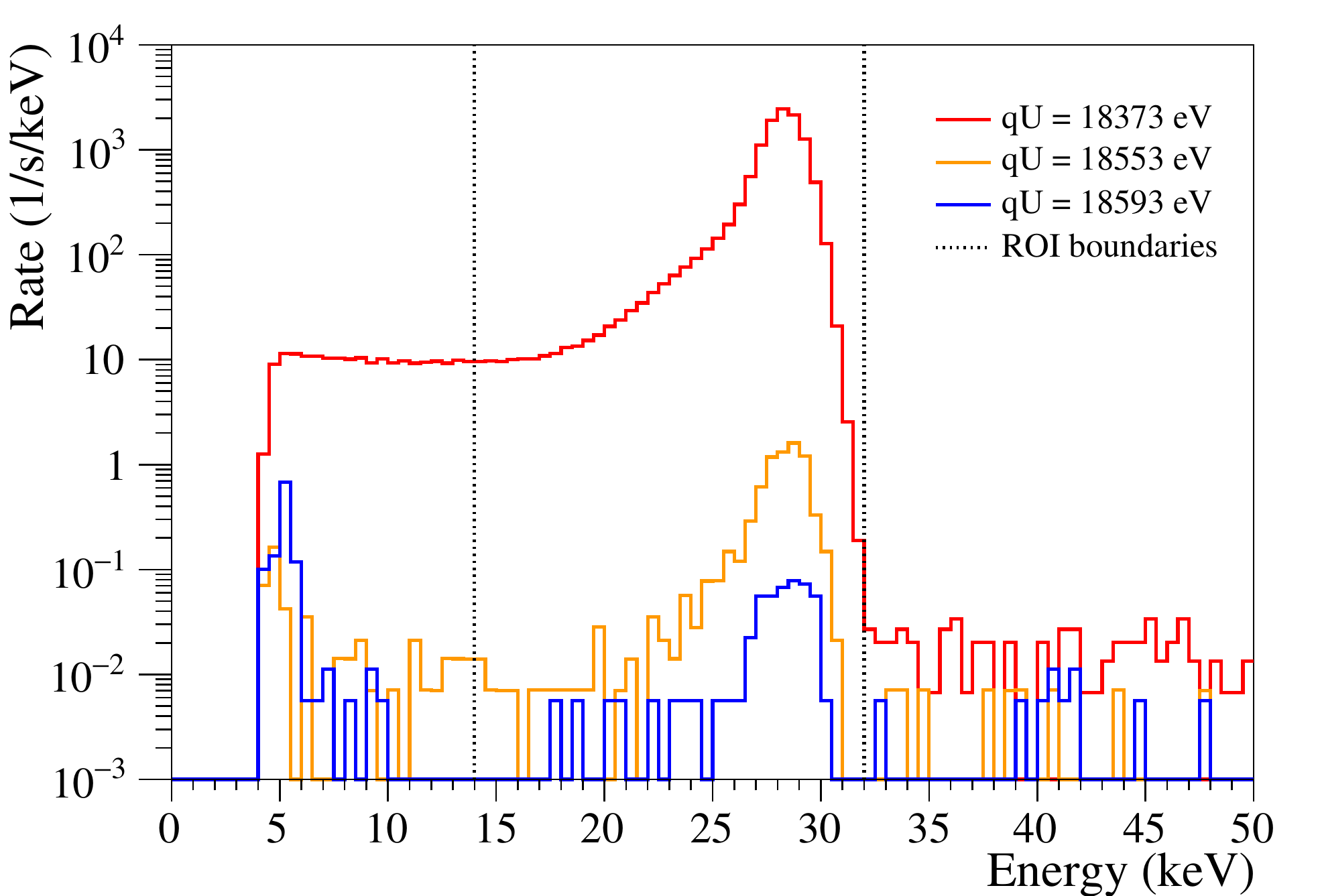}
        \caption{FPD energy spectra recorded during KNM1 at three different scan steps. At $qU = \ezero{} + $ \SI{20}{\electronvolt} (blue), the retarding potential eliminates all \belec{}s. The backgrounds visible in the spectrum consist of electrons from the main spectrometer (energies up to the peak at \SI{28}{\kilo \electronvolt}), electrons from the detector system (distributed across the energy range), and detector electronics noise (below \SI{7}{\kilo \electronvolt}). At $qU = \ezero{} - $\SI{20}{\electronvolt} (orange), the spectrum additionally incorporates signal events from tritium $\beta$- decays. Deep in the \bspec{}, at $qU = \ezero{} - $ \SI{200}{\electronvolt} (red), rates are high and pile-up events become visible above \SI{32}{\kilo \electronvolt}. Events within the ROI, demarcated by the two vertical dotted lines, are included in high-level analysis.}
        \label{fig:FpdEnergySpectra}
    \end{figure}

    Electrons that transit the spectrometer (Sec.~\ref{sec:responsefunctionmodel}) receive an additional \SI{10}{\kilo\electronvolt} of kinetic energy from the post-acceleration electrode, and \SI{120}{\electronvolt} from the bias voltage applied to the FPD. For a retarding potential around \SI{18.6}{\kilo\volt}, this results in a broad peak in the FPD energy spectrum at around \SI{28}{\kilo\electronvolt} (Fig.~\ref{fig:FpdEnergySpectra}). Background electrons and \belec{}s  share this characteristic energy spectrum in the FPD, since the primary background during KNM1 arises from low-energy electrons that are created inside the main spectrometer and then accelerated by the retarding potential (Sec.~\ref{sec:background}). The FPD energy scale is calibrated with a $^{241}$Am gamma source every two weeks.

     Our ROI is defined as [\SI{14}{\kilo\electronvolt}, \SI{32}{\kilo\electronvolt}], as measured by the FPD (Fig.~\ref{fig:FpdEnergySpectra}). The upper bound of the ROI is determined simply from the peak position and the peak width; the lower bound is determined for stability and robustness. In contrast to earlier studies that considered backgrounds originating near the detector~\cite{Amsbaugh:2014uca}, the choice of a low-energy KNM1 ROI lower bound does not reduce the signal-to-background ratio, since an energy cut cannot differentiate between \belec{}s and main-spectrometer background. A cut far away from the peak, where the spectrum shape derivative is small, improves stability against fluctuations of energy scale and resolution. Consequently, corrections for peak-position dependence on retarding potential are  negligible.

     The specific lower bound of the ROI, \SI{14}{\kilo\electronvolt}, was chosen so as to cancel two effects that arise from charge sharing, in which energy from a single incident electron is divided between two neighboring pixels. If a pixel loses more than half the event charge, its loss from the ROI decreases the effective rate; if a pixel receives more than half the event charge, its inclusion in the ROI increases the effective rate. With the FPD threshold set at half the peak energy, these two effects exactly compensate each other.

    \subsection{Data pipeline}
    \label{sec:data_pipeline}

    Following each pixel trigger (Sec.~\ref{sec:electron_counting}), the DAQ records the trigger timestamp from a \SI{20}{\mega\hertz} clock and the energy information as raw ADC counts integrated over the shaping time of the trapezoidal filter. A scan is divided into scan steps. Each scan step is defined by its HV set point, and its duration is determined according to the measurement-time distribution of the scan (Sec.~\ref{sec:MTD}).  Prior to acquisition start at each scan step, handshakes between the DAQ and the HV control system ensure that the HV read-back value has reached the set-point value within a defined accuracy of \SI{50}{\milli\volt}, as measured by a four-point moving average over the last \SI{8}{\second}. The inter-spectrometer electron catcher is inserted and removed during this change of scan steps, so that it does not obstruct the beamline during data-taking. A series of pulse-per-second (PPS) pulses from a precision clock synchronized to the Global Positioning System (GPS) defines both the start and stop times of scan steps, providing boundary time accuracy better than \SI{1}{\nano\second}. The \SI{50}{\nano\second} digitization timestamps are also phase-locked to \SI{10}{\mega\hertz} pulses from the same precision clock. The readout system is capable of handling a pixel rate of \SI{100}{kcps} and a total rate of \SI{3}{Mcps}. Therefore, no deadtime is expected for the actual tritium scan, which has a maximum count rate of \SI{7}{kcps}. A typical two-hour scan produces roughly \SI{120}{MB} of data.

    Immediately after completion of a scan, data files are processed automatically. This processing  includes the transfer to storage computers, time-wise event sorting, conversion to offline data formats, and indexing into a run database, followed by automated user-side analysis including the reduction of data in user-specified data files. Except for the handshakes between the DAQ and HV systems, slow-control channels are independent from the tritium scans. Each slow-control sensor has a defined recording interval, typically between \num{2} and \SI{10}{\second}. This is a heterogeneous system for which timestamps are  taken from computer timestamps synchronized to the Network Time Protocol (NTP). In the off\-line analysis, special care is taken for synchronization among different slow-control channels, as well as between the DAQ and slow controls.

    An intermediate data layer, consisting of user-side shared data storage with version management, splits the data analysis chain. The first half of the chain covers analysis at the event and time-series levels, and the second half provides higher-level analysis including model fitting. For each scan, results of the first-level analysis are summarized in digest files that contain analyzed FPD counts with efficiency corrections, individual scan steps, calibrated slow-control values (including LARA isotopolog concentration and column density, and analyzed rates extracted from $\upbeta$-induced x-ray spectroscopy and the forward beam monitor), and data-quality flags.  Some experimental parameters, such as beamline alignment information and magnetic- and electric-field values determined by measurements and simulations, are shared across all scans in a given measurement period; each such period is summarized in a digest file containing the values of these parameters.

    During data-taking, acquisition occasionally began before the HV readback values achieved stability due to minor synchronization errors. The first two seconds of every scan step were removed from the data to address these issues. Count-rate, livetime, efficiency, and stability calculations are performed after these data-quality cuts.

    \subsection{Acquisition of the integral \bdecay{} spectrum}
    \label{sec:MTD}

    KATRIN measures the integral tritium \bdecay\ spectrum by sequentially applying different retarding energies $qU$, or equivalently HV settings, to the main spectrometer and counting the rate of transmitted \belec{}s, $R(qU)$, with the FPD. Our choice of the scan steps -- that is, the HV set points and the measurement time at each set point -- maximizes the sensitivity for \mtwonue\ by focusing on a narrow region where the impact of the neutrino mass on the spectrum is most pronounced. The location of this region depends on the experimental conditions; in the KNM1 campaign, it lies at $\ezero - $\SI{14}{\electronvolt}~\cite{Kleesiek:2018mel}.
    \begin{figure}[!t]
        \centering
        \includegraphics[width=0.48\textwidth]{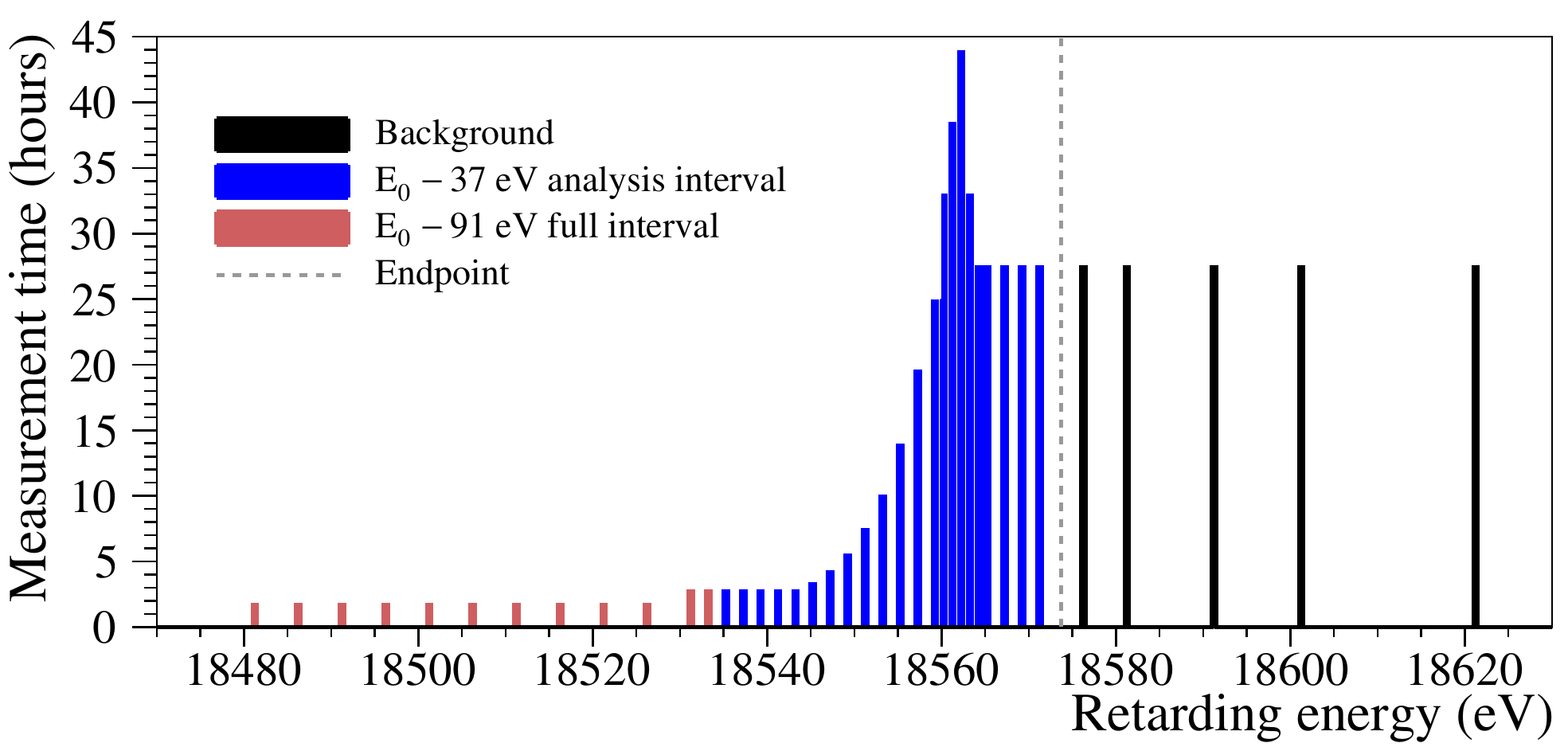}
        \caption{Cumulative measurement-time distribution for KNM1. The \num{27} scan steps of the [E$_{0} -$\SI{37}{\electronvolt}, E$_{0} +$\SI{49}{\electronvolt}] analysis interval are shown in blue below $E_0$, and in black above $E_0$. The most sensitive region to the signal of the neutrino mass is approximately \SI{14}{\electronvolt} below the endpoint, where most of the measurement time is spent.}
        \label{fig:MTD}
    \end{figure}

    Figure~\ref{fig:MTD} shows the measurement-time distribution used during this campaign, developed using a nominal value of $\ezero{} = $~\SI{18574}{\electronvolt}. The spectrum is scanned repeatedly over the range $\lbrack \ezero - $\SI{91}{\electronvolt}, $\ezero + $\SI{49}{\electronvolt}$\rbrack$ by sequentially applying the non-equidistant HV values (each constituting one scan step) to the main spectrometer. A  complete set of measurements at all \num{39} scan steps is defined as a \textit{scan}. Each scan over this energy range takes approximately \SI{2.5}{\hour} and is performed in alternating upward and downward directions. This mitigates the effects of any time-dependent drifts of the slow-control parameters. As explained in Sec.~\ref{sec:fit-range}, the analysis interval is limited to an energy range of $\lbrack \ezero - $\SI{37}{\electronvolt}, $\ezero + $\SI{49}{\electronvolt}$\rbrack$, consisting of \num{27} scan steps. A brief, additional scan step at $\ezero - $\SI{201}{\electronvolt} is used for rate-stability monitoring.

    For each tritium scan, we apply quality cuts to relevant slow-control parameters to select a data set with stable run conditions.
    As Sec.~\ref{sec:general_analysis} describes in detail, data from all active detector pixels are summed, effectively converting the detector wafer into a single, uniform pixel for analysis. Furthermore, all 274~scans are combined by summing counts from like scan steps, forming a single spectrum for fitting.

\begin{figure}[t]
        \centering
        \includegraphics[width=0.48\textwidth]{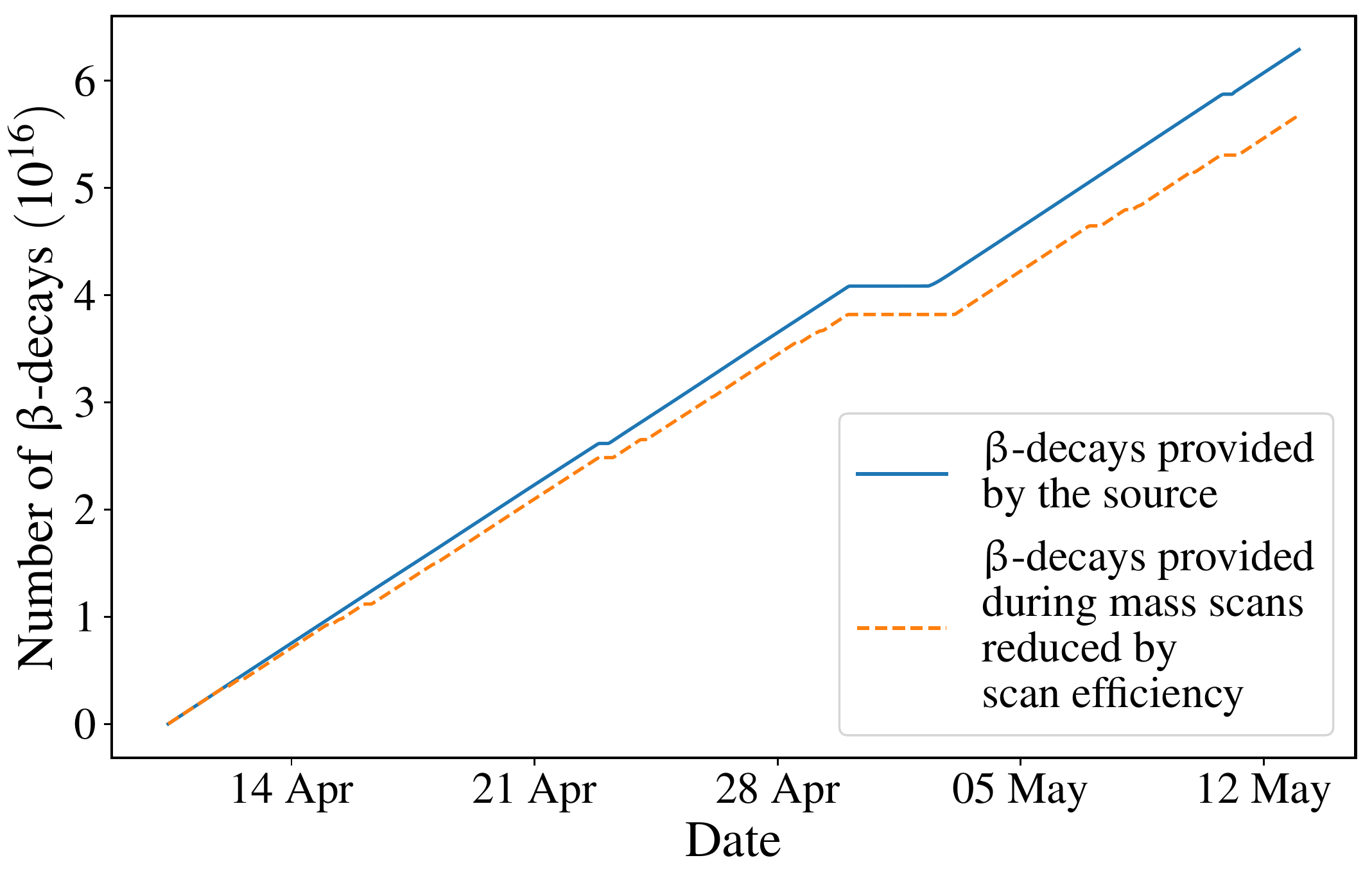}
        \caption{Integrated luminosity over KNM1. Compared to the accumulated number of $\upbeta$ decays delivered by the tritium source (blue line), the collection efficiency during neutrino-mass scans is slightly reduced by calibration runs and by the time it takes for the retarding potential to settle between scan steps (orange line). Dates are in 2019.}
        \label{fig:integrated_luminosity}
    \end{figure}

    The \num{27} scan steps within the analysis interval cover a total measurement time of \SI{521.7}{\hour}, corresponding to \SI{2.03e6}{events}. Table~\ref{tab:knm1_overview_figures} summarizes key operational parameters and figures for events and scans, covering both the full interval and the analysis interval.
The evolution of the integrated \bdecayh{} luminosity over the course of KNM1 is displayed in \cref{fig:integrated_luminosity}.

    \begin{table}[htb]
        \centering
        \caption{Summary of data acquisition for the KNM1 measurement campaign. (See text for details.)}
        \label{tab:knm1_overview_figures}
        \begin{tabular}{ll}
            \toprule
            \multicolumn{2}{c}{\emph{Scan overview}}\\
            \midrule
            Number of \bspec{} scans & \num{274} \\
            Net (total) time per scan & \SI{2}{\hour} (\SI{2.5}{\hour}) \\
            Energy range (full interval) & from $\ezero -$\SI{91}{\electronvolt} \\
                        & to $\ezero +$\SI{49}{\electronvolt} \\
            Energy range (analysis interval) & from $\ezero -$\SI{37}{\electronvolt} \\
                        & to $\ezero +$\SI{49}{\electronvolt} \\
            Number of scan steps & \num{39} \\
            \quad in signal region	 (full interval) & \num{34} \\
            \quad in signal region	 (analysis interval) & \num{22} \\
            \quad in background-only region (both intervals) & \phantom{2}\num{5} \\
            Source activity & \SI{2.45e10}{\becquerel} \\
            Energy resolution at \SI{18.6}{\kilo\electronvolt} & \SI{2.8}{\electronvolt} \\
            \toprule
            \multicolumn{2}{c}{\emph{Event ensemble}} \\
            \midrule
            Accumulated measurement time & \\
            \quad in full interval (39 scan steps) & \SI{541.7}{\hour} \\
            \quad in analysis interval (27 scan steps) & \SI{521.7}{\hour} \\
            Accumulated number of counts & \\
            \quad in full interval (39 scan steps) & \num{12.2e6} \\
            \quad in analysis interval (27 scan steps) & \phantom{2}\num{2.03e6} \\
            \qquad accumulated signal & \phantom{2}\num{1.48e6} \\
            \qquad accumulated background  & \phantom{2}\num{0.55e6} \\
            \bottomrule
        \end{tabular}
    \end{table}

    \section{Tritium-spectrum modeling}
    \label{sec:spectrum_modelling}

    The KNM1 analysis relies on a model of the measured spectrum, which convolves the theoretical \bspec{} (outlined in this section) with the experimental response function (details in Sec.~\ref{sec:responsefunctionmodel}). We first describe the general theory of \bdecayh{} in Sec.~\ref{sec:beta-spec-basics}, along with some straightforward corrections. To account for the physics of KATRIN's molecular source (T$_2$ with some HT and DT), we then address the molecular final-state distribution (FSD) in detail in Sec.~\ref{subsec:FSD}. Since an error in the FSD variance across our measurement interval will (to first order) shift the extracted, squared neutrino-mass value according to Eq.~\ref{eq:sigma} in the previous section, we have invested substantial effort in checking and extending our treatment of the FSD.

    \subsection{Theoretical \texorpdfstring{$\upbeta$}{beta}-spectrum of molecular tritium}
    \label{sec:beta-spec-basics}

    In KATRIN's molecular source, the \bdecay{} parent in Eq.~\ref{eq:beta_decay} becomes $\TT{}$, with a molecular decay product \HeT{}.
    To model the resulting differential \bspec{}, we begin with a point-like Fermi interaction, which causes the weak decay, and then apply the sudden approximation, in which the Coulomb interaction of the \belec{} with the remaining molecular system \HeT{} is neglected. The validity of this approximation was demonstrated in Refs.~\cite{Saenz1997_PartI,Saenz1997_PartII}.

    Choosing the center-of-mass coordinate frame to align with the momentum of the neutrino and integrating over the experimentally unresolved neutrino and electron directions and neutrino energy, the decay rate into the nuclear and molecular configuration $f$ of the daughter \HeT{} at a given electron kinetic energy $E$ reads~\cite{Saenz1997_PartI}
    \begin{align}
        \label{eq:diffspec_single_state}
        \begin{split}
            \betaRate_f(E)
            & =
            \frac{\matrixElementGen{\transitionMatrix_f(E)}}{2\pi^3}
            \phaseSpace{(E+m_e)}{m_e} \\
            &
            \cdot \phaseSpace{\neutrinoEnergy(E)}{m_\nu}
            \heavyside(\neutrinoEnergy(E) - m_\nu)
            ,
        \end{split}
    \end{align}
 \noindent in natural units with $c = \hbar = 1$. $m_e$ and $m_\nu$ are the electron and neutrino masses, respectively; $\neutrinoEnergy(E)$ has the form of the neutrino energy after energy conservation has been enforced by the Heaviside function $\heavyside(\neutrinoEnergy(E) - m_\nu)$. $\matrixElementGen{\transitionMatrix_f}$ is the transition matrix element to the nuclear and molecular state $f$. Since the derivation of the decay rate is performed in the center-of-mass frame, which almost perfectly coincides with that centered on the decaying molecule, there is no need  to integrate over the recoil momentum of the molecule; the recoil kinetic energy is naturally added as a constant energy loss.

 $\matrixElementGen{\transitionMatrix_f}$ may be factorized in the sudden approximation as
    \begin{align}
        \matrixElementGen{\transitionMatrix_f}
        =
        \matrixElement{weak} \matrixElement{lep} \matrixElement{mol}
        \label{eq:T_factorization}
    \end{align}
 \noindent where $\matrixElement{weak}$ is independent of the electron energy for the superallowed tritium \bdecay{}. Similarly, the leptonic part $\matrixElement{lep}$ is independent of the electron energy in the sudden approximation. As is customary, however, the Fermi function $F(E, Z' =2)$ (as given in Ref.~\cite{Simpson1981}) is included in this factor. This allows a partial incorporation of the influence of the Coulomb interaction during the decay by accounting for the charge of an isolated ${}^3$He daughter nucleus, leading to an effectively Coulomb-distorted sudden approximation. Meanwhile, $\matrixElement{mol}$ is equal to the probability $\zeta_f$ that \HeT{} populates the unresolved set of molecular electronic, vibrational, and rotational states with energy $V_f$. Since the motion of the center of mass of \HeT{} must balance the neutrino and electron momenta, $\matrixElement{mol}$ theoretically depends on the electron energy after the integrations are performed. The KNM1 analysis interval is narrow enough to neglect this dependence.

    After evaluating $\matrixElementGen{\transitionMatrix_f}$ according to Eq.~\ref{eq:T_factorization}, summing over the possible final nuclear states, and explicitly summing over the included range of molecular states, we obtain
    \begin{align}
        \begin{split}
            R_{\mathrm{\upbeta}}(E)
            & =
            \frac{\CouplingSq{F} \cos^2\cabibbo }{2 \pi^3} \matrixElementGen{M_{\mathrm{nucl}}} F(E, Z' =2) \\
            &
            \cdot \phaseSpace{(E+m_e)}{m_e} \\
            &
            \cdot \sum_{f \in mol}
            \zeta_f \,
            \phaseSpace{\neutrinoEnergy(E)}{m_\nu}
            \heavyside(\neutrinoEnergy(E) - m_\nu)
            .
        \end{split}
        \label{eq:tritium_spec_diff}
    \end{align}
\noindent   The prefactors include the energy-independent quantities $\Coupling{F}$ (the Fermi constant), $\cabibbo$ (the Cabibbo angle), and $\matrixElementGen{M_{\mathrm{nucl}}}$ (the nuclear matrix element). 
Meanwhile,
    \begin{align}
        \neutrinoEnergy(E)
        =
        \reducedEndpoint - V_f - E
        ,
    \end{align}
\noindent    where the reduced endpoint $\reducedEndpoint$ represents the total maximum electron kinetic energy in the case of a massless neutrino. While $\reducedEndpoint$ is retrieved from the fit during the neutrino-mass analysis (Sec.~\ref{sec:spectralfit}), the internal molecular excitation energies $V_f$ and the corresponding population probabilities $\zeta_f$ come from computation (see Sec.~\ref{subsec:FSD}). The values of all constants are as in Ref.~\cite{Kleesiek:2018mel}.

    Beyond the molecular effects discussed in detail in Sec.~\ref{subsec:FSD}, theoretical corrections to the tritium \bdecay{} spectrum arise at the particle, nuclear, and atomic levels (see Ref.~\cite{Mertens:2014nha} for details). Of these, we include only the radiative corrections~\cite{PhysRevC.28.2433} in this work; these have by far the largest effect on the high-energy tail of the \belec{} spectrum.

    Finally, the electron spectrum $R_{\mathrm{\upbeta}}$ is Doppler-broadened due to the finite motion of tritium molecules in the source. To account for this effect, we replace each discrete final state with a Gaussian centered at the final-state energy $V_f$, normalized to $\zeta_f$ and with a standard deviation of $\SI{94}{meV}$ according to the Doppler broadening at \SI{30}{\kelvin}. Effects due to the bulk gas flow are negligible.  

For effects that give rise to continuous modifications of the spectrum, such as the molecular final-state distribution and Doppler broadening, a mistake in the modeled variance will introduce a bias on the extracted neutrino-mass squared according to Eq.~\ref{eq:sigma}.

    \subsection{Final-state distribution (FSD)}
    \label{subsec:FSD}

    Within the sudden approximation, the \bdecay{} effectively corresponds to a sudden change of the nuclear charge of one of the tritium nuclei. This induces electronic and vibrational excitations of the daughter molecular ion \HeT{}, possibly including its dissociation and/or ionization. Furthermore, the departing \belec{} and neutrino induce external (translational) and internal (rotational, vibrational, and -- to a smaller extent, neglected here -- electronic) excitations.

    Since only the energies of the \belec{}s are analyzed by KATRIN, the undetected energy associated with the remaining molecular system must be computed \textit{ab initio} by first solving the Schrödinger equation for the initial and final molecular systems, and then computing the transition probabilities $\zeta_f = \matrixElement{mol}$ to the molecular daughter states $f$ thus found. Earlier calculations either focused on lower temperatures than KATRIN's \SI{30}{\kelvin}~\cite{Saenz2000}, thus artificially constraining the population of initial molecular states, or did not include all the tritium-containing isotopologs~\cite{Doss2006}. In the following, we provide only a minimal description of the new computations carried out for the initial gas states relevant to KATRIN; a detailed publication is in preparation~\cite{Katrin:FSD}. The theoretical prediction of the dissociation probability of the daughter \HeT{} ion, following \bdecay{}, has recently been experimentally verified~\cite{trims:2020}.

    \subsubsection{Solutions to the molecular Schrödinger equation}

    As in previous works, these computations adopt two fundamental approximations. First, the Coulomb-distorted version of the sudden approximation neglects the interaction of the $\upbeta$ electron with all but the daughter nucleus \He{} in the \bdecay{}. Second, the Born-Oppenheimer approximation allows a separate treatment of the electronic and nuclear motions that define the full, internal molecular Schrödinger equation.

    Our solution of the Schrödinger equation describing the nuclear motion uses the isotopolog-independent Born-Oppenheimer electronic potentials generated according to Ref.~\cite{Saenz1999} and presented explicitly in Ref.~\cite{DossThesis}. Mass-dependent corrections are applied for the electronic ground states of specific isotopologs -- $\TT$, $\DT$, $\HT$, \HeT{}, \HeD{}, and \HeH{} -- and the potential curves are extended up to an internuclear separation of $20 \, \bohr$, with $\bohr$ the Bohr radius. Because of the rotational symmetry of the corresponding Schrödinger equation, the solutions for nuclear motion are expanded as products of spherical harmonics and radial functions. They are then augmented by the rotational barrier for non-zero initial angular momenta $J_i$.

    The electronic ground state of the daughter molecule supports about 300 rotational/vibrational bound states and a large number of predissociative resonances in the dissociation continuum. We have therefore adopted a new approach for solving the nuclear motion in these electronic potentials.  Expanding the radial part in $B$-spline functions and adopting vanishing boundary conditions at the end of the radial grid, the solution of the Schrödinger equation is turned into a generalized matrix eigenvalue problem and requires only the diagonalization of a very sparse matrix. The spectral density and energy range of the resulting discretized spectrum may be controlled by the size of the adopted spherical box and the number of $B$-splines.

    \subsubsection{Energy-resolved FSD}

    With the newly obtained nuclear-motion solutions, and the isotopolog-independent Born-Oppenheimer electronic overlaps defined in Refs.~\cite{DossThesis} (final electronic ground state $n=1$) and~\cite{Saenz1999} (final electronic states $n \in \intInterval{2}{6}$), the transition probabilities between the initial and final states of interest in the KNM1 analysis interval can be obtained by integrating the matrix elements over the internuclear separation vector. The transition operator, which can be expanded into spherical Bessel functions, depends on this vector.

\begin{figure}
        \centering
        \includegraphics[width=.47\textwidth]{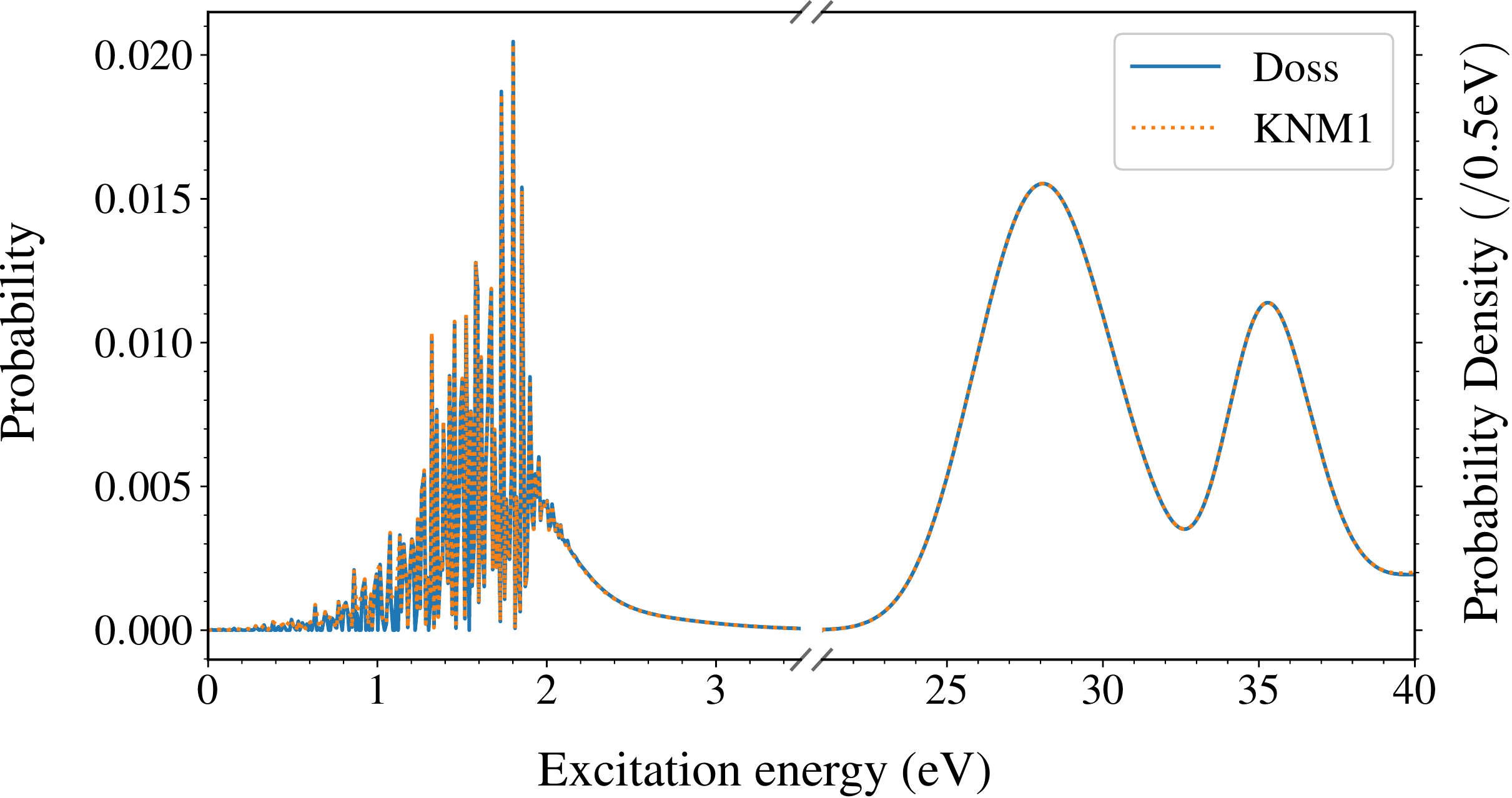}
        \caption{Final-state distribution (FSD) for transitions from $\TT$ in its electronic and vibrational ground state but initial rotational angular momentum $J_i =$ \num{1} (likeliest for a source at $\SI{30}{K}$) to the first \num{13} (KNM1) or $6$ electronic states (Doss~\cite{Doss2006}) of \HeT{}, as defined by Refs.~\cite{Kolos1985,Saenz1999}. The transitions to the electronic ground state (below \SI{4}{\electronvolt}) and its bound rovibrational states are given as probabilities (left axis) while the dissociative continuum is naturally plotted as a density (right axis); the total transition probability to the electronic ground state is \SI{57.4}{\percent}. The broad features at about \SI{28}{\electronvolt} and \SI{35}{\electronvolt} are dominated by the $n=2$ and $n=3$ states.}
        \label{fig:FSD_T2_J1_n1_13_vs_Doss}
    \end{figure}

    Compared to earlier work, our new calculation extends the results of Ref.~\cite{Doss2006} from the first \num{6} to the first \num{13} bound electronic states, and employs more accurate molecular masses than Refs.~\cite{Saenz2000,Doss2006}. These more accurate masses are used in the Hamiltonian, in the fraction of the recoil momentum imparted onto the spectator nucleus -- which selects the population of the states due to the molecular \bdecay{} via the transition operator \mbox{--,} and in the recoil energy of the whole molecular system. For the electronic excited final states $n \in \intInterval{1}{6}$, we have been able to reproduce the results of Refs.~\cite{Saenz2000,Doss2006} for the published initial states of $\TT$ ($J_i \in \intInterval{0}{3}$), $\DT$ ($J_i \in \intInterval{0}{1}$), and $\HT$ ($J_i =$ \num{0}), when using the old kinematic inputs. Figure~\ref{fig:FSD_T2_J1_n1_13_vs_Doss} shows a comparison of the current distribution with Ref.~\cite{Doss2006} for transitions from the most populated $\TT$ initial state at $T =$ \SI{30}{\kelvin}. The new distribution of transitions to the electronic ground state is $\sim$\SI{3}{\milli\electronvolt} lower on average than that in Ref.~\cite{Doss2006}; this difference mostly originates from the updated recoil momentum as a consequence of the more accurately determined endpoint. Electronic final states with $n >$ \num{6}, combined with the electronic continuum, contribute negligibly -- at the \num{e-4} level -- to the KNM1 analysis interval, with its lower bound at \ezero{} -- \SI{37}{\electronvolt}. In our new calculation, these have been adapted for energy-scale changes from the calculations in Ref.~\cite{Saenz2000}. The $n >$ \num{6} bound states were omitted in Ref.~\cite{Doss2006}, explaining the slightly higher transition probabilities of the new distribution around \SI{40}{\electronvolt}.

    For the KATRIN analysis, we consider all $J_i \in \intInterval{0}{3}$ for all three decaying isotopologs and weight their respective contributions based on the source temperature. The Boltzmann distributions are calculated at \SI{30}{\kelvin}. However, for the homonuclear $\TT$ molecule, the resulting $J_i$ probability must be multiplied by nuclear-spin probabilities characteristic of \SI{700}{\kelvin}.  The molecules in the tritium loop dissociate when they arrive at the permeator (Sec.~\ref{sec:setup}), which is operated at 700 K. After diffusion through the permeator, the atoms recombine into molecules with an ortho-para ratio of  \num{0.75}, characteristic of that temperature. The time for natural conversion to a lower-temperature ortho-para ratio is many orders of magnitude longer than the $\mathcal{O}(\SI{1}{\second})$ passage time of the molecules through the \SI{30}{\kelvin} region of the injection capillary and source tube, so the $\TT$ gas retains an ortho-para ratio of \num{0.75}.

Weighting based on the relative concentrations of $\TT$, $\DT$ and $\HT$, as measured during KNM1, is performed at a subsequent stage of the analysis.

    \section{Response function modeling}
    \label{sec:responsefunctionmodel}

    The observed KNM1 tritium integral spectrum $R(qU)$ is the convolution of the differential \belec\ spectrum \rbeta\ from Eq.~\ref{eq:tritium_spec_diff} with the instrumental response function $f(E - qU)$, with an added energy-independent background rate $R_\mathrm{bg}$:
    \begin{equation}
        R(qU) = A_\mathrm{s}\,\cdot\,N_\mathrm{T,eff}\int R_{\mathrm{\upbeta}}(E) \cdot  f(E - qU)~ dE +  R_\mathrm{bg}~.
        \label{eq:int_spec}
    \end{equation}
    Here, $N_\mathrm{T,eff}$ denotes the effective number of tritium atoms in the source, as adjusted by the detector efficiency and by the solid-angle acceptance of the setup $\Delta \Omega/4\pi = (1- \cos{ \theta_\mathrm{max}})/2$, where $\theta_\mathrm{max} \approx \ang{50.5}$ as discussed below. $A_s$ is the signal amplitude.

\begin{figure}[t!]
        \centering
        \includegraphics[width=0.45\textwidth]{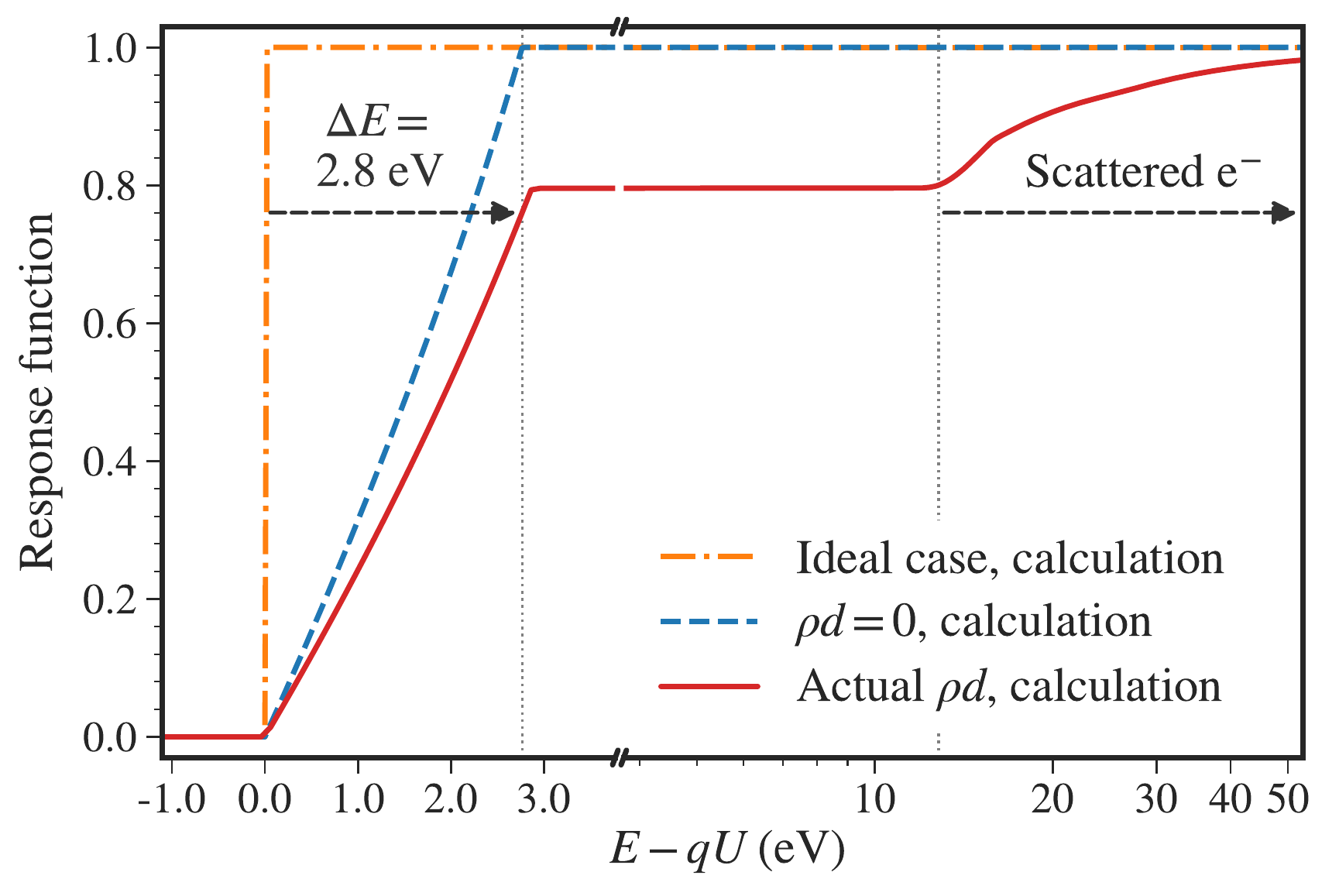}
        \caption{Response function with infinitesimal filter width and no scattering (orange dash-dotted line); actual transmission function of the MAC-E filter without source scattering (blue dashed line); and actual transmission function along with source scattering (solid red line). Electrons emitted with \num{0} $< \theta \le \theta_\mathrm{max}$ need additional surplus energy $E - qU$ to overcome the potential barrier. At higher surplus energies, electrons that lost energy due to source scattering can pass the filter as well, and the transmission probability increases.}
        \label{fig:response}
    \end{figure}

    As shown in Fig.~\ref{fig:response}, the response function $f(E-qU)$~\cite{Kleesiek:2018mel} describes the probability of transmission of an electron with initial energy $E$  through the beamline as a function of its surplus energy $E-qU$ relative to the retarding potential $U$.  Below, we discuss its calculation in detail. First, Sec.~\ref{sec:response} defines the response function and describes the effects of the beamline electromagnetic fields on the \belec{}s. We then treat the inelastic scattering cross section for \belec{}s (Sec.~\ref{ch:InelScattCross}) and develop a model of energy loss experienced in flight through the KATRIN apparatus (Sec.~\ref{sec:eloss-function}). 

    \subsection{Response and transmission functions}
    	\label{sec:response}

The transmission condition for any electromagnetic configuration of the KATRIN MAC-E filter determines whether an electron with starting energy $E$ and starting angle $\theta$ is transmitted through a retarding potential $U$:

\begin{equation}
    \mathcal{T}(E,\theta,U) =\left\{ \begin{array}{ll}
        1 &\text{if}\quad \displaystyle E \, \left(1 - \sin^2 \theta \cdot \frac{B_\mathrm{min}}{B_\text{S}} \cdot \frac{\gamma\!+\!1}{2} \right) \\
        &\qquad\qquad - qU > 0\\
        0 &\text{else}
    \end{array}
    \right. \, .
    \label{eq:trans_condition_function}
\end{equation}

\noindent Here, $\theta = \angle(\vec{p},\vec{B})$ is defined as the initial pitch angle of the electron, the polar angle of its momentum relative to the magnetic field: $p_\perp^2 = E \sin^2\theta \cdot (\gamma+1) \cdot m_\mathrm{e}$. The Lorentz factor $\gamma$ arises from its relativistic motion and has a maximum value of about 1.036 at \ezero{}. Meanwhile, $B_\mathrm{min} = \SI{0.63}{mT}$ is the magnetic field in the analyzing plane, $B_\mathrm{max} = \SI{4.23}{T}$ the maximum field of the beam line, and $B_\mathrm{S} = \SI{2.52}{T}$ the source magnetic field. 

Only electrons with sufficient surplus energy satisfy the transmission condition and are included in the measured integral spectrum. The KATRIN main spectrometer achieves a magnetic-field ratio $B_\mathrm{min}/B_\mathrm{max} \approx 1/6700 \approx \Delta E / E$, corresponding to a filter width (energy resolution) of $\Delta E = \SI{2.8}{\electronvolt}$ at \SI{18.6}{\kilo\electronvolt}. The maximum acceptance angle $\theta_\mathrm{max} = \arcsin{\sqrt{B_\mathrm{S}/B_\mathrm{max}}} \approx \ang{50.5}$ limits the range of pitch angles contributing to the integral spectrum. The magnetic fields and the retarding potential are provided by detailed field calculations using the {\sc Kassiopeia} software~\cite{Furse:2016fch}. To compute the precise electromagnetic fields across the analyzing plane, we use an as-built geometry of the beamline magnets with a detailed three-dimensional model of the main spectrometer. The resulting transmission conditions can be included in the model individually for each active pixel.

    The detailed response function of the KATRIN apparatus is calculated from Eq.~\ref{eq:trans_condition_function}, as modified by energy losses $\epsilon$ between source and analyzing plane~\cite{Kleesiek:2018mel}:
    \begin{align}
        \begin{split}
        f(E-qU)
        & = \int_{\epsilon=0}^{E-qU} \! \int_{\theta=0}^{\theta_\mathrm{max}} \mathcal{T}(E-\epsilon, \theta, U) \sin\theta \\
        &\cdot \sum_s P_s(\theta) \, f_s(\epsilon) \, d\theta\, d\epsilon ~ .
        \end{split}
        \label{eq:transm_func_det}
    \end{align}
    
\noindent For an ensemble of electrons, $f(E-qU)$ depends on the acceptance angle $\theta_\mathrm{max}$ and the amount of neutral gas the electrons pass in the WGTS, which is described by the scattering probability $P_s(\theta)$ and the inelastic-scattering energy-loss function $f_s(\epsilon)$ for a given number of scatters $s$. As Sec.~\ref{sec:eloss-function} will discuss in detail, we measure $f(E-qU)$ using monoenergetic electrons with small angular spread, and thus deduce $f_s(\epsilon)$. Briefly, these electrons are produced in the e-gun with surplus energies $E-qU$ spanning a \SI{50}{\electronvolt} range. They follow the magnetic-field lines and pass through the integral column density $\rho d$ of the source. This allows us to observe single ($s =$ \num{1}) and multiple ($s >$ \num{1}) electron scatterings in the source. The scattering probability $P_s(\theta = 0^\circ)$ (Eq.~\ref{eq:scatterProb}) follows a Poisson distribution with the expected number of scatterings given by the product of the effective column density $\rho d$ and the inelastic-scattering cross section $\sigma$ (Sec.~\ref{ch:InelScattCross}).  

In an isotropic source like the WGTS, electrons are emitted with an angular distribution $\omega(\theta) d\theta = \sin\theta d\theta$, and we can define an integrated transmission function $T(E,U)$:
    
    \begin{align}
        \nonumber
    T&(E,U) = \int_{\theta=0}^{\theta_\text{max}} \; \mathcal{T}(E,\theta,U) \cdot \sin \theta \, d\theta \\
      & = 
        \left\{\begin{array}{ll}
        0 &, \epsilon<0 \\
        1 - \sqrt{1-\frac{E-qU}{E} \frac{B_\text{S}}{B_\mathrm{min}} \frac{2}{\gamma\!+\!1}} \, &, 0\leq E-qU \leq \Delta E \\
        1 - \sqrt{1-\frac{B_\text{S}}{B_\text{max}}}  &, E-qU >\Delta E
  	\end{array}
    \right. \, .
    \label{eq:response:tf_simple}
\end{align}

Although analysis of non-isotropic e-gun data requires the full expression in Eq.~\ref{eq:transm_func_det}, the neutrino-mass analysis in this work exploits the isotropic nature of the tritium $\upbeta$-source and uses the simplified response function

    \begin{align}
        \begin{split}
        f(E-qU) 
        & = \int_{\epsilon=0}^{E-qU} \int_{\theta=0}^{\theta_\mathrm{max}}  T(E-\epsilon, U) \\
        &\cdot \sum_s P_s(\theta) \, f_s(\epsilon) \, d\theta\, d\epsilon ~ .
        \end{split}
        \label{eq:transm_func}
    \end{align}

\noindent In principle, the response function is slightly modified due to the dependence of the path length, and therefore the effective column density, on the pitch angle of the $\upbeta$-electrons~\cite{Kleesiek:2018mel}. The resulting effect on the measured endpoint is small compared to the overall uncertainties of the electric potential of the source, and this effect is not taken into account in the current analysis.  Synchrotron energy losses of $\upbeta$-electrons in the high magnetic field in the source and transport systems are included as an analytical correction to the transmission function ~\cite{Kleesiek:2018mel}.

    \subsection{Inelastic-scattering cross section}
    \label{ch:InelScattCross}

    The theoretical total inelastic-scattering cross section of electrons with ${\rm T_2}$ molecules in the high-energy Born approximation can be written as~\cite{inokuti1971,liu1973,liu1987}:
    \begin{equation}
        \sigma_{\rm inel}(E)=\frac{4\pi a_0^2}{(E_{\rm nr}/R_\mathrm{H})}\left[M_{\rm tot}^2 \cdot \ln\left(4 c_{\rm tot}\cdot \frac{E_{\rm nr}}{R_\mathrm{H}}\right) -0.01 \right], \label{eq:FerencCrossSection}
    \end{equation}
    where $R_\mathrm{H} =$ \SI{13.606}{\electronvolt} is the Rydberg energy, $a_0^2 =$ \SI{28.003e-18}{\centi\metre\squared} the Bohr radius squared, and $E_{\rm nr}$ denotes the non-relativistic kinetic energy of the electron: $E_{\rm nr}=0.5\, \me \beta^2$, with $\beta^2=1-m^2_{\rm e}/(\me+E)^2$ and $E$ the relativistic kinetic energy of the electron. At the spectral endpoint for molecular tritium \bdecay{}, we take $E = E_0 =$ \SI{18.575}{\kilo\electronvolt} and $E_{\rm nr} =$ \SI{17.608}{\kilo\electronvolt}.

    The dominant parameter $M_{\rm tot}^2$ can be calculated reliably and with high accuracy, since it is a special electron expectation value for the ground-state hydrogen-molecule wave function. For the three isotopologs, we have~\cite{KW64,Pachucki}: $M_{\rm tot}^2[{\rm H_2}] =$ \num{1.5497}, $M_{\rm tot}^2[{\rm D_2}] =$ \num{1.5404} and $M_{\rm tot}^2[{\rm T_2}] =$ \num{1.5363}. The calculation of the subdominant parameter $c_{\rm tot}$ is more difficult, and we use the 1987 value of Liu~\cite{liu1987}: $c_{\rm tot} =$ \num{1.18}. With these numbers, we obtain $\sigma_{\rm inel}[{\rm T_2}](E_0) =$ \SI{3.64e-18}{\centi\metre\squared}, with an estimated uncertainty of \SI{0.5}{\percent}. It must be noted that this theoretical cross section differs from the measured value, \SI{3.40\pm0.07e-18}{\centi\metre\squared}~\cite{Aseev2000}, by \SI{7}{\percent} ($\SI{3.5}{\sigma}$). However, it is $\rho d \sigma$, directly measured by the e-gun as described in Sec.~\ref{ch:columnDensityDetermination}, which is used in the neutrino-mass analysis -- not $\sigma$ as a separate input.

    \subsection{Energy-loss function}
    \label{sec:eloss-function} 

    Electrons traversing the WGTS can scatter elastically or inelastically from tritium molecules before being analyzed in the main spectrometer. (Here, ``elastic'' scattering refers to interactions that do not change the electronic state of the molecule.) While elastic scattering only causes a small broadening of the measured response function ($\sim$\SI{0.03}{\electronvolt}), inelastic scattering can result in energy losses from  $\sim$\SI{11}{\electronvolt} up to $E/2$, where the lower bound is associated with the lowest electronic excitations in T$_2$.

    Small inelastic energy losses, in particular, can move electrons emitted at energies close to the endpoint (the sensitive region for m$_\nu^2$) into a region still within the analysis interval extending \SI{37}{\electronvolt} below the endpoint.  
 Precise knowledge of the energy loss spectrum is, therefore, a crucial input for the KATRIN response function. During planning, its uncertainty was estimated to be one of the dominant systematics of the experiment~\cite{KDR2004}. A detailed paper on the energy-loss determination is in preparation~\cite{KATRIN:eloss2020}.

    Various electronic excitations, in combination with rotational and vibrational states of the T$_2$ molecule, result in a rich spectrum up to the ionization threshold at \SI{15.486}{\electronvolt}~\cite{PhysRevA.60.3013}. Prior to this work, there were no calculations of the energy-loss spectrum with the required accuracy. We therefore measured the energy-loss function with the e-gun installed in the rear system of the KATRIN beamline. In contrast to \belec{}s originating within the source, these calibration electrons start with an adjustable kinetic energy chosen close to the endpoint of the tritium $\upbeta$ spectrum and traverse the full length of the source. The dependence of the energy-loss function on the kinetic energy of the electrons can be neglected within the small fit window around the endpoint at $\sim$\SI{18.6}{\kilo\electronvolt}.

    The e-gun uses a pulsed ultraviolet laser to create photoelectrons from a gold layer deposited onto the front face of an optical fiber. These electrons are then accelerated in an electric field with an adjustable angle to the local magnetic field lines. The electron energy is continuously scanned, in alternating directions, between \SI{5}{\electronvolt} below and \SI{55}{\electronvolt} above the main-spectrometer energy threshold $qU$.

\begin{figure}[t!]
        \centering
        \includegraphics[width=0.49\textwidth]{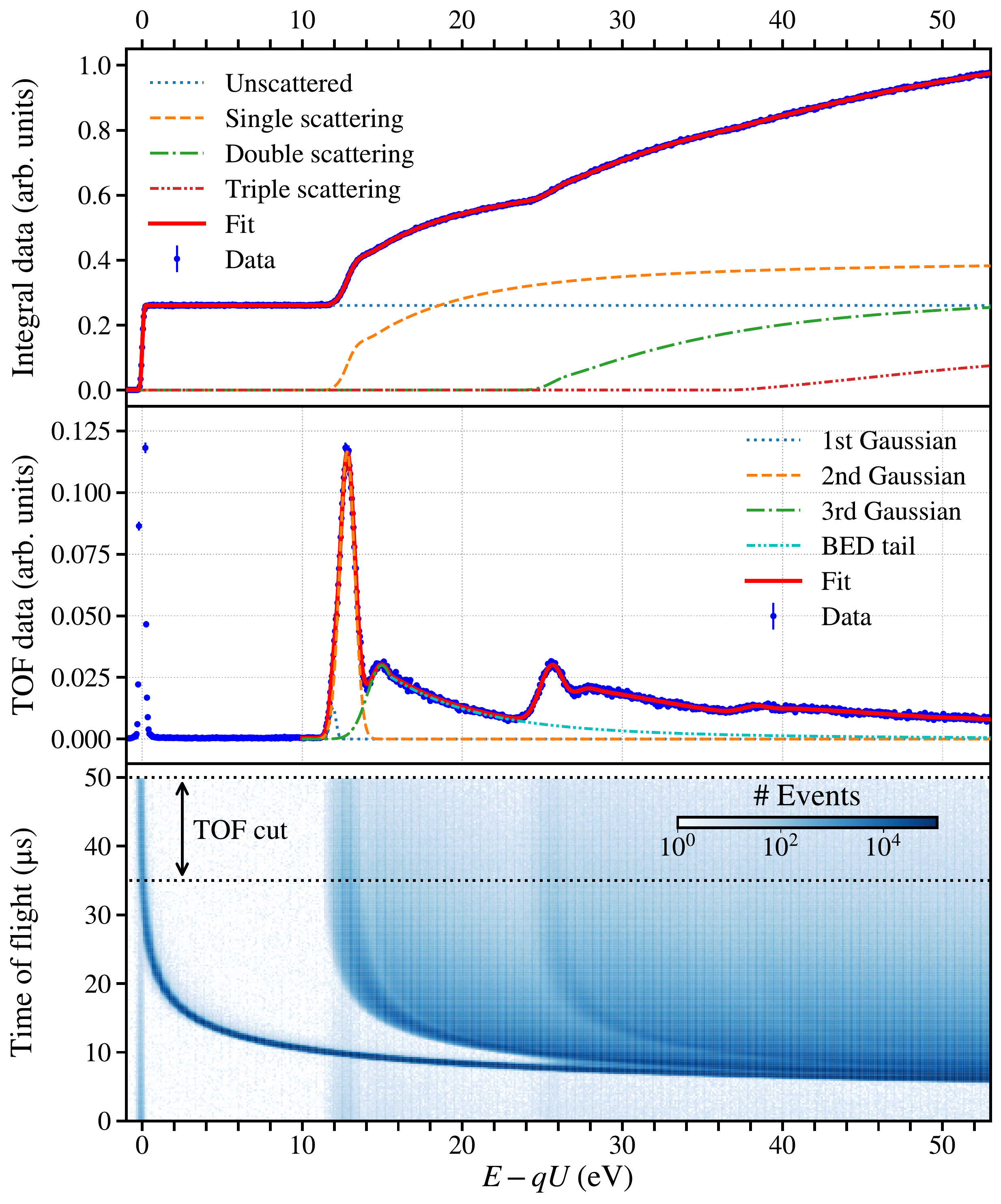}
        \caption{Energy-loss function from e-gun data; e-gun electrons must cross the entire \SI{10}{\meter} source at a pitch angle $\theta = 0$. Top panel: Integral measurement, showing data, fit function and individual components for single and multiple scatters. Center panel: Differential measurements obtained with the TOF cut, showing data, fit function and the four sub-components of the semi-empirical energy-loss parametrization: three Gaussians and a binary-encounter-dipole (BED) tail. The peak at \SI{0}{\electronvolt}, not included in the fit, is the measured spectrum of unscattered electrons. Bottom panel: Time of flight versus surplus energy data, showing the TOF selection region used to obtain the differential spectrum in the center panel.}
        \label{fig:eloss}
    \end{figure}

    The e-gun was operated in two different modes: a fast mode with a \SI{100}{\kilo\hertz} laser repetition rate to obtain a quasi-continuous electron beam used to record integral spectra as shown in Fig.~\ref{fig:eloss}~(top panel)
    and a slower mode with a \SI{20}{\kilo\hertz} repetition rate, in which the electron start times were synchronized with the DAQ to record time-of-flight (TOF) spectra as shown in Fig.~\ref{fig:eloss}~(center panel).

    This TOF information allows us to record a differential energy spectrum by applying a TOF cut on individual events~\cite{Bonn:1999}. Electrons with energies close to $qU$ take significantly longer to reach the detector since they are decelerated to almost zero kinetic energy near the analyzing plane. Selecting electrons with flight times between \SI{35}{\micro\second} and \SI{50}{\micro\second}, as illustrated in Fig.~\ref{fig:eloss}~(bottom panel), effectively turns the main spectrometer from a high-pass filter into a narrow band-pass filter with a width of $\sim$\SI{0.02}{\electronvolt}. Apart from effects of multiple scattering and finite energy resolution, this method provides direct access to the electron energy-loss spectrum.

    The energy-loss function is parametrized by a semi-empirical model using three Gaussians to describe the three groups of lines created by excitations of the $(2p\sigma^1\Sigma^+_u)$, $(2p\pi^1\Pi_u)$ and $(3p\pi^1\Pi_u)$ molecular states around \SI{12.6}{\electronvolt}~\cite{Geiger:1964dn} and the binary-encounter-dipole (BED) model~\cite{KIM:1994tf} to describe the continuous ionization tail at energy losses above \SI{15.5}{\electronvolt} (Fig.~\ref{fig:eloss}~center). The model has nine parameters given by the mean, width, and strength of each Gaussian. The normalization of the tail is chosen such that one obtains a smooth continuation of the Gaussian part of the model at the ionization energy.

    To fit the measured TOF spectra, the model function is first convolved several times with itself, to account for multiple inelastic scatterings in the source, and then with the measured spectrum of unscattered electrons (peak at \SI{0}{\electronvolt} in Fig.~\ref{fig:eloss} center). This spectrum of electrons which have not undergone inelastic scattering naturally includes the effects of elastic scattering and the filter width of the main spectrometer. The resulting curves for single and multiple scattering are then weighted with the Poisson-distributed scattering probabilities and summed. The expectation value of this Poisson distribution is a nuisance parameter in the fit. A combined fit of TOF spectra taken at different column densities must also account for differences in the e-gun laser intensity between the individual measurements, leading to changes in the count rate. Additional normalization factors are therefore included as nuisance parameters in the fit. Finally, additional background components are included in the fit. Background electrons produced by the impact of positive ions onto the photocathode of the e-gun, for example, do not exhibit a TOF structure and appear in the differential spectrum as a small additional component with the shape of an integral energy-loss spectrum. The scaling factors of this background are additional nuisance parameters.

    We performed a combined fit to four TOF datasets measured at different column densities. Each dataset contains about \num{12} hours of data, resulting in $\sim$\num{6e5} events surviving the TOF cut. The nine model parameters of interest are shared between all datasets, whereas each dataset has its own nuisance parameters as described above.

    The resulting best-fit parametrization is shown in Fig.~\ref{fig:eloss} top and center for the integral and differential data, respectively. The same energy-loss function describes all four datasets well and the fit has a reduced $\chi^2$ close to one.

    Uncertainties used in this work are of a statistical nature only. However, more advanced combined fits that also take into account the integral energy-loss measurements yield the same parameter values within their statistical uncertainties.

    Systematic uncertainties in the energy-loss determination are largely canceled by alternating up- and downward scans. A study of systematic effects on the parameter uncertainties has been undertaken using a Monte Carlo (MC) approach and taking into account disturbances like column-density drifts, background events, detector pileup and the binning of the continuous voltage ramp. These systematic uncertainties are negligible for the KNM1 analysis.

    An improved parametrization of the energy-loss function and its uncertainties is under investigation for future, more sensitive neutrino-mass campaigns.

    \section{Background}
    \label{sec:background}

The rate of background events during KNM1 was dominated by the two steady-state mechanisms described in Sec.~\ref{sec:bkg-steady}. In Sec.~\ref{sec:bkg-durationdep}, we also consider a background dependent on the duration of the corresponding scan step. 

\subsection{Steady-State Background}
\label{sec:bkg-steady}

    The steady-state background originates from excited or unstable neutral atoms which can propagate freely in the ultra-high-vacuum environment of the main spectrometer. It has two primary causes.

    First, a significant part of the steady-state background arises from hydrogen Rydberg atoms sputtered from the inner spectrometer surfaces by $^{206}$Pb recoil ions following $\upalpha$ decays of $^{210}$Po. These processes follow the decay chain of the long-lived $^{222}$Rn progeny $^{210}$Pb, which was surface-implanted from ambient air (activity $\sim$\SI{1}{\becquerel\per\square\metre}) during the construction phase. A small fraction of these Rydberg atoms is ionized by black-body radiation when propagating through the magnetic flux tube. The resulting sub-\si{\electronvolt} scale electrons are accelerated to $qU$ by the MAC-E-filter, adding a Poisson component to \rbg{}.

   The second significant steady-state background mechanism originates with $\upalpha$ decays of single $^{219}$Rn atoms ($t_{1/2} =$ \SI{3.96}{\second}) emanating from the non-evaporable-getter pumps. Each decay releases a large number of electrons up to the \si{\kilo\electronvolt} scale. If the decay occurs in the magnetic flux tube, these electrons are stored due to their significant transverse momenta. They subsequently produce secondary electrons by scattering on the residual gas until they have cooled to energies of a few \si{\electronvolt}, when they can escape; both primary and secondary electrons contribute to \rbg{} at $qU$~\cite{Frankle:2011xy}. Since several background electrons may originate from each $^{219}$Rn decay in the magnetic flux tube, this background source is not purely Poissonian. Liquid-nitrogen-cooled copper baffles at the ports to the getter pumps mitigate this effect by preventing $^{219}$Rn from diffusing into the sensitive volume~\cite{Harms2015, Goerhardt:2018wky}. Due to the formation of a thin layer of H$_2$O covering the baffle surface, the retention of $^{219}$Rn was hampered such that \rbg{} retains an observable non-Poissonian component during KNM1.

    In KNM1, the overall steady-state background rate, \rbg{}, is continuously measured through the energy-independent part of the spectrum R($\langle qU \rangle$). The whole spectrum is fitted, leading to a value over the \num{117} selected pixels of $R_\mathrm{bg} = \SI{0.293\pm0.001}{cps}$ that is largely constrained by the \num{5}~scan steps above the expected \ezero{}. This value is consistent with data from independent background runs. Full fit results are given in Sec.~\ref{sec:fit-results}.

    \begin{figure}[t!]
        \centering
        \includegraphics[width=0.45\textwidth]{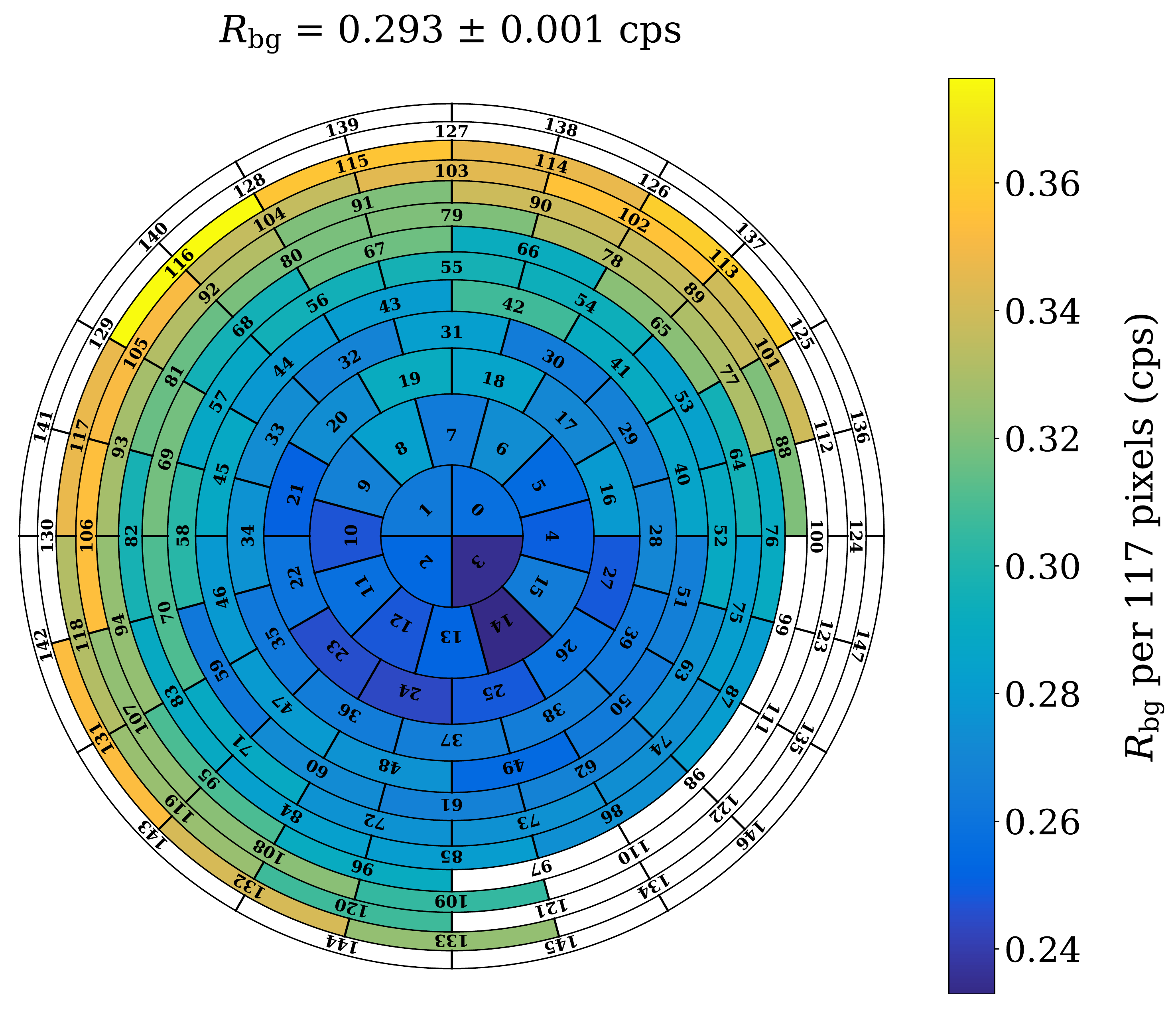}
        \caption{Distribution of pixel-wise, steady-state background rates during the first neutrino mass campaign for the \num{117} included pixels. The background rate increases radially from the center by about \SI{50}{\percent}. White pixels are excluded from the analysis (Sec.~\ref{sec:electron_counting}).}
        \label{fig:bgSpatialUniformity}
    \end{figure}

    The background is not distributed uniformly across the detector, as shown in Fig.~\ref{fig:bgSpatialUniformity}. The decrease of \rbg{} towards smaller radii can be explained by radiative de-excitation of the Rydberg atoms as they propagate inside the main spectrometer. Further from the spectrometer wall, fewer Rydberg atoms are therefore available for ionization by the thermal radiation.

    \begin{figure}[t!]
        \centering
        \includegraphics[width=0.45\textwidth]{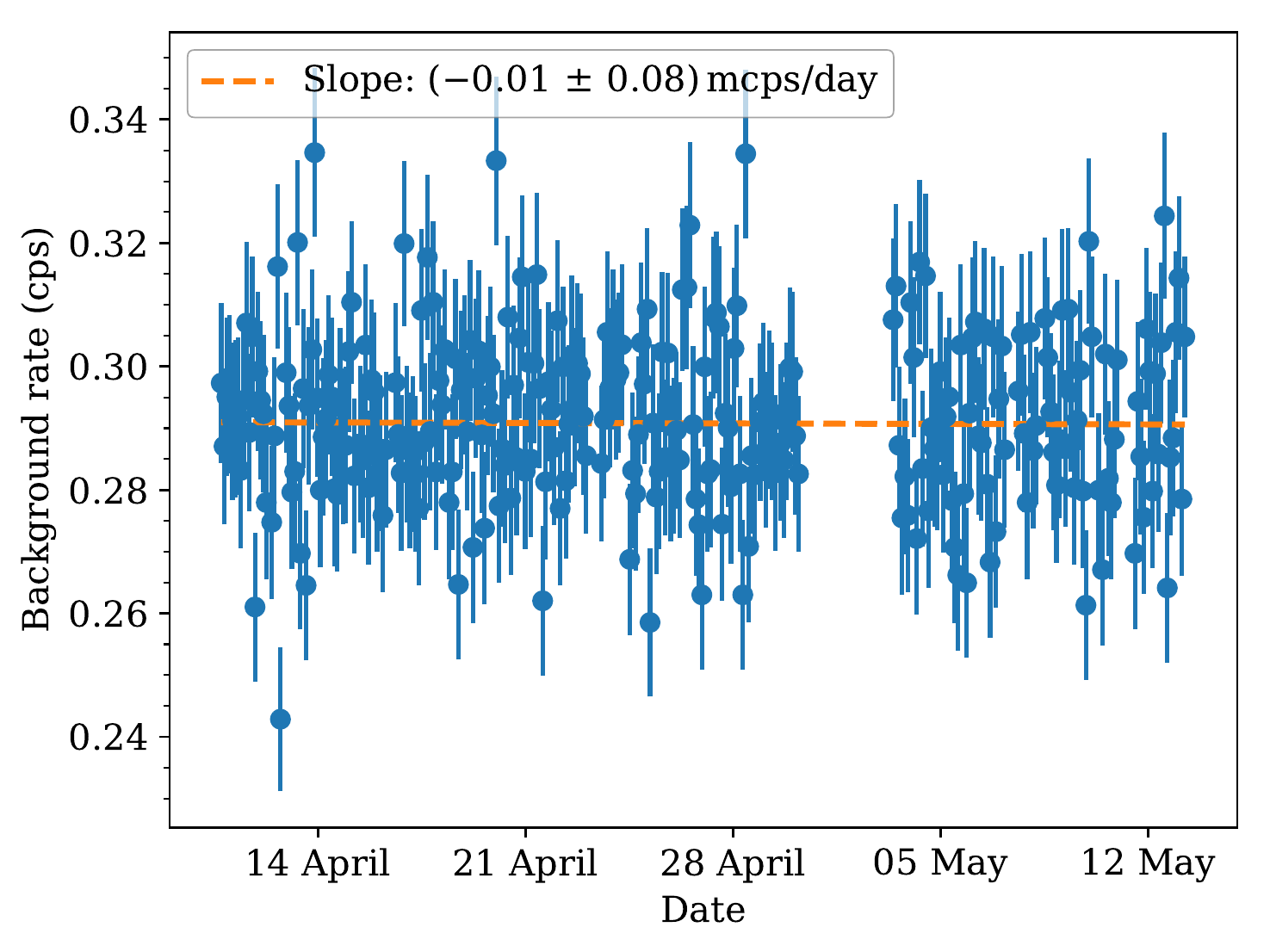}
        \caption{Evolution of the background rate during KNM1, measured via the five background-region scan steps above the expected endpoint \ezero{}. The slope of a linear fit to the data is compatible with zero, indicating the long-term stability of the background.}
        \label{fig:bgTimeStability}
    \end{figure}

    The steady-state background was monitored for each $\upbeta$-scan with the five dedicated background-region scan steps. Figure~\ref{fig:bgTimeStability} shows the time evolution of these background measurements during KNM1. A linear fit was applied to the data in order to test the long-term stability of the background. The slope of \SI{-0.01\pm0.08}{\milli cps\per\day} is compatible with a background that is stable over long time scales.

    \begin{figure}[t!]
        \centering
        \includegraphics[width=0.45\textwidth]{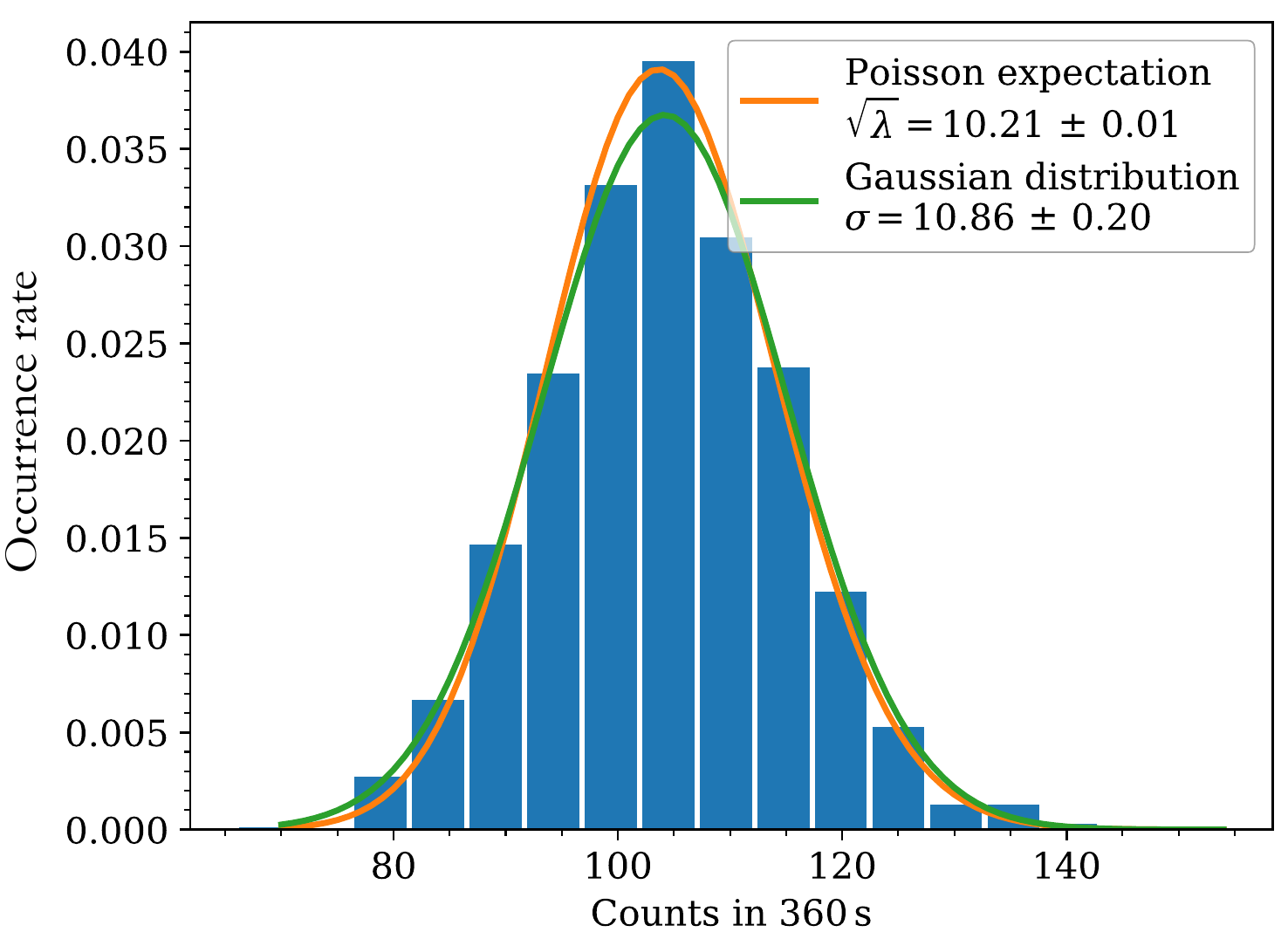}
        \caption{Background-event distribution for the five background-region scan steps during KNM1. The fit of a Gaussian profile to the measured data yields a width of \SI{10.86\pm0.20}{counts}. This constitutes an increase of \SI{6.4}{\percent} with respect to the expectation, from pure Poisson statistics, of \SI{10.21\pm0.01}{counts}.}
        \label{fig:bgNonPoisson}
    \end{figure}

    The non-Poissonian component of \rbg{} causes a broadening of the event distribution of the five background-region scan steps, amounting to \SI{6.4}{\percent} compared to the prediction from pure Poisson statistics (Fig.~\ref{fig:bgNonPoisson}).

    \begin{figure}[t!]
        \centering
        \includegraphics[width=0.45\textwidth]{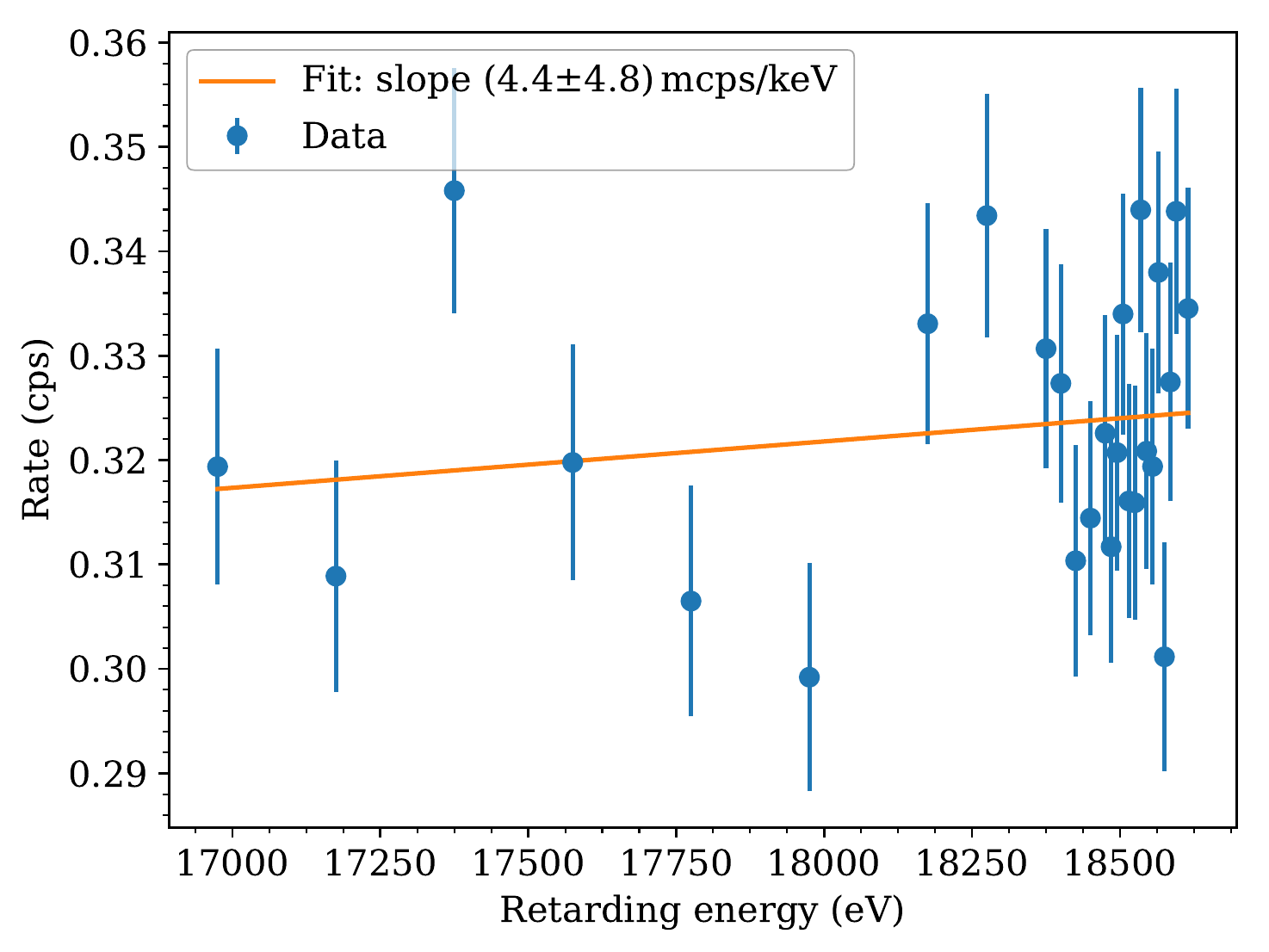}
        \caption{Background measurement without tritium source in a region near \ezero{}. The slope of a linear fit to the data is compatible with zero, supporting our assumption that the background is independent of  $qU$. Fits to the immediate \ezero{} region, where scan steps are more evenly spaced, also find no significant trend in $qU$.}
        \label{fig:bgSlope}
    \end{figure}

	Our model predicts a background that is independent of $qU$ near \ezero{}. To test this expectation, we performed a dedicated background-only measurement, without an active tritium source, in June 2018. As shown in Fig.~\ref{fig:bgSlope}, $qU$ was scanned in \num{26} steps over an interval of \SIrange{16.975}{18.615}{\kilo\electronvolt}. We then fit a line with a free slope parameter to these data. The resulting best-fit slope, \SI{-2.2\pm4.3}{\milli cps\per\kilo\electronvolt}, is compatible with zero, and we take its uncertainty as an overall uncertainty on our assumption of a $qU$-independent background (Sec.~\ref{sec:syst-bkg}).

\subsection{Background Dependence on Scan-Step Duration}
\label{sec:bkg-durationdep}

With both the pre-spectrometer and main spectrometer held at negative retarding potentials, a Penning trap inevitably forms in the strong magnetic field of the grounded inter-spectrometer region. Electrons trapped in this region slowly lose energy by ionizing residual gas molecules. The resulting ions may escape into the main spectrometer, where they can create background electrons when their own collisions with the residual gas or the vessel wall release ionization electrons, Rydberg atoms, or photons. The intense WGTS feeds the Penning trap when \belec{}s produce positive ions on their way into the pre-spectrometer; these ions sputter Rydberg atoms from the pre-spectrometer walls, and the Rydberg atoms in turn produce low-energy ionization electrons that fill the trap~\cite{KATRIN-Penning:2020}. This mechanism may also play a role in main-spectrometer backgrounds, when \belec{}s scatter further downstream and the resulting ions strike the main-spectrometer walls.

During each transition to a new scan step, an electron catcher is briefly inserted into the beamline to remove stored electrons from the Penning trap. At higher pre-spectrometer potential, this has been shown to provide a statistically significant reduction in the baseline background~\cite{KATRIN-Penning:2020}. However, since the electron catcher is inserted only at the beginning of a scan step, the Penning trap continues to fill until a new electron-catcher actuation at the beginning of the next scan step. The corresponding rise of the background rate is strongly influenced by surface conditions and by the achieved pressure between the spectrometers. In principle, however, this mechanism can produce a background that effectively increases in rate for longer-duration scan steps (see measurement-time distribution in Fig.~\ref{fig:MTD}). This effect was observed in a subsequent KATRIN scientific run, but for KNM1 -- the initial science run, with pristine surfaces and lower column density -- no statistically significant dependence on scan-step duration was observed. Section~\ref{sec:syst-bkg} will address the impact on the neutrino-mass measurement.

    \section{Assembling spectral data for KNM1}
    \label{sec:general_analysis}
    Data are acquired in a sequence of $\mathcal{O}$(\SI{2}{\hour}) scans and the integral spectrum (Eq.~\ref{eq:int_spec}) is recorded with the FPD. In the final analysis (Sec.~\ref{sec:spectralfit}), the spectral fit uses four free parameters: the signal amplitude $A_{\mathrm{s}}$, the effective \bdecayh\ endpoint \ezero , the background rate $R_\mathrm{bg}$,  and the squared neutrino mass \mtwonue. In this analysis we leave \ezero\ and $A_{\mathrm{s}}$ unconstrained, which is equivalent to a ``shape-only'' fit. The 4-parameter fit procedure over the averaged scan steps $\langle qU \rangle$ compares the experimental spectrum R($\langle qU \rangle$) to the model $R^\mathrm{model}(\langle qU \rangle)$.

    Spectra from all of the scans and pixels have to be combined in the final analysis without loss of information. In the following we describe the strategy applied to combine all these data prior to the final spectral fit to extract the effective neutrino mass.

    \subsection{Pixel combination}
    During KNM1, the electric potential and magnetic field in the analyzing plane of the  main spectrometer were not perfectly homogeneous, but varied radially by about \SI{140}{\milli\volt} and \SI{2}{\micro\tesla}, respectively, and to a much smaller extent azimuthally. The pixelation of the detector allows us to account for these spatial dependencies. Each pixel has a specific transmission function and records a statistically independent tritium $\upbeta$-electron spectrum. In this analysis, we combine these pixel-wise spectra into a single effective pixel by adding all counts and assuming an average transmission function for the entire detector. The averaging of fields leads to a negligible broadening of the spectrum which does not affect the filter width, and carries a negligible bias of $\mathcal{O}(\SI{E-3}{\electronvolt})$ on \mtwonue.

\begin{figure}[t!]
        \centering
        \includegraphics[width=0.45\textwidth]{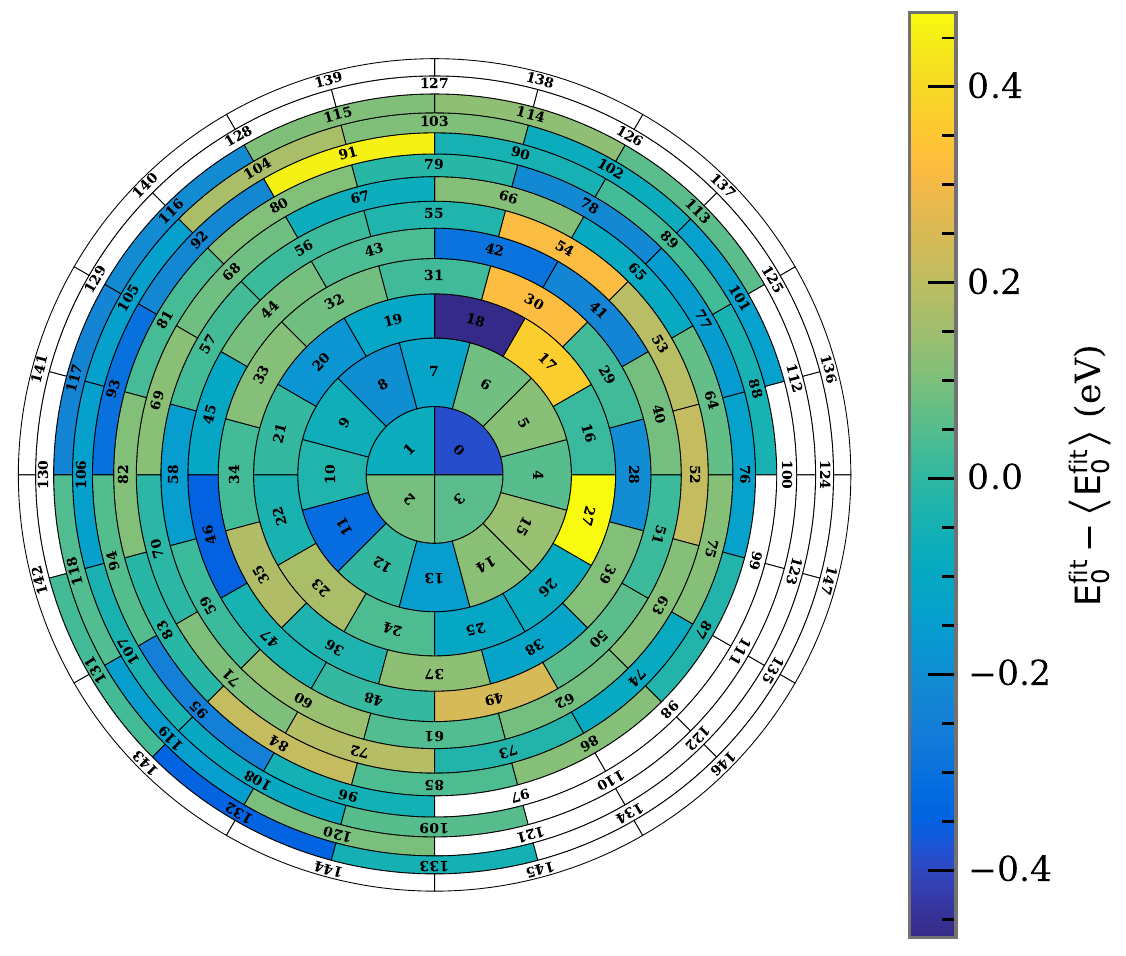}
        \caption{Distribution of the endpoint $E^\text{fit}_0$ over the detector pixels. No spatial inhomogeneity beyond statistical fluctuations is observed, justifying the merge of the data of all \num{117} pixels for the subsequent analysis. White pixels are excluded from the analysis (Sec.~\ref{sec:electron_counting}).}
        \label{fig:stackedsinglepixel_endpoint}
    \end{figure}

    Combining all 274 scans that passed data-quality cuts, single-pixel fits were performed resulting in an endpoint $E^{\text{fit}}_0$ for each pixel, as shown in \cref{fig:stackedsinglepixel_endpoint}. We find no systematic spatial (\textit{i.e.} pixel) dependence of $E^{\text{fit}}_0$. The standard deviation from the mean endpoint is \SI{0.16}{\electronvolt}, which is consistent with statistical fluctuations. This indicates a good description of the electric potential and magnetic field in the analyzing plane, and the absence of a significantly spatially dependent electron starting potential. We therefore merge the data of all~\num{117}~selected pixels used in the analysis (Fig.~\ref{fig:KNM1PixelSelection}).

    \subsection{Scan combination (stacking)} \label{sec:stacking}

\begin{figure}[tbp]
        \centering
        \includegraphics[width=0.45\textwidth]{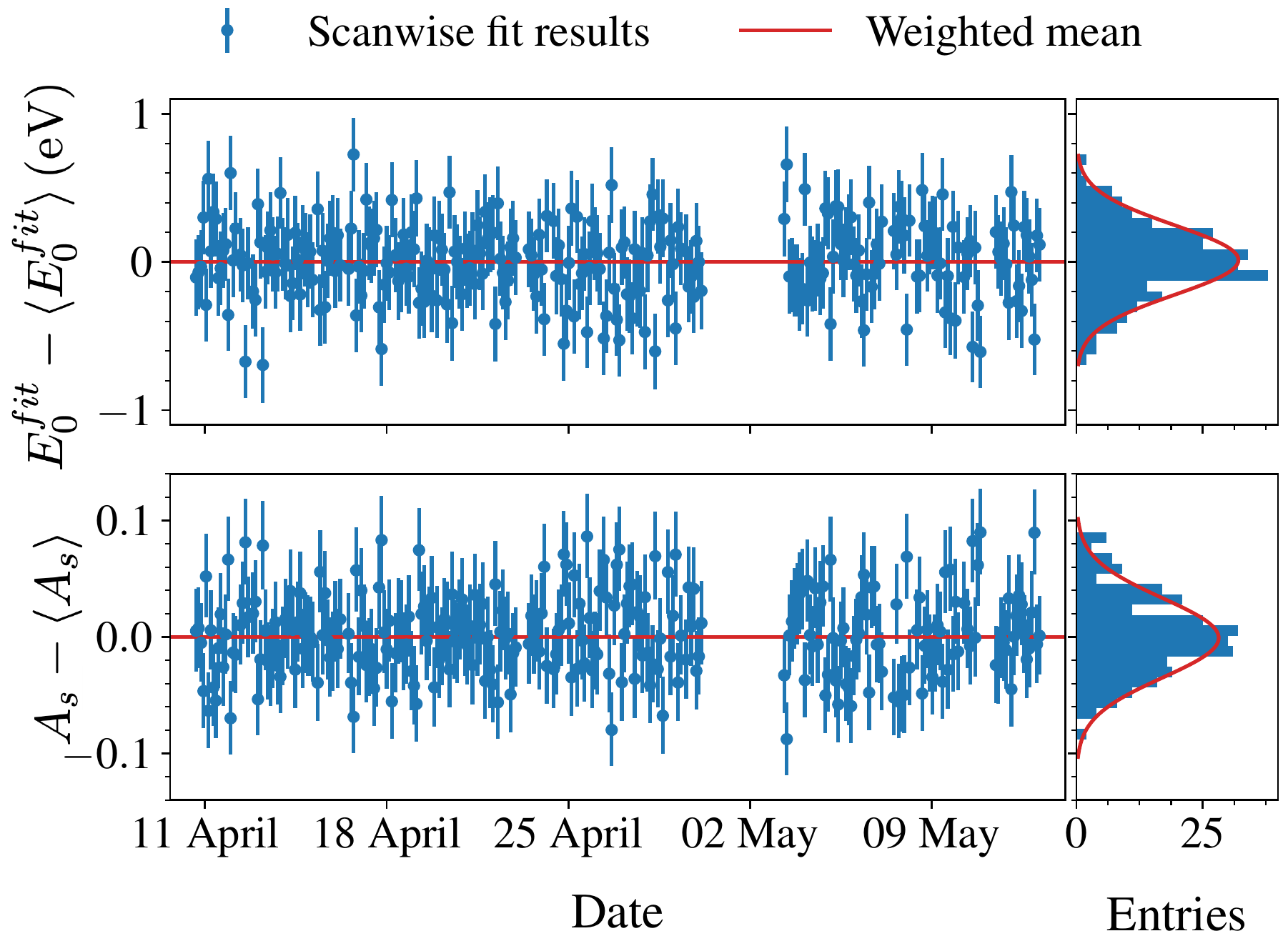}\label{fig:singlescansfit}
        \caption{Evolution of two fit parameters, the endpoint $E^{\text{fit}}_0$ (upper panel) and the signal normalization $A_{\mathrm{s}}$ (lower panel), as functions of time during the whole KNM1 data-taking period. Each plot shows the deviation of the fit parameter, evaluated on a per-scan basis, from its weighted mean during KNM1.}
        \label{fig:runwise-fitparam}
    \end{figure}

    Combining all pixels in a uniform fit, we can now consider the stability of the fit parameters with respect to possible temporal variations. We investigate all four free parameters in the fit. For single scans of \num{2} hours, the accumulated statistics are not sufficient to significantly constrain the neutrino mass. Therefore, the neutrino mass is fixed to zero. The \num{274} fit values show excellent stability over the course of a month (Fig.~\ref{fig:runwise-fitparam}). The standard deviation from the mean endpoint is \SI{0.25}{\electronvolt}, which is again consistent with statistical fluctuations.

    In order to constrain the neutrino mass, the statistics of all \num{274} scans must be combined. Based on our stability results, we achieve this by merging the data of all \num{274}~scans into a single stacked, integral spectrum. In the underlying process, the events at like scan steps are summed and the corresponding retarding-potential values are averaged over all scans. This procedure yields one high-statistics integral spectrum with the same number of scan steps as a single scan. Since this method does not correct for scan-to-scan variations of slow-control parameters, it relies on good time stability and excellent reproducibility of the individual HV settings from scan to scan. The Gaussian spread of these HV settings is on average $\sigma =$ \SI{34\pm1}{\milli\volt} (better than 2~ppm) (Sec.~\ref{sec:ap-potentials}). The scan stacking results in a minor systematic effect, which is included in the analysis.

    \subsection{Resulting integral spectrum }

\begin{figure}[h!]
        \centering
        \includegraphics[width=0.45\textwidth]{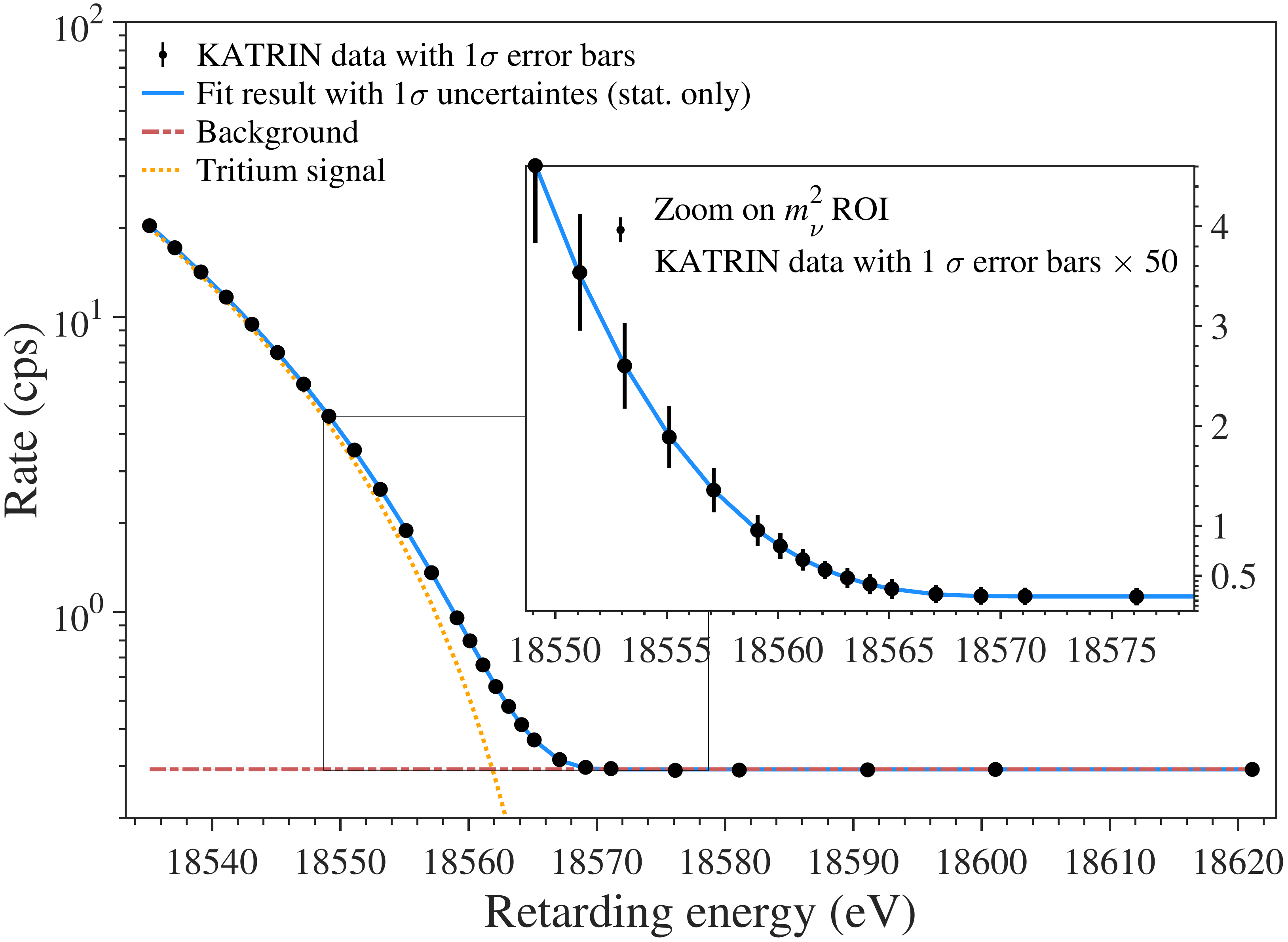}
        \caption{Display of the measured KNM1 endpoint \bspec{} after scan and pixel combination, superimposed to the best-fit model (Sec.~\ref{sec:spectralfit}).}
        \label{fig:experimentalData}
    \end{figure}

    The resulting stacked integral spectrum, R($\langle qU \rangle$), is displayed in Fig.~\ref{fig:experimentalData}. It comprises a total of \num{2.03e6} events, with \num{1.48e6} \bdecay\ electrons below \ezero\ and a flat background ensemble of \num{0.55e6} events in the \SI{86}{\electronvolt} analysis interval, $\lbrack \ezero - $\SI{37}{\electronvolt}, $\ezero + $\SI{49}{\electronvolt}$\rbrack$.

    \section{Systematic uncertainties}
    \label{sec:systematic_uncertainties}

    Systematic uncertainties generally arise from parameter uncertainties that enter into the calculation of the integral spectrum, and  from instabilities of experimental parameters. The KNM1 analysis heavily relies on a precise description of the spectral shape, including all relevant systematic effects and a robust treatment of their uncertainties. Any erroneously neglected effect or uncertainty can lead to a systematic shift of the deduced neutrino mass~\cite{Otten2008}. The individual systematics are described in detail below. A summary of these systematic uncertainties is given in \cref{tab:systematics_input}, while their ultimate impacts on the \mtwonue{} uncertainty budget are collated in Table~\ref{tab:systematics_breakdown-final}.

        \begin{table}[]
        \centering
        \caption{Summary of systematic uncertainties used as input for the neutrino-mass inference. Details of each entry (including those neglected in this analysis) are given in the text.}
        \label{tab:systematics_input}
        \begin{tabular}{lll}
            \toprule
            Effect &  Description & \SI{1}{\sigma} uncertainty \\
            \midrule
            Background & Rate over-dispersion & \SI{6.4}{\percent} \\
                       & Slope & \SI{5.0}{\milli cps\per\kilo\electronvolt} \\
                       & Rate dependence on &  \\
                       	&  \quad scan-step duration & neglected \\
                       \midrule
            Source effects & Expected number of &  \\
            					& \quad scatterings ($\rho d\sigma$) & \SI{0.85}{\percent} \\
                              & Energy-loss function & $\mathcal{O}$(\SI{1}{\percent}) \\
                              & $\beta$ starting potential & neglected \\
                       \midrule
            Scan fluctuations & Column density & \SI{0.8}{\percent} \\
                              & Tritium isotopologs & \SI{0.4}{\percent} \\
                              & High voltage & \SI{2}{ppm} \\
                       \midrule
           Magnetic fields   & Source & \SI{2.5}{\percent} \\
                              & Analyzing plane & \SI{1.0}{\percent} \\
                              & Maximum & \SI{0.2}{\percent} \\
                       \midrule
            Final states      & Normalization: & \\
            					& \qquad Ground state & \SI{1.0}{\percent} \\
                              & Variance: & \\
                              & \qquad Ground state & \SI{1.0}{\percent} \\
                              & \qquad Excited states & \SI{4.0}{\percent} \\
                       \midrule
            Detector efficiency & High-voltage dependence & neglected \\
            \bottomrule
       \end{tabular}
    \end{table}

    \subsection{Tritium concentration}

    The concentration of the tritium isotopologs in the source affects the model in two different ways.

    First, the total activity is directly correlated to the tritium purity described in \cref{eq:tritium_purity}. The absolute number does not impact the neutrino-mass measurement, as the signal normalization is a free fit parameter. Changes during a given scan, however, could introduce a slight spectral distortion which would bias the measurement. As described in Sec.~\ref{sub:source_composition}, the tritium purity was measured continuously by the LARA laser-Raman spectroscopic system. The precision was determined from the shot noise $\sqrt{N_x}$ of the Raman signal and then propagated to $\varepsilon_{\mathrm{T}}$ and $c_x$; the resulting precision of better than \num{2e-3} for each scan was reported in Ref.~\cite{Aker:2019uuj}. Scan-to-scan fluctuations of the tritium purity amount to \num{0.39e-3} after accounting for anti-correlations between the isotopologs.  

    Second, each of the three tritium isotopologs also has a slightly different FSD. Systematic uncertainties on their relative fractions, mainly determined by the trueness of the LARA calibration, thus propagate into the spectral shape. The impact on \mnuesq{} from this effect is less than \SI{2e-4}{\electronvolt\squared} and is thus negligible for KNM1.

    \subsection{Column density and expected number of scatterings}
    The determination of the expected number of scatterings, $\rho d \sigma$, is described in Sec.~\ref{ch:columnDensityDetermination}. The total uncertainty on $\rho d\sigma$ arises from three separate contributions: the limited precision of single column-density measurements made with the e-gun; uncertainty on the throughput measurement, arising from fluctuations of the gas throughput and imperfect reproducibility of the flow meter; and the scaling of the inelastic-scattering cross section to a lower electron energy via Eq.~\ref{eq:FerencCrossSection}. This last operation is necessary because the e-gun is operated at an energy of \SI{18.78}{\kilo\electronvolt}, well above \ezero{}, for measurements of the column density -- but the \belec{}s, at lower energies, have a slightly different scattering cross section. We take \SI{18.575}{\kilo\electronvolt} as a representative value for our observed \belec{}s; the variation of the inelastic-scattering cross section within the analysis interval is negligible.

    Taking these three contributions into account leads to a total systematic uncertainty on $\rho d \sigma$ of less than \SI{0.85}{\percent} for all scan steps.

    \subsection{Electron starting potential}
    \label{sec:syst_startingpot}

Spatial inhomogeneities and temporal fluctuations of the starting potentials of the \belec{}s would lead to a shift of the neutrino mass according to Eq.~\ref{eq:sigma}. As discussed in Sec.~\ref{sec:starting_potential}, the intrinsic width of the \krm{} L$_3$-$32$ line is a diagnostic tool to investigate these effects, probing the plasma-potential distribution.

		In the KNM1 analysis, we treat the fitted Gaussian line broadening in the presence of a T$_2$ plasma as a conservative upper limit for the inhomogeneity of the plasma potential, yielding a negative \mnuesq{} shift with magnitude less than \SI{0.013}{\electronvolt\squared}.

    Electrons undergo inelastic scattering as described in Sec.~\ref{sec:responsefunctionmodel}. The $s$-fold scattering probabilities for each \belec{} depend on the longitudinal position of its creation. As a result, the populations of \belec{}s with different scattering multiplicities also have different distributions of starting positions, and therefore different distributions of starting potentials if the plasma potential is inhomogeneous. Analysis of the positions of the krypton L$_3$-$32$ lines of unscattered and singly scattered electrons shows that a plasma-induced mutual shift of these positions cannot be larger than \SI{70}{\milli\volt}. 
    The corresponding additional \mnuesq-shift can be neglected for KNM1. 
	We thus conclude that the effective L$_3$-$32$ broadening parameters given above serve as a very conservative upper limit of plasma effects in the neutrino-mass analysis.

    In addition to the \krm{} spectroscopy method, radial plasma inhomogeneities can be inferred directly from the neutrino-mass data by radial evaluation of \ezero{}. The spectral fit from twelve separate detector rings (see Fig.~\ref{fig:KNM1PixelSelection} for detector structure) revealed a slope of \SI{-2\pm5}{\milli\volt\per ring}, consistent with a slope of zero.

	A full propagation of the plasma model and its uncertainty was not included in the KNM1 analysis, primarily due to the immaturity of the plasma model as applied to the low KNM1 column density. Adding this $\mathcal{O}$(\SI{-0.01}{\electronvolt\squared}) uncertainty in quadrature to the total systematic uncertainty does not yield significant leverage on the total budget.

The neutral-gas density strongly affects the charge densities from secondary electrons and ions, as well as other plasma parameters. For this reason, we are currently investigating the effect of different column densities, gas temperatures, source magnetic-field strengths, and changing boundary conditions on plasma parameters.
This will inform the consideration of plasma effects in the data analysis for upcoming campaigns, in which the gas throughput will be higher by a factor of up to four.

    \subsection{Detector efficiency}
    Although numerous physical and detector effects can reduce the detector efficiency, any effects which do not depend on the retarding potential $U$ will not affect the KNM1 fit results due to the overall, free scaling parameter for each spectrum and the uniform, all-pixel fit.

    The overall FPD detection efficiency within the ROI has been estimated by both simulation and commissioning analysis to be approximately \SI{95}{\percent}, with an uncertainty of a few percent, and per-pixel variations of about the same size. For KNM1, the ROI is fixed regardless of $U$ (Sec.~\ref{sec:electron_counting}). However, the shape of the FPD energy spectrum changes with $U$, primarily due to the $\beta$-electron energy threshold at $qU$. Additional distortions are due to energy- or rate-dependent detector effects: energy loss in the dead layer, charge sharing among pixels, pileup, and back-scattering of electrons and their subsequent reflection back toward the FPD by local electric and magnetic fields.

    We have studied the effects of these spectral shape changes using a reference spectrum for each pixel, acquired at $U_0 =$ \SI{-18375}{\volt}. For each scan step at $U_i = U_0 + \Delta U$, the reference spectrum is shifted by the corresponding $q\Delta U$ and a count correction is calculated. As $|U|$ decreases, the corrections become larger, with a maximum size of about \SI{0.05}{\percent}. We estimate the error relative to these correction factors at less than \SI{0.05}{\percent}, determined by comparing spectral shapes at nearby $U$ values. In the KNM1 analysis, we apply these corrections to FPD counts while neglecting the corresponding uncertainty.

    Pileup events also result in event loss, since the energy is erroneously reconstructed above the upper bound of the ROI. We assume that pileup events arise from random coincidences; each coincidence produces a total energy deposit that is an integer multiple of \SI{28.6}{\kilo\electronvolt}, within the shaping time of the trapezoidal filter. We calculate and apply the corresponding correction to the event rate for each pixel and scan step, up to a maximum correction factor of \SI{0.02}{\percent} at low $|U|$ and, correspondingly, high rate. Our conservative estimate of the relative error on these correction factors is less than \SI{18}{\percent}, based on the shape of the measured FPD energy spectrum and a simulation of the trapezoidal filter. This error is negligible.

    Our final consideration is electron backscattering from the FPD. The majority of backscattered electrons are reflected back to the FPD, either by magnetic fields in the detector system, or by the electric potentials of the post-acceleration electrode or main spectrometer. Even with multiple backscatters, the electron returns to the same pixel each time, always arriving well within the shaping time of the trapezoidal filter, so that the detector does not register the event as separate hits. Our spectral-shape calculations include the resulting reconstructed-energy shifts, due to multiple transits of the detector dead layer and hits distributed within the shaping time. However, an additional correction is in principle needed for those few backscattered electrons which have enough energy to surmount the $qU$ threshold and escape towards the source. Simulations show that the resulting event loss is less than \SI{0.01}{\percent} for the KNM1 analysis window. This effect is therefore neglected in this analysis.

\subsection{Final-state distribution}
    The uncertainty estimation on the FSD is based on differences between the theoretical \emph{ab initio} calculations from Saenz~\emph{et al.}~\cite{Saenz2000} and Fackler~\emph{et al}~\cite{fackler:1985}. The difference between the calculations for the ground-state variance is found to be small, of $\mathcal{O}$(\SI{1}{\percent})~\cite{Bodine:2015sma}. However, the descriptions of the electronic excited states and the electronic continuum exhibit larger discrepancies.

    We conservatively estimate the uncertainty on the variance of the ground state (excited states and continuum) to be \SI{1}{\percent} (\SI{4}{\percent}). The uncertainty on the normalization of the ground to excited-state populations is taken as \SI{1}{\percent}.

    Our narrow analysis interval, extending \SI{37}{\electronvolt} below \ezero{}, is dominated by electrons from the ground-state distribution. Consequently, the uncertainty on the FSD only contributes on the order of $\mathcal{O}(10^{-2})\,$\si{\electronvolt\squared} to the total systematics budget on \mtwonue{} within our analysis interval.

    \subsection{Response function} \label{sec:system-resp-func}
    Response-function-related systematic uncertainties are connected with the electromagnetic fields that define the transmission function (Eq.~\ref{eq:transm_func}) and with the energy-loss function.  The electromagnetic fields are computed from a simulation of the beamline magnets and the main-spectrometer vessel.

    \paragraph{Magnetic fields}
    Systematic uncertainties on the magnetic field at the analyzing plane arise from residual magnetic fields in the spectrometer hall, \textit{e.g.}\,due to magnetized materials, and from model imperfections. A sensor network was used to compare measured fields at the spectrometer vessel to simulation results. Our assessment of the maximum deviation yields a conservative systematic uncertainty of $\Delta B_\mathrm{min} / B_\mathrm{min} =$ \SI{1}{\percent}.

    The maximum magnetic field, located at the exit of the main spectrometer, was measured in 2015 at the center of the magnet bore~\cite{Arenz:2018jpa} and compared to simulations. We include a conservative systematic uncertainty of $\Delta B_\mathrm{max} / B_\mathrm{max} =$ \SI{0.2}{\percent}.

    The source magnetic field was measured in 2009 by the manufacturer with Hall probes on the central axis and compared to simulations. We include a conservative systematic uncertainty of $\Delta B_\mathrm{S} / B_\mathrm{S} =$ \SI{2.5}{\percent}.

    \paragraph{Electric potentials}
Since any offset of the simulated retarding potential at the analyzing plane is compensated by the free endpoint parameter, no additional systematic uncertainty is assigned for the spectral fit.

    \paragraph{Energy-loss function}
        The uncertainty of the energy-loss parametrization is obtained from fits to the measurements described in Sec.~\ref{sec:eloss-function}. For each of the \num{9} parameters describing the energy-loss function, an individual fit uncertainty is determined. As stated in Sec.~\ref{sec:eloss-function}, the contribution of systematic effects is about one order of magnitude lower than the uncertainties related to the current statistics of the e-gun measurements. As a result, only statistical fit uncertainties are considered for this analysis. Correlations between the energy-loss parameters are taken into account, reducing the overall uncertainty of the energy-loss function with respect to the uncorrelated case.

    The systematic effect on $m_\nu^2$ due to the uncertainties of the energy-loss function is determined to be below \SI{0.01}{\electronvolt\squared}.

    \subsection{Background}
    \label{sec:syst-bkg}

    The steady-state background enters the uncertainty budget in two independent ways: rate and shape.

    The background rate distribution, as shown in \cref{fig:bgNonPoisson} shows an over-dispersion of \SI{6.4}{\percent} compared to the Poisson expectation. This enters the analysis as an additional uncorrelated uncertainty on the background rate, effectively increasing the statistical error in the region with $E > \ezero{}-15$~eV.

    As described in Sec.~\ref{sec:bkg-steady}, we expect the background to be flat with respect to the retarding potential. In this analysis we assess the slope uncertainty via a slope parameter, which makes a first-order correction to the constant expectation. Based on the dedicated measurement in June~2018 (Fig.~\ref{fig:bgSlope}), the slope parameter is consistent with zero, within an uncertainty of \SI{5}{\milli cps \per \kilo \electronvolt}. In the final spectral fit (Sec.~\ref{sec:spectralfit}), we use a central value of  \SI{0}{\milli cps \per \kilo \electronvolt}.
       
	A Penning-induced background (Sec.~\ref{sec:bkg-durationdep}) may increase over the course of each scan step, effectively introducing a higher background for scan steps with longer duration. Since longer scan steps are concentrated near \ezero{} $-$ \SI{14}{\electronvolt} (Sec.~\ref{sec:MTD}), the net effect is a shape distortion of the background shape. An analysis of KNM1 scan steps yields a best-fit linear time slope of \SI[separate-uncertainty = true]{-3.8(44)}{\micro cps \per \second}, which would result in a systematic uncertainty of \SI{0.15}{\electronvolt\squared} on the squared neutrino mass. This systematic was not taken into account in the spectral fit (Sec.~\ref{sec:spectralfit}), but would not alter the statistics-dominated final uncertainty.

    \subsection{Stacking}
    \label{sec:stacking-HV}
    The averaging of the scan steps within the stacking techniques introduces a small bias on \mtwonue{} and \ezero{}. In order to quantify these biases, we construct an Asimov dataset~\cite{Cowan:2010js} by simulating \num{274} statistically unfluctuated ``MC twin'' spectra, incorporating the actual variation of slow-control parameters (including measured high-voltage values, isotopic compositions, and column densities) between scans. Later on, the MC spectra are combined into a single integral spectrum through the stacking procedure, as described in Sec.~\ref{sec:general_analysis}. As a last step, we fit this stacked MC spectrum. Comparing this fit result to the MC truth yielded a $1\sigma$ stacking uncertainty of \SI{14e-2}{\electronvolt\squared} in one analysis approach (Sec.~\ref{sec:covariance-matrix}), and \SI{5e-2}{\electronvolt\squared} in the other (Sec.~\ref{sec:mcprop}), as shown in Table~\ref{tab:systematics_breakdown-final} further below. The discrepancy between the two approaches arises from different treatments of the individual contributions to this subdominant uncertainty; the stacking method and error treatment will be optimized in the analysis of future neutrino-mass campaigns, in which scan-to-scan fluctuations are also expected to be smaller.

    \subsection{Neutrino-mass fit range}
    \label{sec:fit-range}
    The full spectrum was recorded over a large energy range down to \ezero{} $-$ \SI{91}{\electronvolt}. Several systematic uncertainties, like those related to inelastic scattering and the FSD, increase further away from the endpoint, while the statistical uncertainty decreases. The optimization of the neutrino-mass fit range is performed using MC twin simulations of KNM1 (Sec.~\ref{sec:stacking-HV}), assuming a zero neutrino mass and using the set of systematics presented earlier in the section (Table~\ref{tab:systematics_input}). The lower bound of the fit interval is then varied between \ezero{} $-$ \SI{91}{\electronvolt} and \ezero{} $-$ \SI{30}{\electronvolt}, and two fits are performed in turn. The first fit considers statistical uncertainty only, while the second fit uses both statistical and systematic errors. For each pair of fits, the systematic uncertainty is deduced by subtracting the statistical uncertainty in quadrature from the total error. As a result, both statistical and systematic uncertainties become equal for the fit range starting at about \ezero{} $-$ \SI{70}{\electronvolt}, and systematic uncertainties become dominant when including data below \ezero\ $-$ \SI{70}{\electronvolt}. Moreover, the overall sensitivity only marginally improves by including data at energies below \ezero\ $-$ \SI{40}{\electronvolt}.

    This study addresses only the dependence of the measurement precision on the fit range. It does not address the accuracy of the determination of the neutrino mass, since the same model is used for the fit and for the MC twins. Indeed,
further than about \ezero{} $-$ \SI{40}{\electronvolt}, the electronic continuum -- with less well-validated modeling -- dominates the FSD (Sec.~\ref{subsec:FSD}).
Therefore, before unblinding the data (Sec.~\ref{sec:blinding_strategy}, below), we fixed the analysis interval to cover the region of \ezero{} $-$ \SI{37}{\electronvolt} (\num{22} scan steps) and \ezero{} $+$ \SI{49}{\electronvolt} (\num{5} scan steps).

    \section{Spectral fit}
    \label{sec:spectralfit}
    In this section we discuss our blinding method (Sec.~\ref{sec:blinding_strategy}) and present two approaches for inferring the value of the neutrino mass squared \mtwonue{} and the endpoint \ezero{} simultaneously, based on fitting the integrated \bspec{} (Eq.~\ref{eq:int_spec}) assembled as described in Sec.~\ref{sec:general_analysis}. In both approaches, the spectrum is fitted using a shape-only analysis with four free parameters. In addition to \mtwonue{} and \ezero{}, these are the signal amplitude $A_{\mathrm{s}}$ and the background rate $R_\mathrm{bg}$.

    The first approach (Sec.~\ref{sec:covariance-matrix}) uses a standard $\chi^2$ estimator and covariance matrices to encode all uncertainties. The second approach (Sec.~\ref{sec:mcprop}), Monte-Carlo propagation, repeats the final fits many times, for each fit choosing randomized input values for the systematic nuisance parameters.

Three analyses were performed, each with its own spectrum calculation and analysis software: two using the covariance-matrix approach, and one using the MC-propagation approach. The analyses were performed blind and give consistent results, as described in Sec.~\ref{sec:fit-results}. The resulting breakdown of systematic uncertainties is given in Table~\ref{tab:systematics_breakdown-final}, below. Section~\ref{sec:upperlimit} uses these spectral results to derive frequentist bounds on the neutrino mass, while Sec.~\ref{sec:bayesian_analysis} uses the same data to derive Bayesian bounds.

       \subsection{Blinding strategy}
    \label{sec:blinding_strategy}

    For the KNM1 analysis we enforced blind analysis procedures to fix data selection, analysis cuts, and model composition before the model was fitted to the data. This standard technique is designed to avoid observer's bias.

    For this first KATRIN \mtwonue{} limit, we employed model blinding rather than data blinding. The fit results are highly dependent on the molecular FSD (Sec.~\ref{subsec:FSD}); in particular, the value  of \mtwonue{} depends on the width of the distribution of transitions to the electronic ground state of the daughter molecule \HeT{}. Using an FSD with too large a width pushes \mtwonue{} towards higher values, while too narrow a width pushes it towards lower values. Indeed, historically, inaccurate FSD models were likely responsible for artificially negative \mtwonue{} results from the Los Alamos~\cite{Robertson:1991vn} and Livermore~\cite{Livermore1995} experiments, a problem which is resolved by using the more modern theory described in Sec.~\ref{subsec:FSD}~\cite{Bodine:2015sma}.

    If we fit the data with a model using an FSD ground-state width that has been picked randomly within a suitable interval,  the true value of \mtwonue{} cannot be retrieved. That is, the analysis is blind to its parameter of interest, while  the remaining three parameters are left essentially unaffected~\cite{Heizmann2018}. The range of possible ground-state widths was chosen so that the sensitivity of the KATRIN blind analysis could not improve upon the results of previous direct \mtwonue{} measurements~\cite{Aseev2011, Kraus2005}. In addition, because the endpoint fit parameter only depends -- to a good  approximation -- on the mean of the FSD, leaving that mean value untouched ensured that the endpoint could still be used during a blind analysis, \textit{e.g.} for comparison with other independent  measurements (Sec.~\ref{sec:endpoint_measurement}).

    In practice, the theoretical electronic ground-state manifold of the FSD was swapped with a Gaussian distribution function, constructed with the true mean and a randomly chosen width.
    To prevent accidental unblinding, the adjusted FSD was provided as
    an independent software module synchronized with the main fitting software.

    The second measure to mitigate biasing is to perform the full analysis, including parameter fitting, using MC-based data sets first, before turning to the experimental data. For each experimental scan $i$ we generate an MC twin (Sec.~\ref{sec:stacking-HV}) from its averaged slow-control parameters to calculate the expected rate $R_{\upbeta}(E)_{i}$ with the corresponding response function $f(E - \langle qU \rangle)_i$ and background rate $R_{\mathrm{bg},i}$. Analyzing the MC twins allows us to verify the accuracy of our parameter inference by recovering the correct input MC values for \mtwonue.

    This MC dataset is used to assess statistical ($\sigma_{\mathrm{stat}}$) and systematic ($\sigma_{\mathrm{syst}}$) uncertainties and to compute our expected sensitivity. It is also used to benchmark the independent analysis codes. At this stage, all model inputs and systematic uncertainties are frozen.

    Before the unblinding via incorporation of the unmodified FSD, a final benchmark was successfully performed on the data with the blinded FSD to verify that the independent analysis codes eventually lead to very consistent results. After this final test, the ``true'' FSD was revealed to the collaboration for the final neutrino-mass analysis of the data. The first, overnight fits -- using the independent analysis codes -- already yielded preliminary, consistent results the very next morning.

    \subsection{Covariance-matrix approach}
    \label{sec:covariance-matrix}
    Here, we report on our results using the covariance-matrix approach to include and propagate systematic uncertainties in the neutrino-mass fit. The spectrum calculation code and methods used for this analysis are described in detail in Ref.~\cite{thesis-lisa}.

    The free fit parameters in our analysis, $\params$, are inferred from the data points $\{R_i\}$ by minimizing the negative logarithm of the ratio of the Poisson likelihood function to the saturated model
    \begin{align}
        \label{eq:negLogLike}
        -2\ln \mathcal{L} (\params) = 2\sum_i &\left[	R_i^{\mathrm{model}}(\params,\nuisance)-R_i + \phantom{\frac{R_i}{R_i^0}} \right. \nonumber \\
        								& \left.	R_i \ln \left(\frac{R_i}{R_i^{\mathrm{model}}(\params,\nuisance)}\right) \right] 
    \end{align}
\noindent \noindent{where the summation is over scan steps $i$.}

    The model points, denoted by $R_i^{\mathrm{model}}$, depend on both the model parameters $\params$ and the systematic nuisance parameters $\nuisance$ (including column density and tritium isotopolog concentrations). In the fit the nuisance terms $\nuisance$ are fixed according to our best knowledge of operational parameters averaged over KNM1.

    Since the \bspec{} measured in this first KATRIN science run comprises a large number of observed events in each scan-step bin, the negative Poisson likelihood function (Eq.~\ref{eq:negLogLike}) is replaced by the standard $\chi^2$ estimator
    \begin{equation}
        \label{eq:chi2}
        \chi^2(\params) = \left(\mathbf{R}-\mathbf{R}^{\mathrm{model}}(\params,\nuisance)\right)^\intercal C^{-1} \left(\mathbf{R}-\mathbf{R}^{\mathrm{model}}(\params,\nuisance)\right).
    \end{equation}
The covariance matrix $C$ describes the correlated and uncorrelated model uncertainties, including both statistical and systematic uncertainties. This fit procedure has been extensively tested by injecting fake neutrino-mass signals in simulated pseudo-experiments. It was verified that the fit results provide an unbiased estimation of the injected parameters.

    Systematic uncertainties on the nuisance parameters $\nuisance$ are propagated using covariance matrices. For this purpose the values of $\nuisance$ are randomized according to their associated probability density functions. Correlations between parameters are taken into account. Subsequently, $\mathcal{O}$($10^{4}$) sample spectra $\{\mathbf{R}_\textrm{sample}\}$ are simulated~\cite{Barlow:213033,DAgostini:1993arp,Aker:2019qfn}. For each sample-spectrum calculation, a different $\nuisance$ is drawn from the set $\{\nuisance_\textrm{sample}\}$.

    The signal normalization $A_{\mathrm{s}}$, being a free fit parameter, is not considered in the uncertainty propagation. Therefore, all fluctuations in $\{\mathbf{R}_\textrm{sample}\}$ that translate solely into an overall signal normalization uncertainty must be eliminated. The transformation of $\{\mathbf{R}_\textrm{sample}\}$ into shape-only sample spectra is achieved by normalizing the statistics of each sample spectrum to the statistics of the average sample spectrum.

    Finally, the shape-only covariance matrix is estimated from $\{\mathbf{R}_\textrm{sample}\}$ using the sample covariance as an estimator. For any set of uncorrelated systematic effects, the associated covariance matrices can be calculated independently of one another. The sum of all matrices encodes the total uncertainties on the model points $\mathbf{R}^{\mathrm{model}}$ and their scan-step-dependent correlations.

    In the fit, $\chi^2(\params)$ is minimized to determine the best-fit parameters $\paramsbest$, whereas the profile of the $\chi^2$ function is used to infer the uncertainties on $\paramsbest$. Once the covariance matrices are pre-calculated, the spectral fit and major diagnoses can be performed within a few hours on a standard personal computer.

\begin{figure}[t!]
        \centering
        \includegraphics[width=0.45\textwidth]{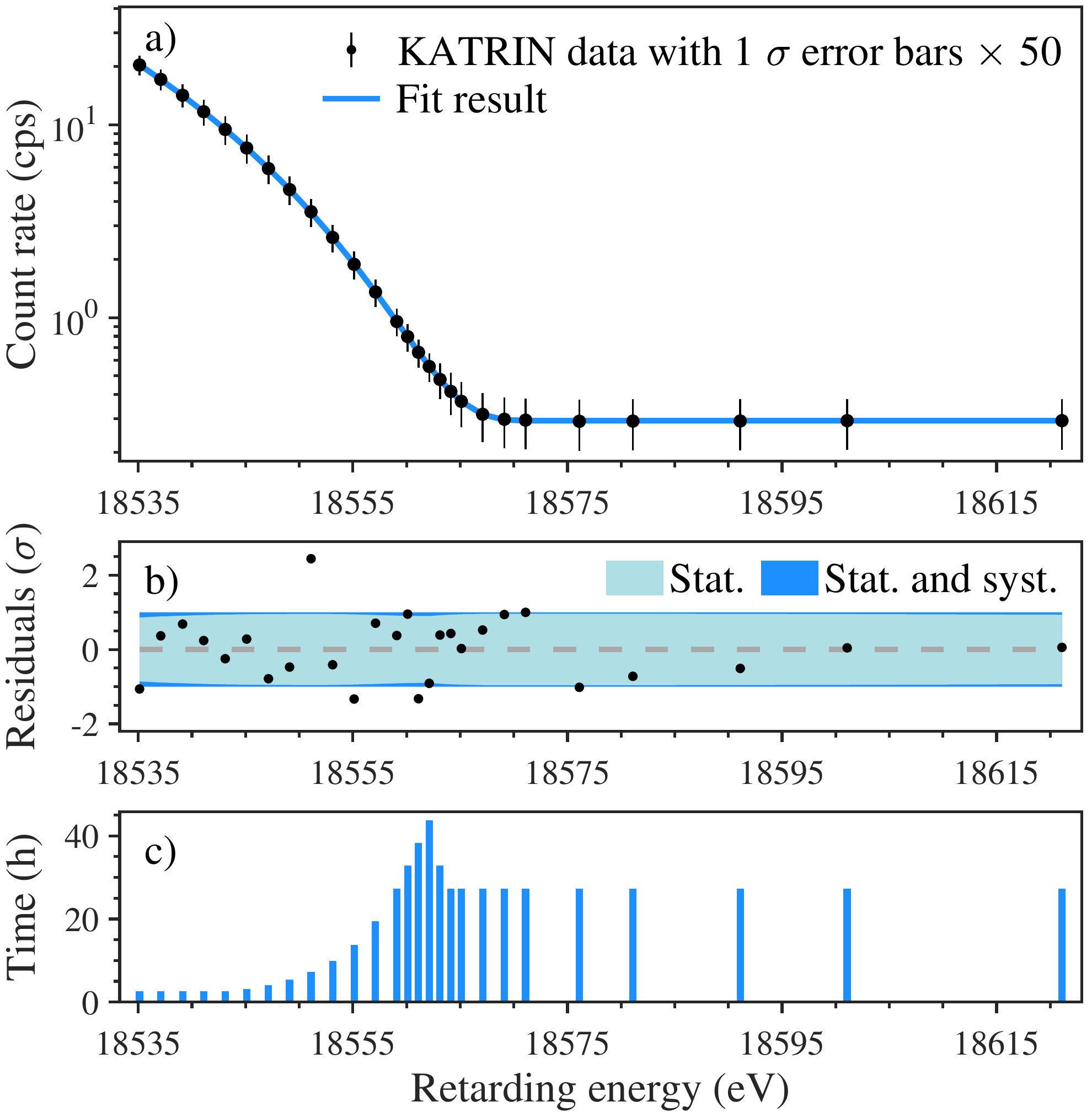}
        \caption{a) Spectrum of \belec{}s $R(\langle qU \rangle)$ over a \SI{86}{\electronvolt}-wide interval from all \num{274} tritium scans and best-fit model $R^\mathrm{model} (\langle qU \rangle)$ (line). The integral \bdecayh\ spectrum extends up to \ezero\ on top of a flat background \rbg. Experimental data are stacked at the average value $\langle qU \rangle $ of each scan step and are displayed with \SI{1}{\sigma} statistical uncertainties enlarged by a factor of \num{50} for visibility. b) Residuals of  $R(\langle qU \rangle)$ relative to the \SI{1}{\sigma} uncertainty band of the best-fit model. c) Integral measurement-time distribution of all \num{27} scan steps; see also Fig.~\ref{fig:MTD}. Figure reproduced from Ref.~\cite{Aker:2019uuj}.}
        \label{fig:beta_spectrum}
    \end{figure}

    The data and results of this fit are displayed in Fig.~\ref{fig:beta_spectrum}. Of the four free parameters, the signal amplitude $A_{\mathrm{s}}$ is unconstrained for the shape-only analysis. The effective \bdecayh\ endpoint \ezero\ can be related to the Q-value after final corrections of the energy scale (Sec.~\ref{sec:endpoint_measurement}). The background rate $R_\mathrm{bg}$ is primarily constrained by the \SI{5}{HV} scan steps above \ezero{}. The squared neutrino mass \mtwonue\ can be varied freely and therefore can take any positive or negative value.

    We find a best-fit value of \mtwonue\ = $(-0.98~ ^{+~ 0.89}_ {-~ 1.06})$ eV$^2$ with a goodness of fit of $\chi^2 =$ \num{21.4} for \num{23} degrees of freedom (d.o.f.). This corresponds to a p-value of \num{0.56}, meaning that there is a probability of \SI{56}{\percent} to retrieve a $\chi^2$-value at least as large as the one obtained.

    The total uncertainty budget of \mtwonue\ is first calculated on an Asimov data set assuming the null hypothesis. Based on the final fit applied to these simulated data, we derive $\mnuesq = \asi{0.00}{0.94}{0.78}{\electronvolt\squared}$. The relative impact of each systematic effect is assessed by performing a series of fits, each one including solely the selected effect in addition to statistical uncertainties (stat+1 test). The statistical uncertainty is then subtracted in quadrature. The same breakdown is then calculated using the unblinded data, and is in excellent agreement with our MC expectations. This data-driven uncertainty breakdown is shown in Table~\ref{tab:systematics_breakdown-final}. As expected, the total uncertainty is largely dominated by \sstat\ (0.94 eV$^2$) as compared to \ssyst\ (0.30 eV$^2$).

        \begin{table}[]
        \centering
        \caption{Uncertainty breakdown obtained from the data in the covariance-matrix (Sec.~\ref{sec:covariance-matrix}) and MC-propagation (Sec.~\ref{sec:mcprop}) approaches. The statistical uncertainty is drawn from data in all cases. Covariance-matrix results are averaged over positive and negative uncertainties. Section references are provided for the two systematics that are neglected in both approaches.}
        \label{tab:systematics_breakdown-final}
        \begin{tabular}{lcc}
            \toprule
            Effect &  \multicolumn{2}{c}{\mtwonue\  uncertainty (\SI{1}{\sigma}; \SI{e-2}{\electronvolt\squared})}\\
            & Cov.\,matrix & MC prop. \\
            \midrule
            Background rate  & 22 & 30 \\
			Scan fluctuations & 14 & \phantom{0}5  \\
            Background slope & \phantom{0}9 & \phantom{0}7  \\
            Final-state distribution & \phantom{0}9 & \phantom{0}2  \\
            Magnetic fields & \phantom{0}7 & \phantom{0}5  \\
			Expected number of & & \\
			\phantom{sss} scatterings ($\rho d \sigma$) & \phantom{0}5 & \phantom{0}5  \\
			Detector efficiency & \phantom{0}2 & Neglected  \\
		    Energy loss 	& $<1$ & $<1$  \\
			Theoretical corrections & $<1$ & Neglected   \\
		\midrule
			Electron starting potential & Neglected & Neglected  \\
			\phantom{sss} (Sec.~\ref{sec:syst_startingpot}) & & \\
			Background dependence on & & \\
			\phantom{sss} scan-step duration & Neglected & Neglected \\
			\phantom{sss} (Sec.~\ref{sec:syst-bkg}) & & \\
			\midrule
			\textbf{All included systematics} & 30 & 31  \\
		\midrule
			Statistical	& 94 & 97  \\
            \bottomrule
       \end{tabular}
    \end{table}

    \subsection{Monte-Carlo-propagation approach}
    \label{sec:mcprop}
    Here we report the fit results using the MC-propagation approach to propagate systematic uncertainties. The spectrum-calculation code used is described in Ref.~\cite{thesis-christian} while the method is adapted from Refs.~\cite{Cousins:1991qz,Harris:2014}.

    In the MC-propagation method, we repeat the fitting process $\sim$ \num{e4} times, each time with newly randomized input values for the systematic nuisance parameters $\nuisance$ that are held fixed during that fit.  Compared to the well-known approach of free nuisance parameters constrained with pull terms, this method has two key advantages for the KATRIN analysis. Foremost, the computationally expensive response function does not have to be recomputed with varying $\nuisance$ during the fit. In addition, the minimization is technically simplified due to the reduced number of free parameters.

    To retrieve an initial estimate of the best-fit values $\paramsdata$ of our four fit parameters $\params$ (that is, $\mnuesq, E_0, A_\text{s}, R_\text{bg}$), we fit the original data with the additional parameters $\nuisance$ fixed to our best knowledge from the experiment. Next, we generate MC spectra assuming the values $\paramsdata$ for our model and a Poisson distribution of the counts. We then fit each of these statistically randomized MC spectra, retrieving one sample of values $\paramsmcstat$ for our free parameters. The resulting distribution of $\{\paramsmcstat\}$ can be used to infer the statistical uncertainty of $\params$.

    Our next step is to assess the systematic uncertainties, beginning by varying the values of $\nuisance$ according to their uncertainties. The model is initialized with the random values $\nuisancemc$. We then fit the randomized model to our reference spectrum, which assumes the best estimate for $\nuisance$ and $\paramsdata$. In principle, the resulting distribution of $\{\paramsmcsyst\}$ reflects the systematic uncertainty, taking into account only the external information on $\nuisance$. However, the data may also contain information to constrain $\nuisance$. To account for this, we also fit the randomized model to the data to retrieve the likelihood value $\likelihood(\paramsmcsyst)$. This likelihood value is used to weight the corresponding sample $\paramsmcsyst$. The resulting weighted distribution $\{\paramsmcsyst\}_\text{weight}$ is then used to retrieve the systematic uncertainty on $\params$ as proposed in Ref.~\cite{Biller:2015}. At this point we would like to note that this systematics-only distribution is solely used to calculate a breakdown of the uncertainties and does not enter into the final confidence interval.

    In the final step, we combine the statistics- and systematics-only steps described above. As in the systematics-only approach, we initialize our model with randomized values for the nuisance parameters $\nuisancemc$. Instead of fitting it to the unfluctuated best estimate, we now fit this model to statistically randomized spectra to retrieve the values $\paramsmctot$ of our parameters of interest. This model is then also fit to the unmodified data spectrum to retrieve the likelihood $\likelihood(\paramsmctot)$. We infer the combined statistical and systematical uncertainty from the distribution of $\{\paramsmctot\}_\text{weight}$, which is weighted by these likelihood values.

    Initially, we apply this method to the MC twin data described in Sec.~\ref{sec:stacking-HV}). From the statistics-only fit, we derive $\mnuesq = \asi{0.00}{0.90}{0.75}{\electronvolt\squared}$. Including the systematic uncertainties described in Sec.~\ref{sec:systematic_uncertainties}, the best-fit value becomes $\mnuesq = \asi{0.00}{0.96}{0.76}{\electronvolt\squared}$. This is only a slight change with respect to the statistics-only analysis.

\begin{figure}[t!]
        \centering
        \includegraphics[width=0.45\textwidth]{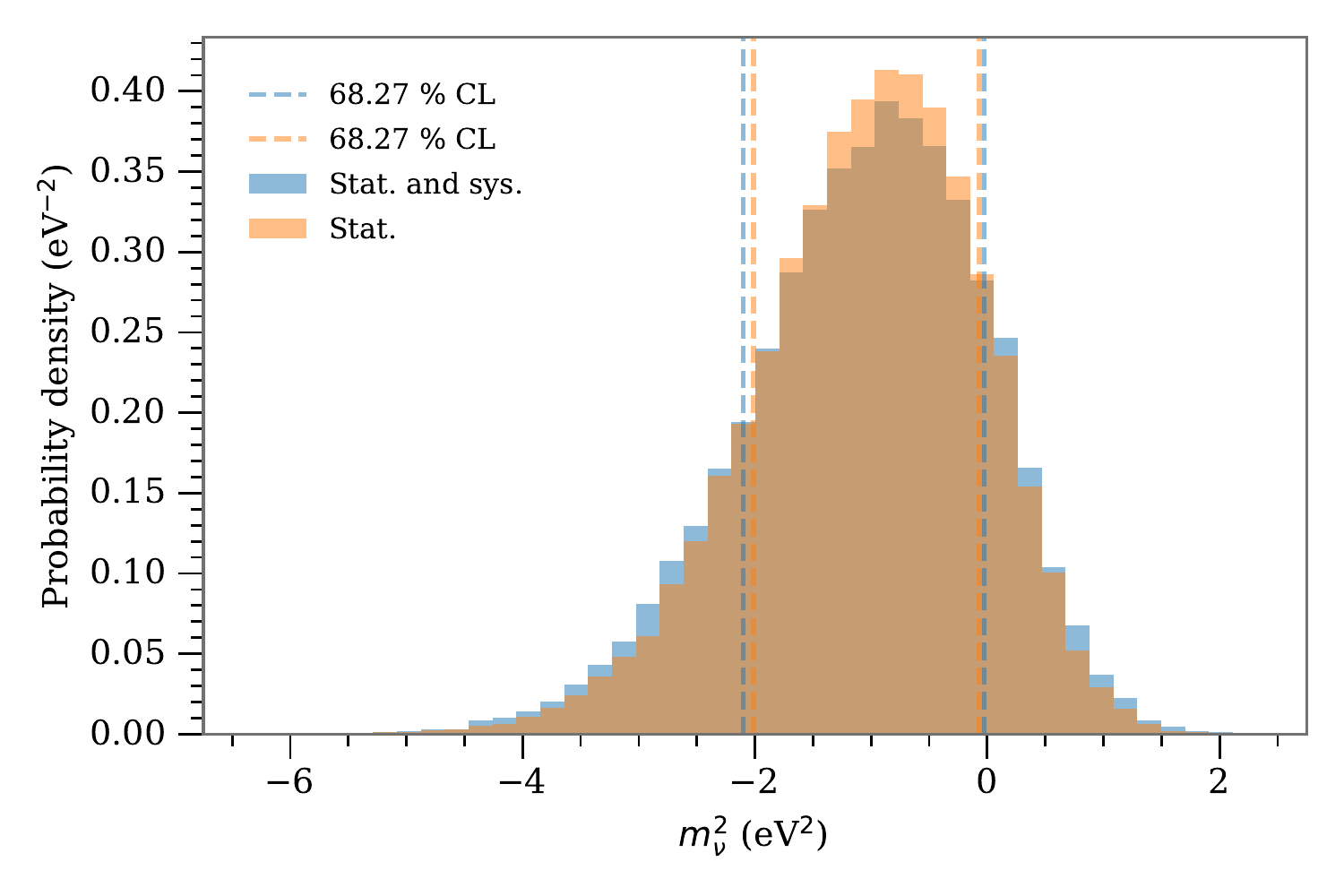}
        \caption{One-dimensional distribution of $\mnuesq$ with statistical uncertainty only (orange) as well as statistical and systematical uncertainty combined (blue). The dashed lines indicate the \SI{1}{\sigma} confidence interval for each case.}
        \label{fig:mcprop_1d}
    \end{figure}

    After freezing the method and inputs on MC spectra, we repeat the analysis on the data.  Here the statistics-only fit to the data gives a best-fit value of $\mnuesq = \asi{-0.94}{1.07}{0.87}{\electronvolt \squared}$ at a goodness-of-fit of \num{-2} $\ln \mathcal{L} =$ \num{23.3} for \num{23} d.o.f., corresponding to a p-value of \num{0.44}. When including systematic uncertainties, we arrive at $\mnuesq = \asi{-0.96}{1.14}{0.93}{\electronvolt\squared}$. The one-dimensional $\mnuesq$ distributions used to derive these values are shown in Fig.~\ref{fig:mcprop_1d}.

    Using the MC propagation of uncertainty, it is possible to analyze the impact of individual systematic effects on the parameters of interest. Table~\ref{tab:systematics_breakdown-final}, further above, shows the uncertainty budget on $\mnuesq$ for KNM1.

    \subsection{Fit results}
    \label{sec:fit-results}

    The results of the two independent methods of Secs.~\ref{sec:covariance-matrix} and~\ref{sec:mcprop} agree to within a few percent of the total uncertainty. As a best-fit value for the squared neutrino mass, we quote \mtwonue\ = $\asi{-1.0}{1.1}{0.9}{\electronvolt\squared}$. This best-fit result corresponds to a \SI{1}{\sigma} statistical fluctuation to negative values of \mtwonue. Assuming the true neutrino mass is zero, the probability to retrieve a best-fit value as negative as ours is \SI{16}{\percent} and is thus fully compatible with statistical expectations. The total uncertainty budget of \mtwonue\ is largely dominated by \sstat\ (\SI{0.97}{\electronvolt\squared}) as compared to \ssyst\ (\SI{0.32}{\electronvolt\squared}). The dominant contributions to \ssyst\ are found to be  the non-Poissonian background from radon and the uncertainty on the background slope. Uncertainties on the column density, energy-loss function, FSD, and magnetic fields play a minor role in the budget of \ssyst. Likewise, the uncertainties induced by fluctuations of $\varepsilon_{\mathrm{T}}$ and HV parameters during a scan are negligibly small compared to \sstat.

\begin{figure}[t!]
        \centering
        \includegraphics[width=0.45\textwidth]{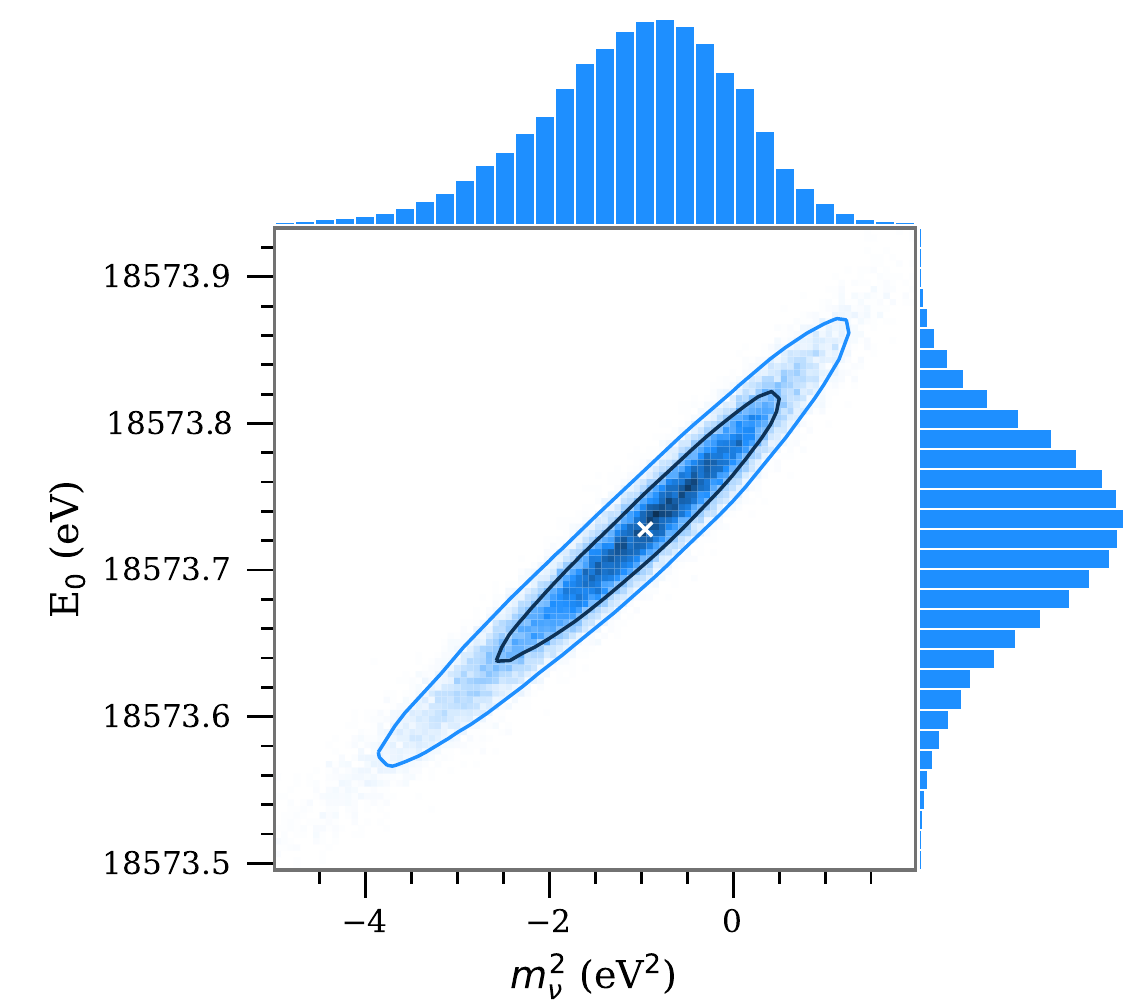}
        \caption{Scatter plot of fit values for the squared neutrino mass \mtwonue\  and the effective \bdecay{} endpoint \ezero{} together with \SI{1}{\sigma} (black) and \SI{2}{\sigma} (blue) contours around the best-fit point (white cross). Results are generated from a large set of pseudo-experiments emulating our experimental data set and its statistical and systematical uncertainties -- each one an individual sample from the MC propagation. Figure reproduced from Ref.~\cite{Aker:2019uuj}.}
        \label{fig:mnu2endpoint}
    \end{figure}

    For the effective $\upbeta$-decay endpoint we find a best fit value of \SI{18573.7 +- 0.1}{\electronvolt}. \Cref{fig:mnu2endpoint} shows the interplay between $\mtwonue$ and $E_0$. The large correlation (0.97) between the two parameters is in line with expectation~\cite{Otten2008,Drexlin2013}.

    For completeness, we report here that our best-fit background rate is $R_\text{bg} = \SI{293 +- 1}{\milli cps}$. The signal-normalization parameter $A_s$ absorbs the rate effects of our systematic uncertainties, and does not have a straightforward interpretation.

    \section{Frequentist bounds on the neutrino mass}
    \label{sec:upperlimit}
    The result of a neutrino-mass experiment is commonly presented in form of a confidence interval for the neutrino mass, or an upper limit if the lower boundary of the confidence interval is zero. These values are used by the community for constraining phenomenological models, developing theoretical predictions, and comparing the results of different experiments, and as input parameters to both terrestrial experiments and cosmological observations.

    There are several methods of constructing the confidence intervals with additional information on the estimated parameter. To account for the physical bound of $m_\nu^2 \geq 0$, despite the fact that $m_\nu^2$ is unconstrained in the fit, we perform full Neyman constructions using the methods of Lokhov and Tkachov and of Feldman and Cousins, for completeness. Both of these methods avoid empty confidence intervals for negative best-fit estimates $\widehat{m}_\nu^2$. In each case, we apply both of our spectral analysis approaches (described in Sec.~\ref{sec:covariance-matrix} and Sec.~\ref{sec:mcprop}) to incorporate statistical and systematic uncertainties into the calculated Monte Carlo quantities. This results in two calculations of each type of confidence interval, which agree with each other in both cases. We briefly compare the Feldman-Cousins and Lokhov-Tkachov methods below. 

\begin{figure*}[t]
        \centering
        \subfloat[Feldman-Cousins]{\includegraphics[width=0.45\textwidth]{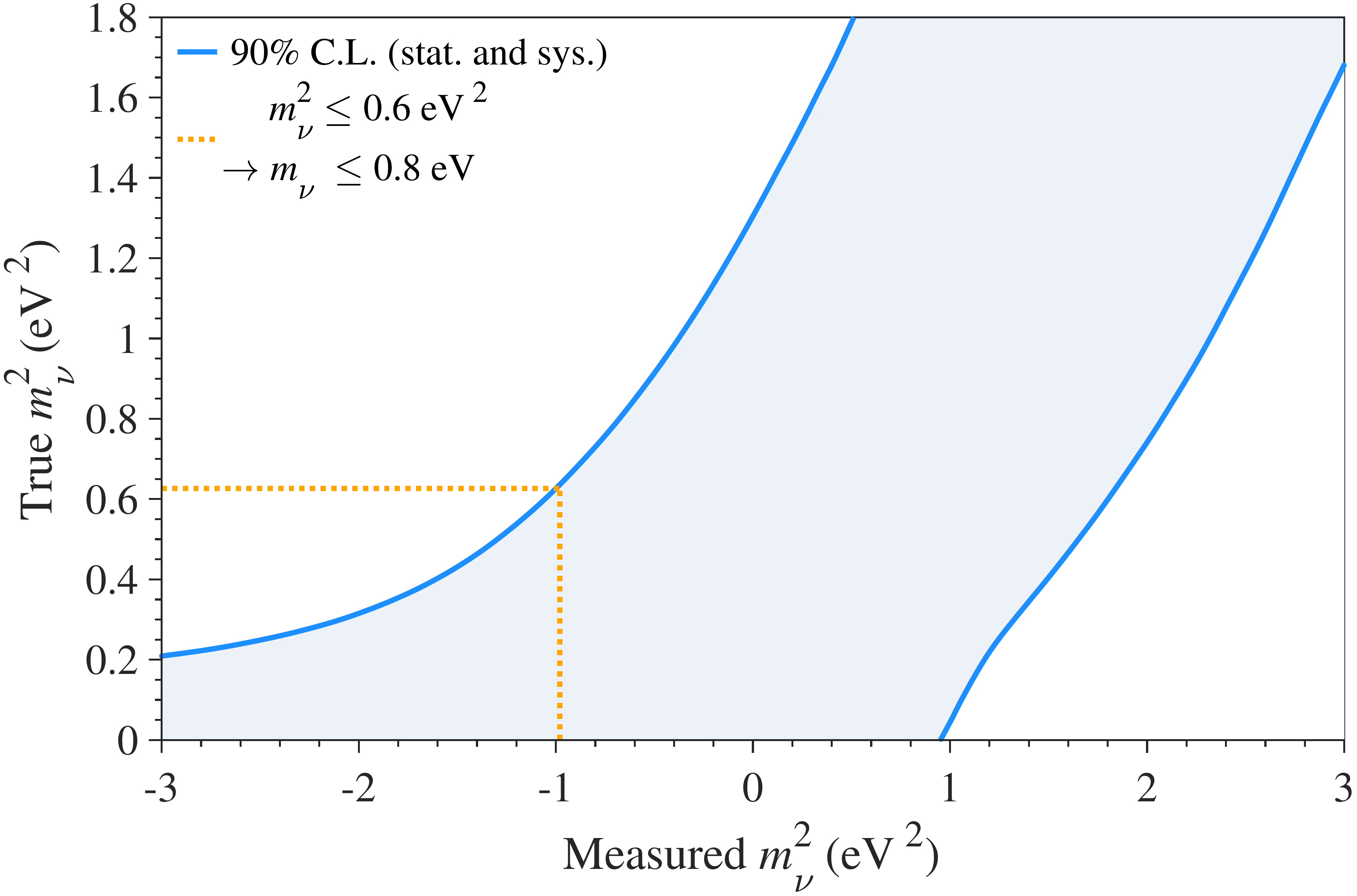}\label{fig:contour_fc}}\hfill
        \subfloat[Lokhov-Tkachov]{\includegraphics[width=0.45\textwidth]{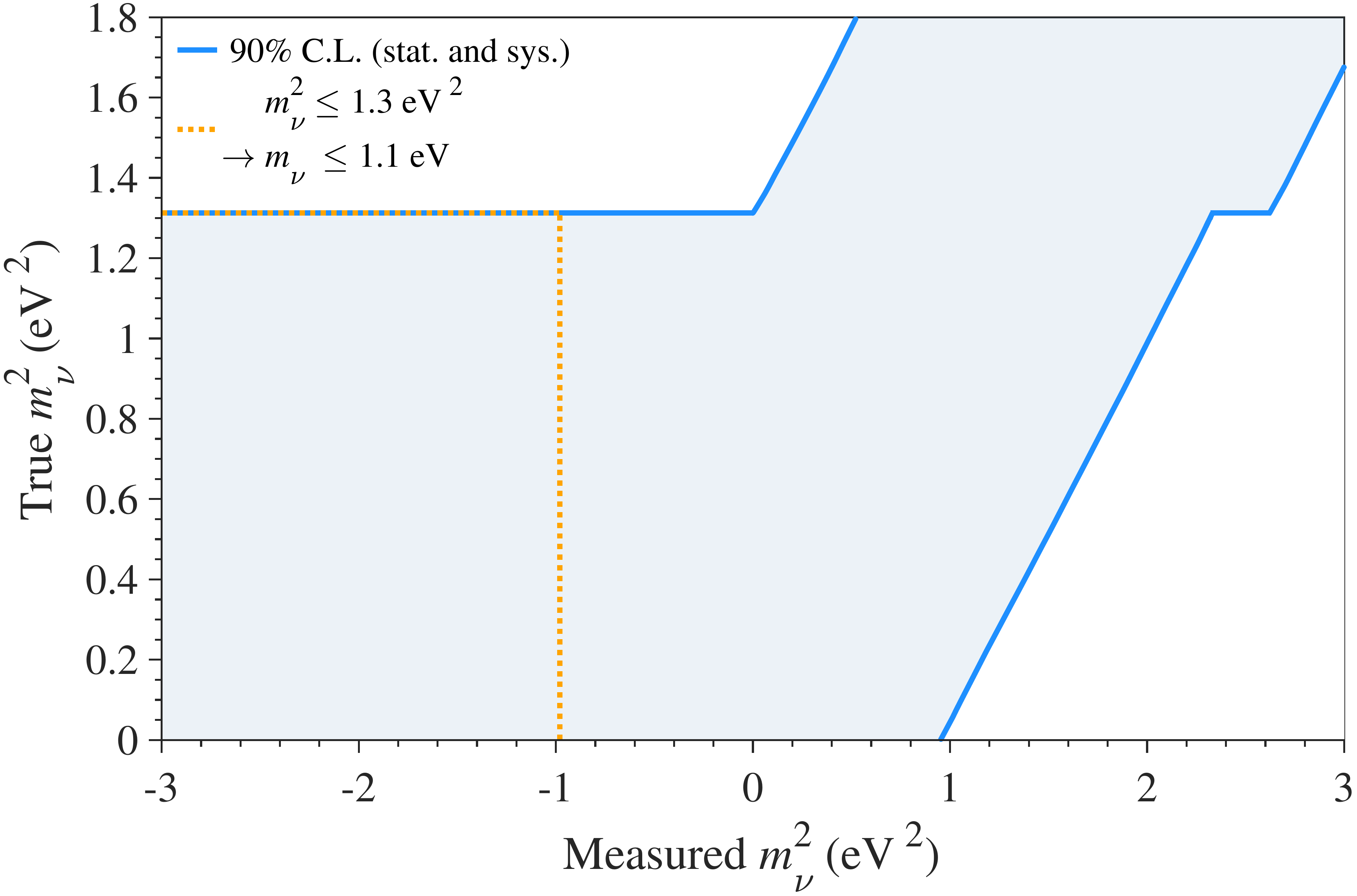}\label{fig:contour_lt}}
        \caption{The confidence belts constructed for $m_\nu^2$ using the methods of Feldman and Cousins \protect\subref{fig:contour_fc} and of Lokhov and Tkachov \protect\subref{fig:contour_lt}. The blue contour represents the \SI{90}{\percent} CL confidence belt, which takes into account statistical and systematic uncertainties. The orange vertical line corresponds to the best-fit estimate $\widehat{m}_\nu^2$, while the horizontal line then defines the upper limit on $m_\nu^2$. The Lokhov-Tkachov method gives an upper limit coinciding with the KNM1 experimental sensitivity to $m_\nu^2$.}
    \end{figure*}

    In the Feldman-Cousins method~\cite{Feldman:1997qc}, the likelihood ratio
    \begin{equation}
        \frac{\mathcal{L}\left(\widehat{m}_\nu^2\,|\,m_\nu^2\right)}{\mathcal{L}\left(\widehat{m}_\nu^2\,|\,\mathrm{max}(0,\,\widehat{m}_\nu^2)\right)}
    \end{equation}
    determines the order in which the estimates $\widehat{m}_\nu^2$ are added to the acceptance region for an assumed value of $m_\nu^2$, thereby constructing the confidence interval. This ordering principle avoids empty intervals, but at the same time results in more stringent limits for negative best-fit estimates that are further from zero, as in \cref{fig:contour_fc}. This yields an excessively strict upper limit in the case of statistical fluctuations in one direction, or in the presence of an unknown systematic bias as seen in most neutrino-mass experiments of the early 1990s (see Fig.~\ref{fig:historyNuMass-Combined}). While our best-fit result is statistically compatible with zero, we decided after unblinding to pursue an alternative approach to ensure a conservative handling of fluctuations.

    Following the prescription of Lokhov and Tkachov~\cite{Lokhov:2015zna}, a new estimator $\widetilde{m}_\nu^2$ can be defined such that
    \begin{equation}
        \widetilde{m}_\nu^2 = \mathrm{max}(\widehat{m}_\mu^2, \, 0).
    \end{equation}
    The estimator is by definition as close as possible to the unknown true non-negative value of the $m_\nu^2$, which is the fundamental aim of the statistical estimation. The confidence interval for the new estimator $\widetilde{m}_\nu^2$ is then constructed according to the Neyman procedure, which guarantees the correct coverage. The non-physical values of the best-fit estimate $\widehat{m}_\nu^2$ are indistinguishable and give the same confidence interval from zero to the experimental sensitivity (Fig.~\ref{fig:contour_lt}). Therefore more negative values of \mtwonue, obtained due to a statistical fluctuation or an improperly treated systematic contribution, do not yield better upper limits. This makes it possible to compare the upper limits of different measurements directly without the need to know the best-fit estimate, as long as $\mtwonue$ is not significantly positive.

    In order to allow the squared-neutrino-mass estimator to become negative in either analysis, the differential spectrum shape must be extended into the unphysical region of $m_\nu^2 <$ \num{0}. In previous experiments~\cite{Aseev2011, Kraus2005} the extension was made by modifying the differential spectrum shape so that the $\chi^2$ function became symmetric around $m_\nu^2 =$ \num{0}. Such a modification depends on the particular shape of the $\chi^2$ function and consequently on the experimental setup. In the present analysis we take the differential spectrum shape in \cref{eq:diffspec_single_state} without any modification for $m_\nu^2 <$ \num{0}. This leads to a $\chi^2$ function with an asymmetric shape, as shown in \cref{fig:mcprop_1d}. The Lokhov-Tkachov method yields the same upper limit for all $m_\nu^2 <$ \num{0}. Therefore, by construction, the upper limit does not depend on a particular choice of the extension.

    Using the Lokhov-Tkachov construction we derive an upper limit of $\mnue <$ \SI{1.1}{\electronvolt} (\SI{90}{\percent} CL) as the central result of this work. For comparison, the Feldman-Cousins method yields the upper limit $\mnue <$ \SI{0.8}{\electronvolt} (\SI{90}{\percent} CL).
We have also derived upper limits at \SI{95}{\percent} CL for comparison to the Mainz~\cite{Kraus2005} and Troitsk~\cite{Aseev2011} Feldman-Cousins results. In the Lokhov-Tkachov method, this becomes $\mnue <$ \SI{1.2}{\electronvolt} (\SI{95}{\percent} CL); using Feldman-Cousins, as was done by Mainz and Troitsk, we find $\mnue <$ \SI{0.9}{\electronvolt} (\SI{95}{\percent} CL).

    \section{Bayesian bound on the neutrino mass} 
    \label{sec:bayesian_analysis}

    Bayesian analysis methods provide an alternative means of handling the unphysical, $m_\nu^2 <$ \num{0} region. We used the MC-propagation model and data framework, described in Sec.~\ref{sec:mcprop}, to set a first limit using Bayesian techniques. Posterior probability distributions were constructed according to Bayes' theorem, using Markov-chain Monte Carlo methods within the Bayesian Analysis Toolkit (BAT)~\cite{Caldwell:2009khb}.
We use uniform priors, flat in probability, for $A_s$, \ezero{}, and \rbg{}; this choice is most straightforward for analysis of stacked spectra. An informative prior, restricting the result to only physically allowed \mtwonue{} values (equal to or larger than zero), is used to ultimately obtain an upper credibility limit on the neutrino mass in a Bayesian interpretation. In the allowed region, this prior is flat in \mtwonue{} space. Future work will investigate alternate choices of prior, including a prior flat in $\mnue$.

    First, we extract statistical uncertainties and compare with other analysis methods using the basic model, including the four-parameter set $\params$ with flat prior probabilities. The global mode (maximum value) of the 4-dimensional posterior for \mtwonue{} is found at \SI{-1.0}{\electronvolt\squared}. The two-sided $1\sigma$ interval, with equal probability on either side, is obtained from the posterior distribution marginalized for \mtwonue{} as [$-2.1, -0.3$]~\si{\electronvolt\squared}.

    Four of the leading systematic uncertainties are included in this analysis, and are incorporated into the fit in various ways. A background slope is included as a fifth free parameter with a Gaussian prior probability centered around zero and a width given by its uncertainty. Non-Poissonian background counts are included by widening the underlying likelihood distribution in each scan step according to background measurements (Sec.~\ref{sec:background}). Variations of the response due to uncertainties in the magnetic field or the column density were too computationally expensive at the time of the analysis. Instead of including these as free parameters in the model, multiple independent fits were parallelized on a computing cluster. Each fit was started with the input systematic fixed at a different value, following a Gaussian distribution with a width given by the parameter uncertainty. The median values of the output posterior distributions were used to obtain parameter estimates with systematic uncertainties. The same results are obtained by combining the Markov chains of the individual fits into a single chain, and subsequently performing the same parameter-estimation procedure.      Additional systematics will be analyzed in future work.

    The present dataset is strongly dominated by statistical uncertainties, and individual systematic effects are largely masked below \SI{0.1}{\electronvolt\squared} by numerical uncertainties. These uncertainties come from the finite number of Markov-chain Monte-Carlo samples and are on the order of \SI{0.006}{\electronvolt\squared} in the $1\sigma$ posterior width.
    Hence, the systematic budget was investigated with Asimov data, artificially increasing the amount of data and thus enhancing each included systematic effect with respect to statistical uncertainties.  

    Taking these four explicitly included systematic uncertainties into account, the most probable \mtwonue{} value was found at \SI{-1.0}{\electronvolt\squared} and the two-sided, $1\sigma$, probability-symmetric interval at [$-2.2, -0.3$]~\si{\electronvolt\squared}. Using Table~\ref{tab:systematics_breakdown-final} to estimate upper bounds on the primary excluded systematics -- scan fluctuations and the FSD -- we find that they affect the total uncertainty on this most probable value by about 1\%.

\begin{figure}[t!]
        \centering
        \includegraphics[width=0.45\textwidth]{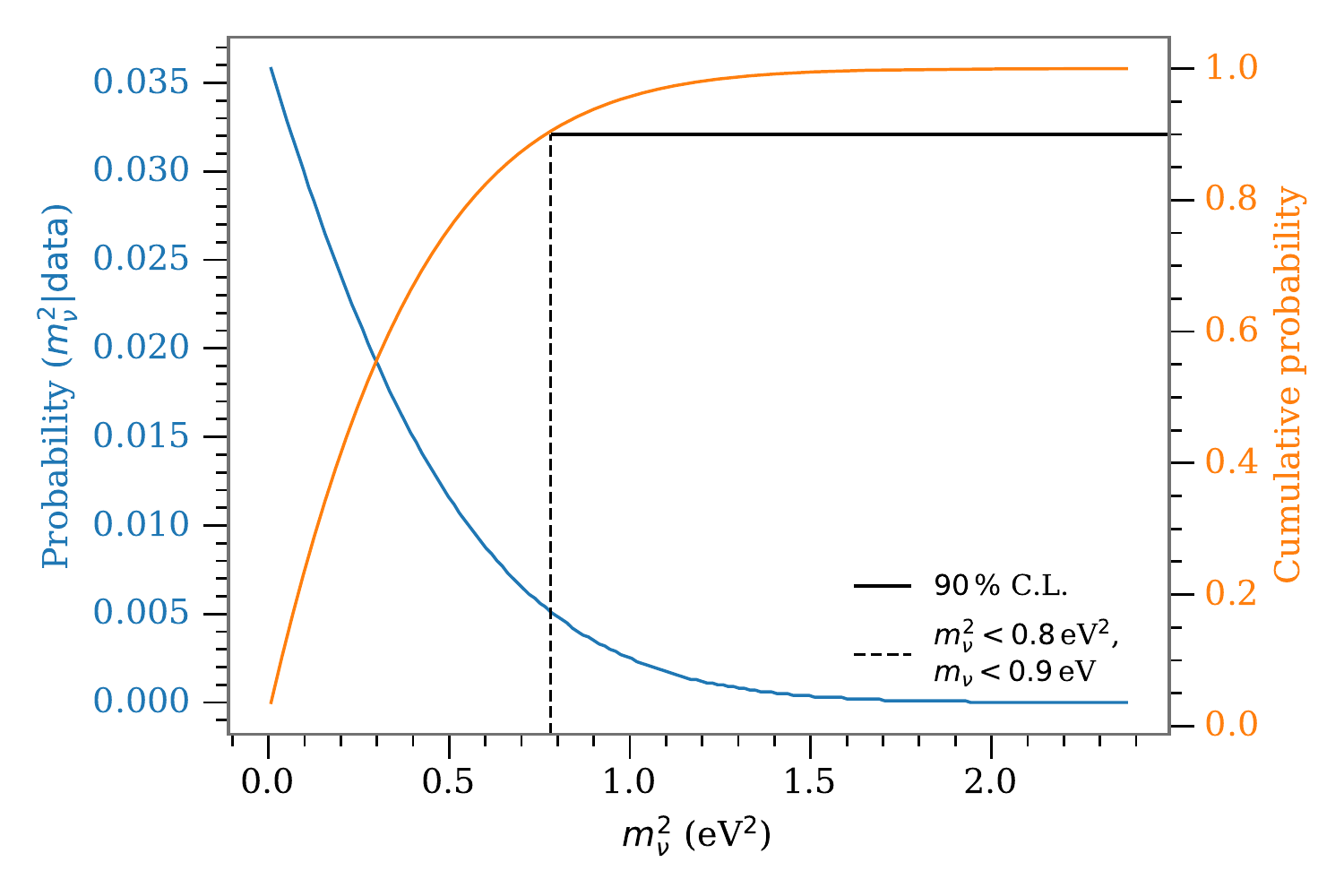}
        \caption{Posterior probability distribution using a flat prior for $m_\nu^2 \ge$ \SI{0}{\electronvolt\squared}  (blue curve). Also shown are the cumulative probability (orange curve) and the \SI{90}{\percent} quantile (solid black line) used for limit setting (dashed black line).}
        \label{fig:LimitPosteriorFitriumBAT}
    \end{figure}

    To determine the limit on the neutrino-mass, we then perform the same fits with a flat prior in $\mtwonue{} \ge 0$. The $\mtwonue{}$ marginalized posterior distribution is shown in Fig.~\ref{fig:LimitPosteriorFitriumBAT}. The best-fit value is found at $\mtwonue{} =$ \num{0}. The \SI{90}{\percent} quantile of the marginalized posterior distribution is at \SI{0.78}{\electronvolt\squared}.
    The Bayesian upper limit is thus $\mnue<$ \SI{0.9}{\electronvolt} (90\% C.I.). The constant prior probability in $m_\nu^2$-space gives equal probability for statistical fluctuations in the data. In our case, the Bayesian 90\% credibility limit is numerically closer than the Feldman-Cousins 90\% confidence limit to the sensitivity of the experiment and to the Lokhov-Tkachov limit, as is often observed in the presence of larger statistical fluctuations.

    As an additional test, the positive flat prior was slightly modified by knowledge from oscillation experiments, allowing only $m_\nu >$ \SI{8}{\milli\electronvolt} (normal ordering) or $m_\nu >$ \SI{50}{\milli\electronvolt} (inverted ordering)~\cite{Esteban:2018azc}. The posterior quantiles show no numerical difference, as is expected with the current data.

    \section{Q-value measurement}
    \label{sec:endpoint_measurement}
    \begin{figure}
        \centering
        \includegraphics[width=0.45\textwidth]{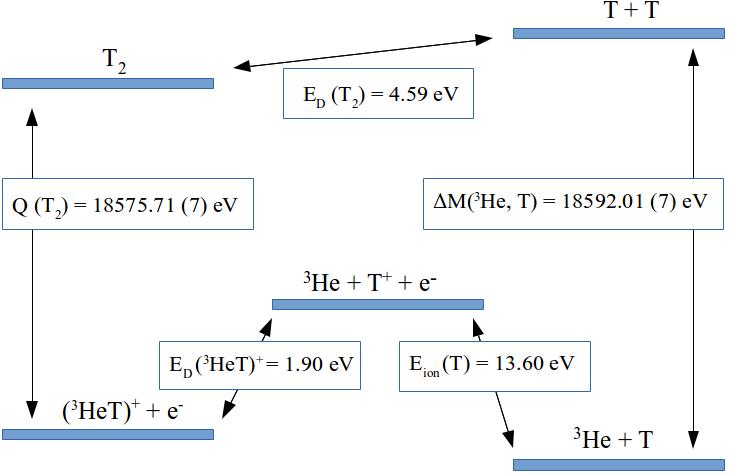}
        \caption{Energy diagram showing the connection between the atomic mass difference $\Delta$M($^3$He,T) and the Q-value of molecular tritium Q(T$_2$).}
        \label{fig:dM_Q_diagram}
    \end{figure}
A consistency check of the energy scale of KATRIN can be performed by extracting the experimental Q-value for molecular tritium from KATRIN data, and comparing it to Q-values based on Penning-trap measurements of the \mbox{$^3$He-T} atomic mass difference. The Q-value represents the amount of kinetic energy released in \bdecay{} for zero neutrino mass; Fig.~\ref{fig:dM_Q_diagram} shows its relationship to the mass difference and the binding energies of the atomic and molecular states involved in T$_2$ \bdecay{}. In equation form, we have:
    \begin{equation}
        \label{eq:mass_difference_q_value}
        \Delta M (^3\mathrm{He},\mathrm{T}) - Q(\mathrm{T}_2) = E_\mathrm{ion} (\mathrm{T}) - E_\mathrm{D} (^3\mathrm{HeT})^+ +E_\mathrm{D} (\mathrm{T}_2).
    \end{equation}{}

   Table~\ref{tab:qvalue-massdiff} summarizes literature values for the relevant energies. Inserting these values into Eq.~\ref{eq:mass_difference_q_value}, we obtain

    \begin{equation}
        Q(\mathrm{T}_2)_{\Delta M} = \Delta M(^3\mathrm{He},\mathrm{T}) - \SI{16.2967}{\electronvolt} = \SI{18575.71\pm0.07}{\electronvolt}
    \end{equation}
\noindent   for the Q-value derived from the measured $^3$He-T atomic-mass difference.

        \begin{table}[]
        \centering
        \caption{Relevant energies for deducing the Q-value of T$_2$ \bdecay{} from measured mass differences. $E_\mathrm{D}$ denotes a dissociation energy and $E_\mathrm{ion}$ denotes an ionization energy. The ionization energy of tritium is calculated as $R_\mathrm{H} \frac{1}{1+ \frac{m_\mathrm{e}}{m_\mathrm{T}}}$, with $R_\mathrm{H}$ the Rydberg constant.}
        \label{tab:qvalue-massdiff}
        \begin{tabular}{ccc}
            \toprule
			Quantity & Value (eV) & Reference \\
			\midrule
			 $\Delta M (^3\mathrm{He},\mathrm{T})$ & \num{18592.01\pm0.07} & \cite{Myers2015} \\
			 $E_\mathrm{D} (\mathrm{T}_2)$ & \num{4.5909} & \cite{Pachucki2019} \\
			 $E_\mathrm{D} (^3\mathrm{HeT}^+)$ & \num{1.8974} & \cite{DossThesis} \\
			 $E_\mathrm{ion}(T)$ & \num{13.6032} & \cite{codata:2018} for $R_\mathrm{H}$ \\
            \bottomrule
       \end{tabular}
    \end{table}

   The KATRIN result for the Q-value in molecular tritium \bdecay{} is derived from the best-fit value of $E_0$ with corrections for the center-of-mass molecular recoil of the $^3$HeT$^+$ daughter ion, as well as the relative offset of the electron starting potential in the source to the work function of the inner electrode of the main spectrometer.

For the effective endpoint, our two fitting methods both obtain a best-fit value of \ezero\ $=$ \SI{18573.7\pm0.1}{\electronvolt} (Sec.~\ref{sec:fit-results}). The recoil of the $^3$HeT$^+$ molecule is given by
    \begin{equation}
        E_\mathrm{rec} = \frac{E_0^2 + 2 E_0 \me}{m_\mathrm{HeT^+}} = \SI{1.720}{\electronvolt}.
    \end{equation}{}

E-gun data were used to investigate the work function of the inner electrode system of the main spectrometer. First, the work function of this electron source was measured with the Fowler method~\cite{Fowler-PhysRev.38.45} to be $\Phi_\mathrm{egun} =$ \SI{4.44\pm0.05}{\electronvolt}. Next, a transmission function was measured with photo\-electrons from the e-gun traveling through an evacuated source. Knowing the energy of the transmission edge and the work function of the e-gun, we can estimate the work function of the inner-electrode system as $\Phi_\mathrm{IE} =$ \SI{4.1\pm0.2}{\electronvolt}.

The \belech{} starting potential inside the tritium source is defined by the cold and strongly magnetized plasma within its boundary conditions at the rear wall and the grounded beam tube (Sec.~\ref{sec:starting_potential}). By assuming that the magnitude of the plasma potential is small, as indicated by the \krm{} measurement campaign, we treat the electron starting potential as mainly defined by the bias voltage and work function of the gold-plated rear wall, especially at small radii.

			The work function of the rear wall was measured with the Fowler method prior to KNM1. Due to the illumination conditions, only the inner two-thirds of its area could be used for the measurement. The resulting raw, mean value from this measurement is $\Phi_\mathrm{RW}^{\mathrm{vac}} =$ \SI{4.29}{\electronvolt}. However, this measurement was performed with an evacuated source. Previous measurements with deuterium gas indicate that the work function changes by about \SI{-100}{\milli\electronvolt} when the rear wall is exposed to hydrogen isotopes in the source, as is the case during tritium operation. This estimate of the \textit{in situ} work function of the rear wall has a large uncertainty, which we estimate at about \SI{\pm200}{\milli\electronvolt}. Further, during KNM1 the rear wall was set to a voltage of $U_\mathrm{RW} =$ \SI{-150}{\milli\volt}, which is numerically equivalent to an increase of the work function by \SI{150}{\milli\electronvolt}. These considerations lead us to estimate an actual rear-wall work function of $\Phi_\mathrm{RW} =$ \SI{4.34\pm0.2}{\electronvolt} during KNM1.

\begin{figure}
        \centering
        \includegraphics[width=0.45\textwidth]{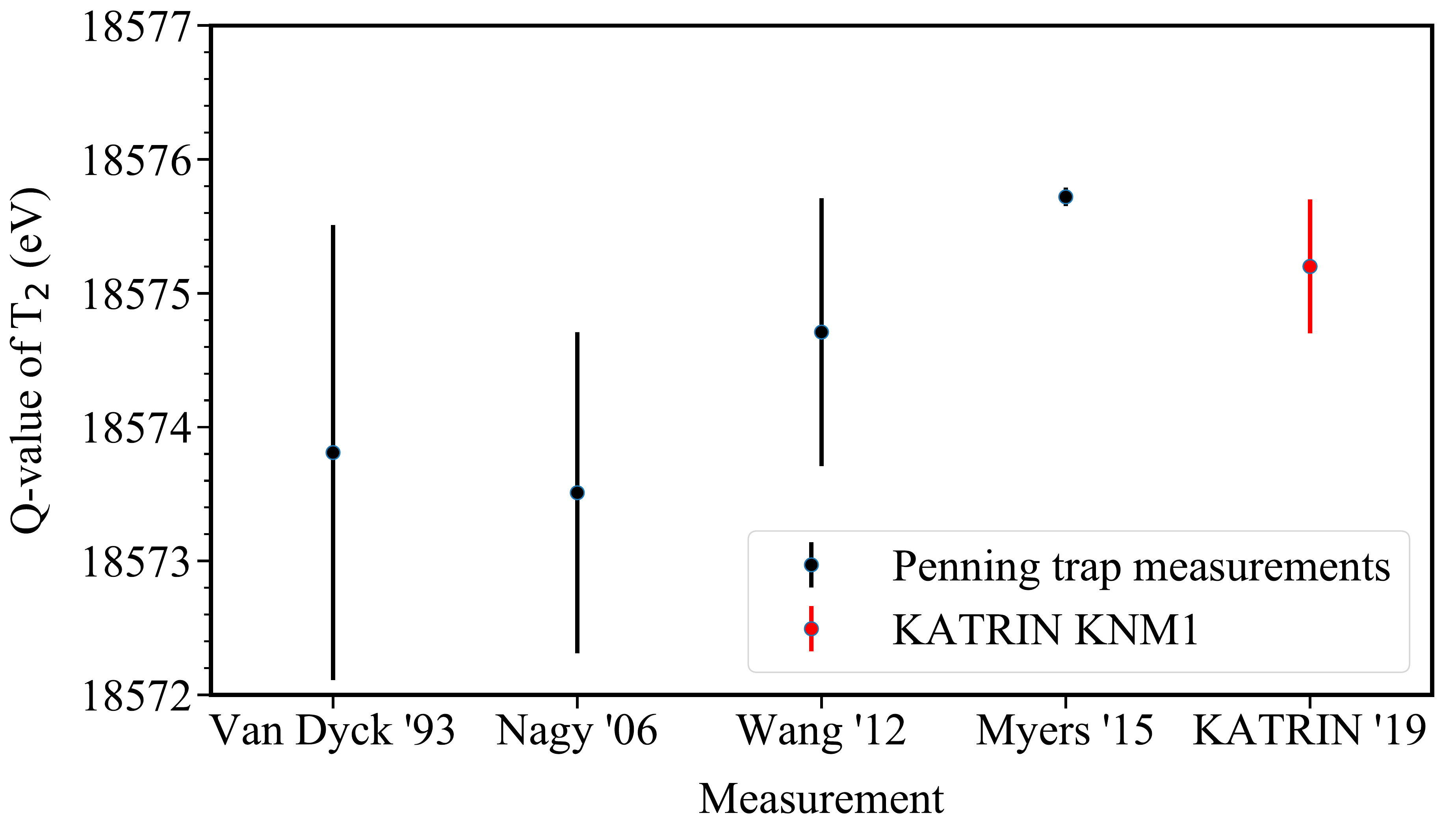}
        \caption{Comparison of the Q-value of molecular tritium found in this work to values derived from Penning-trap measurements of the atomic-mass difference. In chronological order, the values of the Penning-trap measurements are those reported in Refs.~\cite{PhysRevLett.70.2888_van_dyck_mass_difference},~\cite{Nagy_2006_mass_difference},~\cite{Wang_2012_AME12_mass_difference} and~\cite{Myers2015}.}
        \label{fig:Q_value_comparison}
    \end{figure}

			We assume an additional uncertainty of \SI{\pm100}{\milli\volt} for the sum of all involved voltages. The main contribution to this is the uncertainty of the absolute voltage of the main spectrometer, $\Delta U_\mathrm{abs} =$ \SI{\pm94}{\milli\volt}~\cite{Arenz:2018ymp}. The dominant uncertainty for the Q-value determination is the possibility of a plasma potential in the source that differs from the rear-wall potential. We assume an uncertainty of $U_\mathrm{plasma} =$ \SI{\pm400}{\milli\volt} because we cannot directly probe the plasma potential under KNM1 operational conditions. Our final result is then:
    \begin{equation}
        Q(\mathrm{T}_2)_{\mathrm{KNM1}} = E_0 + E_\mathrm{rec} - \SI{0.2}{\electronvolt} \SI{\pm0.5}{\electronvolt} = \SI{18575.2\pm0.5}{\electronvolt}.
    \end{equation}{}
    $Q(\mathrm{T}_2)_{\mathrm{KNM1}}$ and $Q(\mathrm{T}_2)_{\Delta M}$ agree within uncertainties. Figure~\ref{fig:Q_value_comparison} shows a comparison of the obtained Q-value in KATRIN with values derived from Penning-trap measurements. The consistency of the Q-values underlines the robustness of the energy scale in our scanning measurement of the T$_2$ \bspec.

    \section{Results and discussion}
    \label{sec:discussion}

    In this work we have presented the first neutrino-mass measurement campaign of the KATRIN experiment. The acquired high-precision T$_2$ \bdecay{} spectrum, containing a total of \num{2} million electrons in an energy range of
    $\lbrack \ezero - $\SI{37}{\electronvolt}, $\ezero + $\SI{49}{\electronvolt}$\rbrack$, was compared against a model of the theoretical spectrum, incorporating relevant experimental effects such as electromagnetic fields, backgrounds, and scattering. The experiment was operated at a reduced column density. Taking into account both the reduced activity and the reduced scattering probabilities, the \belec{}s recorded in the ROI during our four-week KNM1 campaign correspond to just \SI{9}{days} of measurement time at the full, design source strength.

    The full analysis was carried out applying a multi-stage blinding scheme. All analysis inputs were fixed on MC twin copies of the data; the spectrum model was blinded with a modified molecular final-state distribution; and finally the full analysis was performed using two independent analysis techniques (covariance matrix and MC propagation) which revealed a high degree of consistency.

    We find excellent agreement of the calculated spectrum with the data. The covariance-matrix fit method obtains a goodness of fit of of $\chi^2 =$ \num{21.4} for \num{23} d.o.f. (corresponding to a p-value of \num{0.56}) and the MC-propagation technique finds a goodness-of-fit of $-2 \ln \mathcal{L} =$ \num{23.3} for \num{23} d.o.f. (corresponding to a p-value of \num{0.44}).

    The effective spectral endpoint \ezero{}, which is inferred from the spectral fit alongside \mnuesq, can be related to the nuclear Q-value using the molecular recoil and the offset between the source potential and spectrometer work function. Our analysis gives a Q-value of \SI{18575.2\pm0.5}{\electronvolt}, which is in excellent agreement with measurements based on the $^3$He-$^3$H atomic mass difference~\cite{Myers2015}. While the neutrino-mass result does not depend on the absolute energy scale of the spectrum, this consistency check is still of major importance to our understanding of the obtained spectra.

    The best fit of the squared neutrino mass was found at $\mnuesq = \asi{-1.0}{1.1}{0.9}{\electronvolt\squared}$. The uncertainty is largely dominated by the statistical error of $\sigma_{\text{stat}}(\mnuesq) =$ \SI{0.97}{\electronvolt\squared}.

    If one were to assume the true neutrino mass to be equal to zero, the probability of obtaining this fit result given our total error budget is \SI{16}{\percent}. The best-fit results of the covariance-matrix and MC-propagation techniques agree within \SI{2}{\percent}.

    We have applied three methodologies to derive an upper limit on the neutrino mass, based on the best-fit result. The Lokhov-Tkachov limit construction was developed in particular for direct neutrino-mass experiments~\cite{Lokhov:2015zna}. By construction, in the case of a negative best-fit value of \mnuesq{} it yields the experimental sensitivity as an upper limit. Based on this technique we find $m(\nu_e) <$ \SI{1.1}{\electronvolt} (\SI{90}{\percent} CL). The standard Feldman-Cousins technique for confidence-belt construction~\cite{Feldman:1997qc} yields an upper limit of \SI{0.8}{\electronvolt} (\SI{90}{\percent} CL). Finally, we also apply Bayesian inference methods to the neutrino-mass search, excluding negative values of \mnuesq{} through a flat, positive prior. The Bayesian result is presented in this work for the first time, yielding a \SI{90}{\percent} credibility interval of 0 to \SI{0.9}{\electronvolt}.

\begin{figure}[t!]
        \centering
		\includegraphics[width=0.45\textwidth]{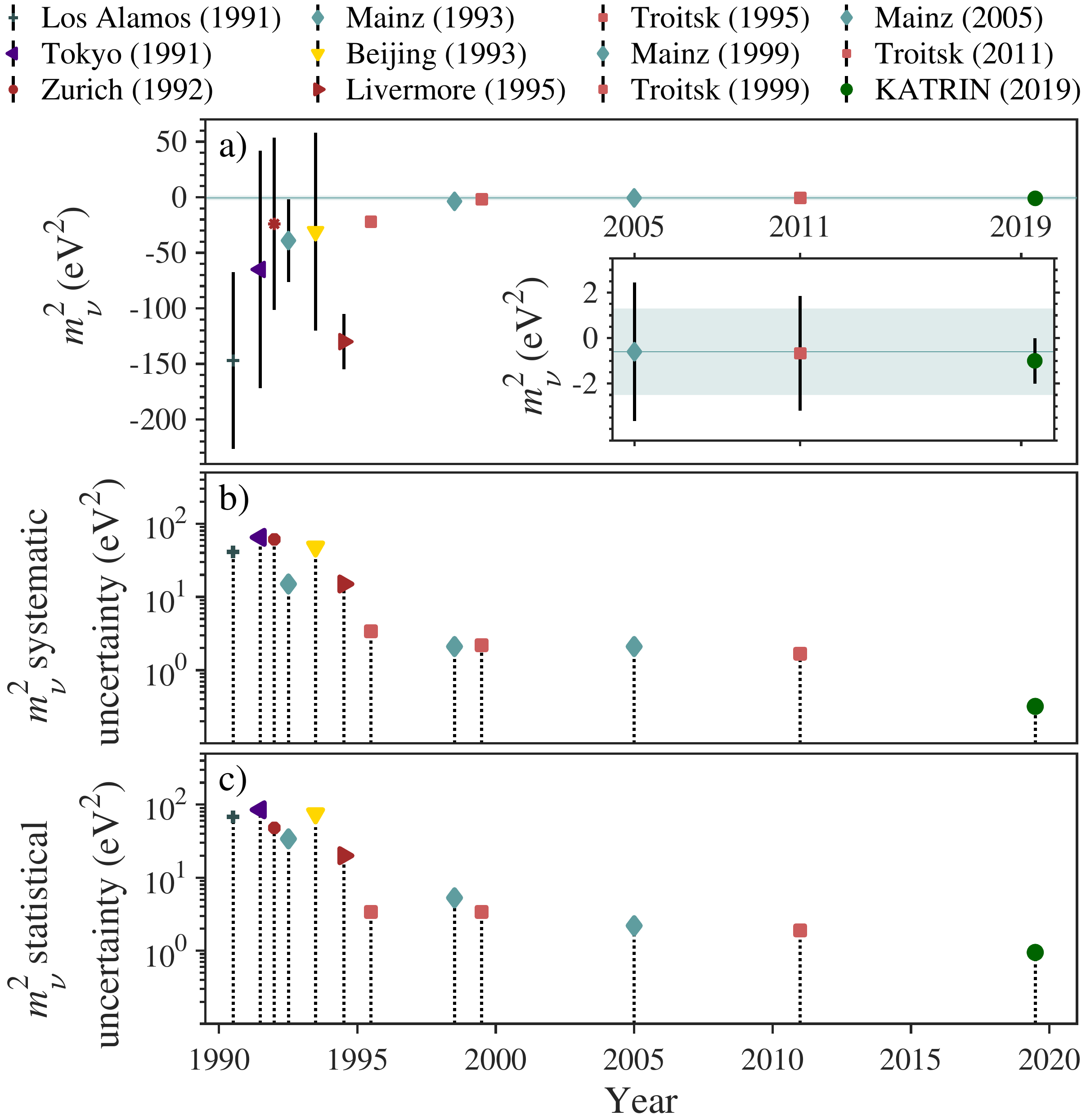}
        \caption{Overview of the neutrino-mass results obtained from tritium \bdecay{} in the period 1990--2019, plotted against the year of publication. (a) Squared-neutrino-mass results; the inset shows a more detailed comparison to results from the most recent experiments, Mainz and Troitsk. (b) Systematic uncertainties and (c) statistical uncertainties, both on the squared neutrino mass. The total uncertainty is reduced by a factor of three.
        The historical measurements plotted here are: Los Alamos (1991)~\cite{Robertson:1991vn}, Tokyo (1991)~\cite{Tokyo1991}, Zurich (1992)~\cite{Zurich1992}, Mainz (1993)~\cite{Mainz1993}, Beijing (1993)~\cite{beijing:1993}, Livermore (1995)~\cite{Livermore1995}, Troitsk (1995)~\cite{Troitsk1995}, Mainz (1999)~\cite{Mainz1999}, Troitsk (1999)~\cite{Troitsk1999}, Mainz (2005)~\cite{Kraus2005}, Troitsk (2011)~\cite{Aseev2011}.}
        \label{fig:historyNuMass-Combined}
    \end{figure}

    The newly obtained upper limit on the neutrino mass improves the previous best direct bounds by a factor of nearly two (Fig.~\ref{fig:historyNuMass-Combined}, top). The effective \SI{9}{days} of measurement time of this first neutrino-mass campaign (out of a total planned measurement time of 1000 days) led to an improvement of the statistical uncertainty on \mtwonue\ by a factor of  two compared to the final results of the Troitsk and Mainz experiments~\cite{Aseev2011, Kraus2005} (\cref{fig:historyNuMass-Combined}, bottom), while the systematic uncertainties are reduced by a factor of six (\cref{fig:historyNuMass-Combined}, center).

    The systematic error budget is expected to improve with future measurement campaigns. Most notably, new means to further suppress the background rate are now in place. These will increase the signal-to-background ratio and at the same time reduce the dominant systematic uncertainties related to the dependence of the background on time and retarding potential. Furthermore, in this first measurement the activity stability suffered from a burn-in phase, in which the structural material was exposed to tritiated gas for the first time. Subsequent to this first campaign, significant improvements of the activity stability have been demonstrated at an increased intensity about four times the KNM1 source strength. Finally, sub-dominant systematic effects, such as uncertainties in the final-state distribution, have been conservatively estimated for this analysis. Our knowledge of these systematics is expected to improve significantly in our future commissioning and measurement phases.

    \section{Conclusion}
    \label{sec:conclusion}
    The new upper limit $\mnue <$ \SI{1.1}{\electronvolt} (\SI{90}{\percent} CL) from KATRIN's first science run improves upon previous work~\cite{Kraus2005,Aseev2011} by almost a factor of two, based on a measuring period of only four weeks while operating at reduced column density -- equivalent to just \SI{9}{days} at nominal source strength.

    In the coming years, KATRIN will soon reach the first sub-\si{\electronvolt} sensitivity, and finally tackle its ultimate design sensitivity of \SI{0.2}{\electronvolt} (\SI{90}{\percent} CL). In addition, the precise measurement of the tritium spectrum allows searches for physics beyond the Standard Model, including right-handed weak currents \cite{Steinbrink:2017} and sterile-neutrino admixtures with masses from the \si{\electronvolt} \cite{Gariazzo:2015, Aker-KSN1:2020} to the \si{\kilo\electronvolt} scale \cite{Mertens:2014nha}.

    KATRIN's model-independent probe of the neutrino mass is of paramount importance for both particle physics and cosmology. In particle physics, this measurement narrows the allowed range of quasi-degenerate neutrino-mass models. In cosmology, it provides laboratory-based input for studies of structure evolution in $\Lambda$CDM and other cosmological models. In the absence of a definitive observation of dark matter, the neutrino-mass scale is  unique as a $\Lambda$CDM parameter that is directly observable in the laboratory.
    
    Upcoming cosmological probes are expected to achieve a determination of the sum of the neutrino masses over the next \num{5} to \num{10} years, making this laboratory measurement particularly important for obtaining a consistent picture of the neutrino as both particle and dark-matter constituent in the universe. This first KATRIN result serves as a milestone towards this goal.

\begin{acknowledgments}
We acknowledge the support of Helmholtz Association, Ministry for Education and Research BMBF (5A17PDA, 05A17PM3, 05A17PX3, 05A17VK2, and 05A17WO3), Helmholtz Alliance for Astroparticle Physics (HAP),  Helmholtz Young Investigator Group (VH-NG-1055), Max Planck Research Group (MaxPlanck@TUM), and Deutsche Forschungsgemeinschaft DFG (Research Training Groups GRK 1694 and GRK 2149, Graduate School GSC 1085 - KSETA, and SFB-1258) in Germany; Ministry of Education, Youth and Sport (CANAM-LM2015056, LTT19005) in the Czech Republic; Ministry of Science and Higher Education of the Russian Federation under contract 075-15-2020-778; and the United States Department of Energy through grants  DE-FG02-97ER41020, DE-FG02-94ER40818, DE-SC0004036, DE-FG02-97ER41033, DE-FG02-97ER41041, DE-AC02-05CH11231, DE-SC0011091, and DE-SC0019304, and the National Energy Research Scientific Computing Center. This project has also received funding from the European Research Council (ERC) under the European Union Horizon 2020 research and innovation program (grant agreement No. 852845).
\end{acknowledgments}

    \bibliography{knm1prd}

\end{document}